%% file: Kr.tex
\documentclass[modern]{aastex62}
\newcommand{\kms}{\ensuremath{\rm{km\,s^{-1}}}}
\newcommand{\fstar}{\ensuremath{F_*}}
\usepackage[utf8]{inputenc}
\usepackage{epsfig}
\shorttitle{ISM Depletions of O, Ge and Kr}
\shortauthors{Jenkins}
\begin{document}
\title{A Closer Look at Some Gas-Phase Depletions in the ISM:\\
Trends for O, Ge and Kr vs. $F_*$, f(H$_2$), and Starlight Intensity\footnote{Based on 
observations with the NASA/ESA Hubble Space Telescope obtained from the Data Archive 
at the Space Telescope Science Institute, which is operated by the Associations of 
Universities for Research in Astronomy, Incorporated, under NASA contract NAS5-26555.~ 
\copyright 2019. The American Astronomical Society. All rights reserved.}}
\author[0000-0003-1892-4423]{Edward B. Jenkins}
\affiliation{Princeton University Observatory\\
Princeton, NJ 08544-1001}
\email{ebj@astro.princeton.edu}
\begin{abstract}
In a survey of archived ultraviolet spectra of 100 stars recorded by the echelle 
spectrograph of the Space Telescope Imaging Spectrograph (STIS) on the Hubble Space 
Telescope (HST), we measure the strengths of the weak absorption features of O~I, Ge~II 
and Kr~I in the interstellar medium.  Our objective is to undertake an investigation that 
goes beyond earlier abundance studies to see how these elements are influenced 
independently by three different environmental properties: (1) values of a generalized 
atomic depletion factor \fstar\ due to condensations onto dust grains (revealed here by the 
abundances of Mg and Mn relative to H), (2) the fraction of H atoms in the form of H$_2$ 
$f({\rm H}_2)$, and (3) the ambient intensity $I$ of ultraviolet starlight relative to an 
average value in our part of the Galaxy $I_0$.  As expected, the gas-phase abundances of all 
three elements exhibit negative partial correlations with \fstar.  The abundances of free O 
atoms show significant positive partial correlations with $\log f({\rm H}_2)$ and $\log 
(I/I_0)$, while Ge and Kr exhibit negative partial correlations with $\log (I/I_0)$ at 
marginal levels of significance.  After correcting for these trends, the abundances of O 
relative to H show no significant variations with location, except for the already-known 
radial gradient of light-element abundances in the Milky Way.  A comparison of Ge and O 
abundances revealed no significant regional enhancements or deficiencies of
neutron-capture elements relative to $\alpha$-process ones.
\end{abstract}
\section{Background}\label{sec: background}

From studies of interstellar absorption lines in the UV spectra of stars in our region of the 
Galaxy, it is well established that interstellar dust sequesters into solid form much of the 
available interstellar gas atoms for elements heavier than helium, and these atomic 
depletions, some of which are profound for some elements, offer additional insights on the 
elemental composition and relative amounts of the dust (Savage \& Sembach 1996).  In 
broadest terms, we know that the severity of element depletions is related to two factors: 
(1) elements that can form highly stable, refractory compounds experience the strongest 
depletions, and (2) depletions appear to increase for regions that have high gas densities, 
indicating that the growth and destruction of grains depend on the local gas environments. 

To understand better variations in the amounts of dust grains and their composition, we 
consider a logarithmic depletion factor for any element $X$ from the gas phase, 
\begin{equation}\label{eqn: depl_def}
[X_{\rm gas}/{\rm H}]=\log\left({N(X)\over N({\rm H~I})+2N({\rm H}_2)}\right)_{\rm 
obs} - \log\left({X\over {\rm H}}\right)_{\rm ref}~,
\end{equation}
where $N(X)$ is the column density of the preferred ionization stage of element $X$, and 
the reference abundance ratio $(X/{\rm H})_{\rm ref}$ can apply to either B-type stars 
(Nieva \& Przybilla 2012) or the Sun (Asplund et al. 2009).  

In a study aimed at improving our understanding of such depletions, Jenkins (2009, 
hereafter J09) devised a unified interpretation for 17 different elements reported in more 
than 100 papers that covered 243 sight lines to different stars.  His interpretation made use 
of an empirically determined principle that, from one region to the next, the logarithmic 
strengths of depletions of different elements followed one another in a linear fashion, and 
this behavior could be characterized with good accuracy in terms of a few simple 
coefficients.  First, the overall severity of depletions for any sight line could be collectively 
characterized by a single scale factor, which he designated as \fstar.  According to this 
construction, any element $X$ responds to changes in \fstar\ in a manner that could be 
described by three coefficients unique to this element through the equation
\begin{equation}\label{eqn: depl_construction}
[X_{\rm gas}/{\rm H}]=B_X+A_X(\fstar-z_X)~,
\end{equation}
where the constants $B_X$, $A_X$, and $z_X$ are unique to each element $X$.\footnote{The 
offset quantity $z_X$ may seem superfluous for a linear equation that needs only two 
coefficients, but its use is intended to make the covariances in the errors for $A_X$ and 
$B_X$ equal to zero, which then simplifies the derivations of uncertainties of any 
relationships that make use of $A_X$ and $B_X$.}  It follows that if all of the missing atoms 
are sequestered into solid form (or free molecules), their abundances relative to hydrogen 
are given by
\begin{equation}\label{eqn: dust_abundance}
(X_{\rm dust}/{\rm H})=(X/{\rm H})_{\rm ref}(1-10^{[X_{\rm gas}/{\rm H}]})~.
\end{equation}

Equations~\ref{eqn: depl_construction} and \ref{eqn: dust_abundance} assign absolute  
values for the depletions and dust abundances, but they rely on the reference abundances 
$\log ({X/ {\rm H}})_{\rm ref}$ in Eq.~\ref{eqn: depl_def} being correct.  One can instead 
focus on differential changes as atoms accumulate in (or depart from) solid forms using the 
relationship
\begin{eqnarray}\label{eqn: differential_grain_comp}
d(X_{\rm dust}/{\rm H})/d\fstar&=&-(\ln 10)(X/{\rm H})_{\rm ref} 
A_X10^{B_X+A_X(\fstar-z_X)}\nonumber\\
&=&-(\ln 10)A_X(X_{\rm gas}/{\rm H})_{\fstar}~,
\end{eqnarray}
which depends only on the how rapidly the depletions change with \fstar\ (through the 
slope coefficient $A_X$) and the measured interstellar medium (ISM) abundance relative to 
hydrogen at a given value of \fstar.

The present study is motivated by two puzzling conclusions that emerged from the 
depletion analysis reported in J09.  The depletions of oxygen and krypton appeared to 
exceed what one might have expected, as will be outlined in the following subsections. We 
now hope to gain new insights by enlarging the sample of sight lines and studying how the 
abundances of O and Kr relate not only to \fstar\ but also to other interstellar gas 
parameters such as the molecular hydrogen fraction $f({\rm H}_2)=2N({\rm H}_2)/[ 
2N({\rm H}_2)+N({\rm H~I})]$ and the intensity of starlight in the ultraviolet.  Another 
quantity that may seem relevant is the average volume density of hydrogen along a sight 
line $\langle n_{\rm H}\rangle=N({\rm H})/d$, where $d$ is the distance to the star.  Many 
studies have shown that $\langle n_{\rm H}\rangle$ correlates strongly with the strengths 
of depletions for those elements that are strongly depleted (Savage \& Bohlin 1979 ; Harris 
et al. 1984 ; Murray et al. 1984 ; Gondhalekar 1985 ; Jenkins et al. 1986 ; Jenkins 1987 ; 
Welsh et al. 1997 ; Snow et al. 2002 ; Cartledge et al. 2004, 2006 ; Jensen \& Snow 2007b, 
2007a).  We are not surprised to find that the quantities \fstar\ and $\langle n_{\rm 
H}\rangle$ are strongly correlated with each other (see Fig.~16 of J09).  In making a choice 
between these two parameters, we regard \fstar\ to be a more direct indicator of the 
maturity of the depletion process in any given sight line.  

\subsection{The Problem with Oxygen}\label{sec: oxygen_problem}

Cartledge et al. (2004) found that [${\rm O_{gas}/H}$] showed a weak but convincing 
downward trend with increasing values of $\langle n({\rm H})\rangle$.  From a qualitative 
perspective, this is not surprising since $\langle n({\rm H})\rangle$ is strongly correlated 
with \fstar, and the relative concentrations of oxygen-bearing compounds such as silicates 
and oxides should increase with greater values of \fstar.  However, as recognized by J09, 
quantitatively speaking the depletion of gas-phase oxygen is surprisingly strong.  For 
instance, for a representative sight line with strong depletions (at $\fstar=1$), we can use 
the coefficients in J09 and Eq.~\ref{eqn: dust_abundance} to arrive at a value $({\rm 
O_{dust}/H})=241$ parts per million (ppm) if $({\rm O/H})_{\rm ref}=575\,$ppm is taken 
from B-star abundances (Nieva \& Przybilla 2012).\footnote{The B-star reference 
abundance adopted here happens to be the same as the one adopted by J09, which was 
based on the present-day solar photospheric abundance $\log ({\rm 
O/H})_\odot+12=8.69$ recommended by Lodders (2003) plus her +0.07~dex correction 
for gravitational settling to arrive at a protosolar abundance.  If such a correction were 
applied to some more recent measurements of the present-day solar photospheric O 
abundance of approximately 8.77 by Ayres et al. (2013) and Steffen et al. (2015), $B_{\rm 
O}$ would revert to a value of $-0.225$, $({\rm O/H)_{ref}}=690\,{\rm ppm}$, and then an 
application of Eqs.~\protect\ref{eqn: depl_construction} and \protect\ref{eqn: 
dust_abundance} for $\fstar=1$ would yield $({\rm O_{dust}/H})=356\,{\rm ppm}$ .}   
This value for $({\rm O_{dust}/H})$ is substantially larger than 170\,ppm taken from the 
corresponding sum of the dust abundances of 113\,ppm for Mg, Si, and Fe multiplied by the 
largest plausible ratio of 1.5 for O for the most favorable combination of silicate and oxide 
compounds, MgSiO$_3$ plus Fe$_2$O$_3$ (Cardelli et al. 1996 ; Whittet 2010).  The 
disparity between $({\rm O_{dust}/H})=241$\,ppm and that consumed by silicates and 
oxides has created a challenge for explaining where some of the oxygen atoms are 
sequestered (Whittet 2010 ; Wang et al. 2015).

A contrasting view was presented by Voshchinnikov \& Henning (2010).  They claimed that 
a large fraction of their determinations of $({\rm O_{dust}/H})$  came out to be 
considerably lower than 170\,ppm, and they argued that there was no problem in 
accounting for all of the O being incorporated into silicates.  However, they adopted a value 
$({\rm O/H})_{\rm ref}=490\,$ppm, and this reduced value affects both of the major terms 
on the right-hand side of Eq.~\ref{eqn: dust_abundance} to create a prediction for a 
removal of only 155\,ppm with the J09 coefficients at $\fstar=1$.

If one distrusts the reference abundances, one can resort to comparing the differential 
depletions of O and Si+Mg+Fe through the use of Eq.~\ref{eqn: differential_grain_comp}.  
At $\fstar=0$ the quantity $d({\rm O_{dust}/H})/d\fstar$ divided by $d({\rm 
Mg_{dust}+Si_{dust}+Fe_{dust}/H})/d\fstar$ equals 2.30, which to within the uncertainties 
is acceptable for O being incorporated into silicates and oxides.  However, the ratio 
increases to 16 at $\fstar=1$, which indicates that in dense regions with strong depletions 
O must bind to some other element that has a high abundance (or to itself in the form of 
O$_2$).  Clues on what processes may facilitate the removal of gas-phase O could emerge 
from a more comprehensive comparison of oxygen abundances with factors other than just 
\fstar.

\subsection{The Problem with Krypton}\label{sec: krypton_problem}

Superficially, one might expect krypton to be an element that is unlikely to show any 
depletions.  It is a noble gas that is chemically inert because its outer valence shell of 
electrons is filled.  Its van der Waals binding with neutral systems is extremely weak and 
easily disrupted.  From these two perspectives, it may seem puzzling that the strengths of 
absorption features of this element’s dominant ionization state, Kr~I, indicated an 
apparent, almost universal deficiency ($\sim 0.3$\,dex) of gas-phase Kr
 (Cardelli \& Meyer 1997 ; Cartledge et al. 2003, 2008 ; Ritchey et al. 2018) 
when compared to its reference abundance relative to hydrogen.  A link between Kr 
depletions and the relative concentrations of solid materials in the ISM was established 
later by J09, who found that Kr depletions appeared to become more severe with 
increasing values of \fstar\ (i.e., $A_{\rm Kr}=-0.166$, but with an uncertainty of 0.103).  
The statistical significance of this trend was improved in the more recent investigation by 
Ritchey et al. (2018) in their study of the interstellar abundances of
$r$-process elements.  

In a related development, there evolved an awareness that an acid-resistant residue of 
some primitive meteorites, known as phase Q,\footnote{Q stands for quintessence, a 
designation originated by Lewis et al. (1975).  This phase typically composes less than 
0.04\% of the mass of a meteorite.} was shown to have significant concentrations of noble 
gases, with fractionations favoring the retention of heavier elements (Schelhaas et al. 1990 
; Amari et al. 2013) .   Laboratory experiments designed to explore this issue indicated that 
noble gases can be made to bind to certain compounds found in meteoritic materials 
(Amberg et al. 1955 ; Yang \& Anders 1982a, 1982b ; Yang et al. 1982 ; Wacker 1989 ; 
Marrocchi et al. 2005).

There has also evolved a recognition that free noble gas atoms could bind to positive ions 
and charged molecular complexes (Holloway 1968 ; Wyatt et al. 1975).  Within the contexts 
of the ISM and protoplanetary disks, the most prominent possibilities are the couplings 
with the partners H$^+$, H$_2^+$, and H$_3^+$ (Pauzat \& Ellinger 2007 ; Pauzat et al. 
2009 ; Theis et al. 2015).  Indeed, emission and absorption features arising from 
interstellar ArH$^+$ have been detected in spectra recorded by instruments on the {\it 
Herschel\/} mission (Barlow et al. 2013 ; Schilke et al. 2014).  Nevertheless, the 
concentrations of ArH$^+$ are small compared to the total abundances of Ar.  Dissociative 
recombinations with free electrons are likely to strongly inhibit any accumulations of these 
complexes by amounts that could cause significant depletions of the atomic forms of noble 
gases.  Electron fractions  $n(e)/N({\rm H_{tot}})\approx 3\times 10^{-4}$ and typical 
destruction reaction rate constants of order $10^{-7}{\rm cm}^3{\rm s}^{-1}$ (Mitchell 
1990)\footnote{One exception is the rate for ${\rm ArH}^++e\rightarrow {\rm Ar+H}$, 
which is $< 10^{-9}{\rm cm}^3{\rm s}^{-1}$ (Mitchell et al. 2005).}  make it doubtful that 
appreciable amounts of Kr can be bound in charged molecular complexes. 

Two other considerations may be factored into the krypton abundance findings.  One is 
that there might exist a real change in the overall abundance of Kr from one region to 
another, as proposed by Cartledge et al. (2008).  The other possibility is that if the Kr and H 
are exposed to ionizing radiation with a sufficiently high energy to penetrate the mostly 
neutral hydrogen, the Kr could be more strongly ionized than H because its ionization cross 
section just above its threshold energy (14\,eV) is much higher than that of H  by about one 
order of magnitude (Sterling 2011).  As a result, the apparent $({\rm Kr_{gas}/H})$  would 
be lowered, much as what one sees with the ratio of Ar~I to O~I in the low-density, partly 
ionized medium (Jenkins 2013).  

\section{Investigation Strategy}\label{sec: strategy}
\subsection{Basic Design}\label{sec: design}

Ultraviolet spectra of stars observed using the echelle spectrographs of the Goddard High 
Resolution Spectrograph (GHRS) and the Space Telescope Imaging Spectrograph (STIS) on 
the {\it Hubble Space Telescope\/} (HST) have enabled observations for the investigation 
of interstellar atomic abundances, including those of O and Kr.  There have been many 
publications of results for these two elements, and references to these works are listed in 
Table~1 of J09.\footnote{More recent measurements have been carried out by Ritchey et 
al. (2018).}  Since the time of the latest of these publications, many new spectra have been 
recorded using the STIS echelle spectrograph in the medium- and high-resolution modes.   
The new spectra were acquired after the repair of STIS during the Fourth Servicing Mission 
(SM4) for {\it HST\/} in 2009, and they allow us to increase the number of samples.   So 
that we can maintain a uniform approach for interpreting the absorption features, none of 
the previously reported measurements of column densities are incorporated in the present 
study; all of the outcomes given here are based on an original analysis using a set of rules 
applied in a consistent manner throughout all cases.

A particularly large volume of the STIS echelle spectra cover the wavelength region from 
1170 to 1372\,\AA, which includes transitions of O~I, Mg~II, Mn~II, Kr~I, and Ge~II.  
These spectra are available from the {\it Mikulsky Archive for Space Telescopes\/} 
(MAST).  Except for Kr, whose absorption is securely detected in only about half of the sight 
lines, the transitions for these elements are strong enough to measure with reasonable 
accuracy, but not so strong that saturation effects seriously compromise the derivations of 
column densities. The strongly depleted elements Mg and Mn allow us to determine \fstar\ 
after their abundances are compared to those of atomic and molecular hydrogen.  While O 
and Kr are the main focus of this investigation, we also consider abundances of Ge to check 
on whether or not enhancements or deficiencies of this element might signify variations in 
the total abundances of neutron capture elements from one location to the next, as 
suggested by Kr data analyzed by Cartledge et al. (2008).  Ge is also a mildly depleted 
element whose correlation behaviors can be instructive when comparing them with those 
of O and Kr.

The remainder of this section and the one that follows describe how target stars and atomic 
transitions were selected and how the absorption features were analyzed.  Tables~\ref{tbl: 
element_outcomes} and \ref{tbl: H_outcomes} in Section~\ref{sec: column densities} 
present the column density outcomes, and the more detailed measurement results appear 
in Table~\ref{tbl: basic_measurements} in the appendix of this paper.  Sections~\ref{sec: 
derivations_fstar}  and \ref{sec: starlight_intens}  describe the derivations of \fstar\ and 
starlight intensity parameters, respectively.  Section~\ref{sec: depletion_trends}   
describes some analyses of the depletion trends, followed by
Section~\ref{sec: interpretation}  that interprets them.
Finally, Section~\ref{sec: regional_variations} describes two investigations that attempted 
to reveal regional changes in abundances after correcting for the local environmental 
conditions that we identified earlier. 

\subsection{Selection of Target Stars}\label{sec: targets}

The first step to identify candidate spectral exposures was to collect all STIS E140H and 
E140M observations of stars listed as of 2016 in the Planned and Archived Exposures 
Catalog (PAEC) maintained by the Space Telescope Science Institute (STScI).  By examining 
the quick-look preview displays in MAST and checking spectral types in the Simbad 
database, useful observations were identified that satisfied all of the following criteria:
\begin{enumerate}
\item The spectrum did not suffer from confusion arising from narrow stellar lines (E140M 
spectra are less tolerant for such confusion than those taken with E140H).  The spectrum 
must also have had a respectable signal-to-noise ratio.  (17 stars had no preview spectra 
available, so they were excluded in the selection process.)
\item The strengths of the two Mg~II lines (see Fig.~\ref{fig: HD170740}) are used to 
indicate that there is a sufficient amount of interstellar gas to make a meaningful 
determination of the fractional abundance of Kr.  We do not exclude cases where the Kr~I 
lines themselves are too weak to see, since the resulting upper limits for $N$(Kr~I) could 
signify a large and interesting deficiency.
\item Stars of spectral type later than B3 were excluded, since the strong stellar 
Ly$\alpha$ feature prevents one from determining $N$(H~I). Initially, stars with 
temperatures corresponding to the B2 and B3 were deemed acceptable if their E(B$-$V) 
values were of order or greater than 0.2, since the interstellar H~I absorption could 
dominate over the stellar feature.  In later stages of selection, stars that had corrections for 
stellar features (described in Section~\ref{sec: H}) that were larger than the apparent 
value of $\log N({\rm H~I})-0.5$ were excluded.  Stars B1 and earlier are acceptable, but 
some luminous stars with strong N~V P~Cygni absorptions depressed the flux at 
1236\,\AA\ to too low a level to see any Kr~I absorption.
\item The star must have had a spectrum covering the region containing H$_2$ lines (at 
around 1100\AA\ and shortward) available from the MAST archive of observations by the 
{\it Far Ultraviolet Spectroscopic Explorer\/} (FUSE).  However, a few exceptions are 
noted in Table~\ref{tbl: basic_measurements}.
\end{enumerate}

The properties of the stars considered in this survey are summarized in Table~\ref{tbl: 
stellar_data}.  Except for the entries in Columns~10 and 11, which will be explained later, 
the meaning of the listed quantities should be self-evident.  Stars with spectra in the 
archive that survived the initial screening but that were later rejected are listed in 
Table~\ref{tbl: rejected_stars}.  Most of these rejections were based on our inability to 
obtain satisfactory corrections for the effects of the stellar Ly$\alpha$ absorption features. 
\vspace{3cm} 
\startlongtable
\begin{deluxetable}{
r	
c	
c	
c	
c	
c	
c	
c	
c	
c	
c	
l	
}
\tabletypesize{\scriptsize}
\tablewidth{0pt}
\tablecaption{Stellar and Interstellar Medium Data\label{tbl: stellar_data}}
\tablecolumns{12}
\tablehead{
\colhead{Target} & \multicolumn{2}{c}{Galactic Coordinates} & \colhead{$d$} & 
\colhead{Source\tablenotemark{a}} & \colhead{$z$} & \multicolumn{2}{c}{Magnitudes} 
&
 \colhead{$E(B-V)$} &\colhead{Depletion} & \colhead{Starlight} & \colhead{Spectral} \\
\cline{2-3} \cline{7-8}
\colhead{Star} & \colhead{$\ell$} & \colhead{$b$} & \colhead{(kpc)} &
\colhead{for $d$} & \colhead{(kpc)} & \colhead{$B$} & \colhead{$V$} & 
\colhead{(mag)} & \colhead{Strength \fstar\tablenotemark{b}} & 
\colhead{$I/I_0$\tablenotemark{c}} & \colhead{Type} \\
\colhead{(1)} &
\colhead{(2)} &
\colhead{(3)} &
\colhead{(4)} &
\colhead{(5)} &
\colhead{(6)} &
\colhead{(7)} &
\colhead{(8)} &
\colhead{(9)} &
\colhead{(10)} &
\colhead{(11)} &
\colhead{(12)}\\
}
\startdata
\input{stellar_data}
\enddata
\tablenotetext{a}{(1) Bowen et al. (2008), (2) Jenkins (2009), (3) Savage et al. (2017), (4) 
this paper.}
\tablenotetext{b}{See Section \ref{sec: derivations_fstar}.}
\tablenotetext{c}{Logarithm of the intensity of starlight capable of ionizing neutral carbon 
for the foreground gas, relative to the Galactic average,
as computed by Jenkins \& Tripp (2011).  See Section~\ref{sec: starlight_intens}}
\end{deluxetable}
\begin{deluxetable}{
r	
l	
}[b]
\tabletypesize{\normalsize}
\tablewidth{0pt}
\tablecaption{Rejected Stars\label{tbl: rejected_stars}}
\tablehead{
\colhead{Star} & \colhead{Reason}
}
\startdata
BD+48~3437&Too many lines were difficult to measure.\\
HD23478&B3, $\log N({\rm H~I)_{stellar}}> \log N({\rm H~I)_{meas.}}-0.5$\\
HD24190&B2~V and no photometric data available\tablenotemark{a}\\
HD27778&B3V; $\log N({\rm H~I)_{stellar}}> \log N({\rm H~I)_{meas.}}-0.5$\\
HD62542&B5V, but see note below\tablenotemark{b}\\
HD72754& Too many lines were difficult to measure.\\
HD102065&Enormous stellar Ly$\alpha$ absorption, even though listed as a B2V star\\
HD114441&B2~V and no photometric data available\tablenotemark{a}\\
HD117111&B2~V and no photometric data available\tablenotemark{a}\\
HD153262&B0/3, so photometric data probably not useful\tablenotemark{a}\\
HD203532&B3~IV; $\log N({\rm H~I)_{stellar}}> \log N({\rm H~I)_{meas.}}-0.5$\\
\enddata
\tablenotetext{a}{Photometric data are needed to compute $N({\rm H~I_{stellar}})$ using 
the method of Diplas \& Savage (1994).}
\tablenotetext{b}{This case is unusual: Mg and Mn are very weak and hard to measure 
against bad stellar lines, but Ge and Kr show up as narrow and very believable
absorptions.  Abundances of Ge and Kr must be high relative to other elements, and their 
measurement outcomes are listed in Table~\protect\ref{tbl: basic_measurements}.
This sightline seems to be noteworthy, even though we could not include it in the general 
statistical analysis.}
\end{deluxetable}
\clearpage
\subsection{Choices for Heavy Element Transitions}\label{sec: transitions}

\begin{figure}[t]
\plotone{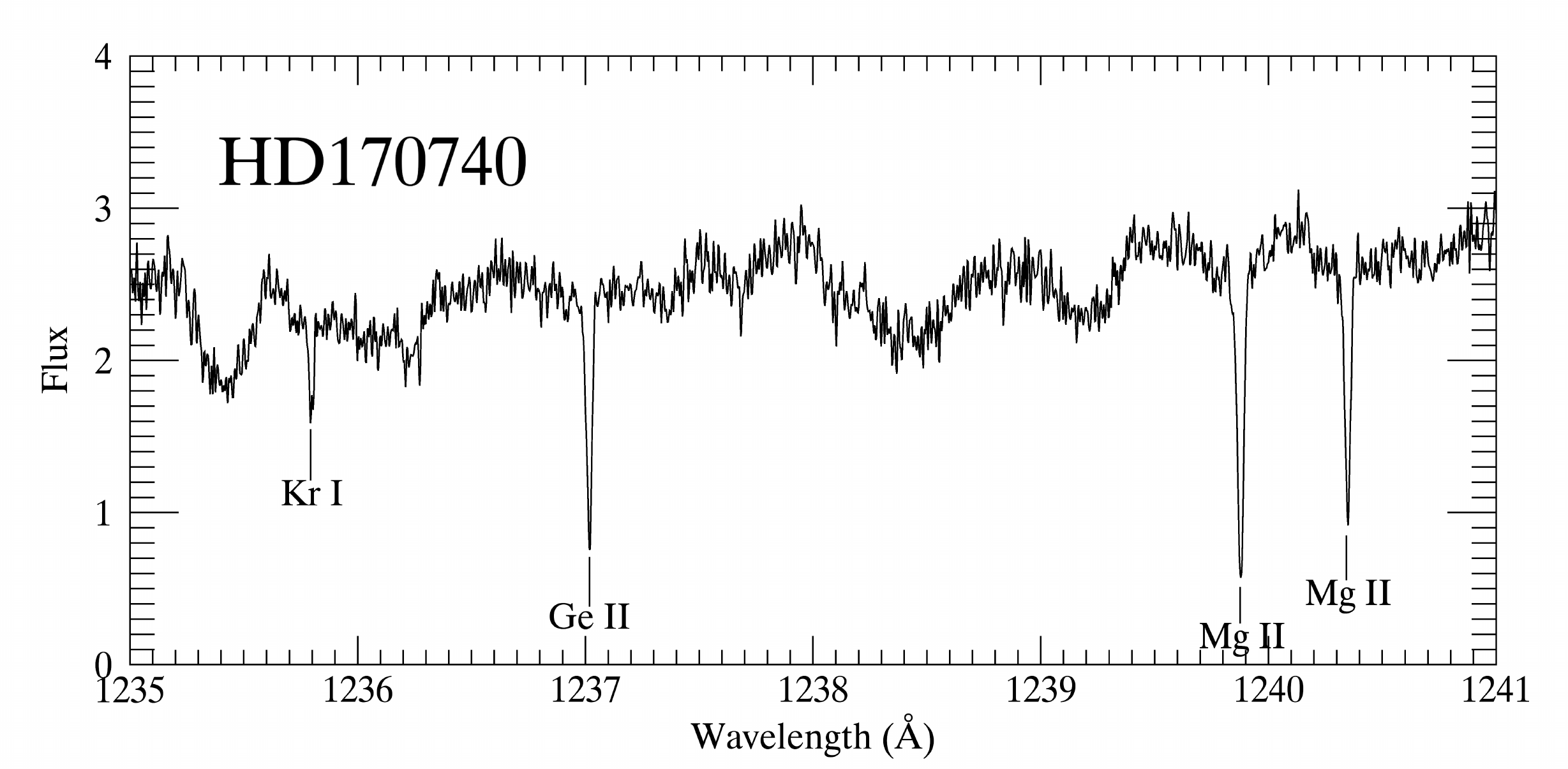}
\caption{A STIS E140H exposure of the spectral region covering the transitions of Kr~I, 
Ge~II, and Mg~II in the spectrum of HD\,170740.\label{fig: HD170740}}
\end{figure}

Table~\ref{tbl: transitions} lists the transitions, their wavelengths, and values of $\log 
f\lambda$ that were used to derive the column densities of the elements under study.  For 
Mn, Kr, and Ge, Cashman et al. (2017) have offered updated $f$-values as suggested 
replacements to those listed in the very popular compilations of
Morton (2000, 2003).  Since the trends reported in J09 had all column densities modified 
so that they were consistent with the Morton listings, Column~5 of the table states the 
difference (${\rm revised}~\log f\lambda-{\rm Morton’s}~\log f\lambda$).  The original 
sources for the $f$-values are listed in Column~4 of the table [except for those given in 
Morton (2003)].  Figure~\ref{fig: HD170740} illustrates the appearance of the Kr, Ge, and 
Mg lines in the spectrum of one star in our survey, HD\,170740.
\begin{deluxetable}{
l	
c	
c	
l	
c	
}
\tablewidth{0pt}
\tablecaption{Atomic Transitions\label{tbl: transitions}}
\tablehead{
\colhead{Atom and} & \colhead{$\lambda$} & \colhead{$\log f\lambda$} &
\colhead{Source of the} & \colhead{Revision in the}\\
\colhead{Ionization State\tablenotemark{a}} & \colhead{(\AA)} & \colhead{} & 
\colhead{$f$-value} & \colhead{Value of $\log f\lambda$\tablenotemark{b}}\\
\colhead{(1)} &
\colhead{(2)} &
\colhead{(3)} &
\colhead{(4)} &
\colhead{(5)}\\
}
\startdata 
O~I\dotfill &1355.598\tablenotemark{c}&-2.805&(Morton 2003)&\nodata\\ 
Mg~II~1\dotfill &1239.925&-0.106&(Morton 2003)&\nodata\\
Mg~II~2\dotfill &1240.395&-0.355\\
Mn~II~1\dotfill &1197.184&2.248&(Toner \& Hibbert 2005)&$-0.166$\\
Mn~II~2\dotfill &1201.118&1.999&&$-0.164$\\ 
Ge~II\dotfill &1237.059&3.033&(Heidarian et al. 2017)&$-0.150$\\
Kr~I\dotfill &1235.838&2.422&(Chan et al. 1992)&+0.020\\
\enddata
\tablenotetext{a}{The Arabic numerals that follow some of the elements link the 
transitions to their measurements reported in Table~\protect\ref{tbl: 
basic_measurements}.}
\tablenotetext{b}{Adopted value of $\log f\lambda$ minus the value reported in Morton 
(2000) or Morton (2003).}
\tablenotetext{c}{Four stars had wavelength coverages that did not include this 
transition.}
\end{deluxetable}
\clearpage

\section{Derivations of Column Densities}\label{sec: column densities}
\subsection{Heavy Elements}\label{sec: heavy elements}

\begin{figure}
\plotone{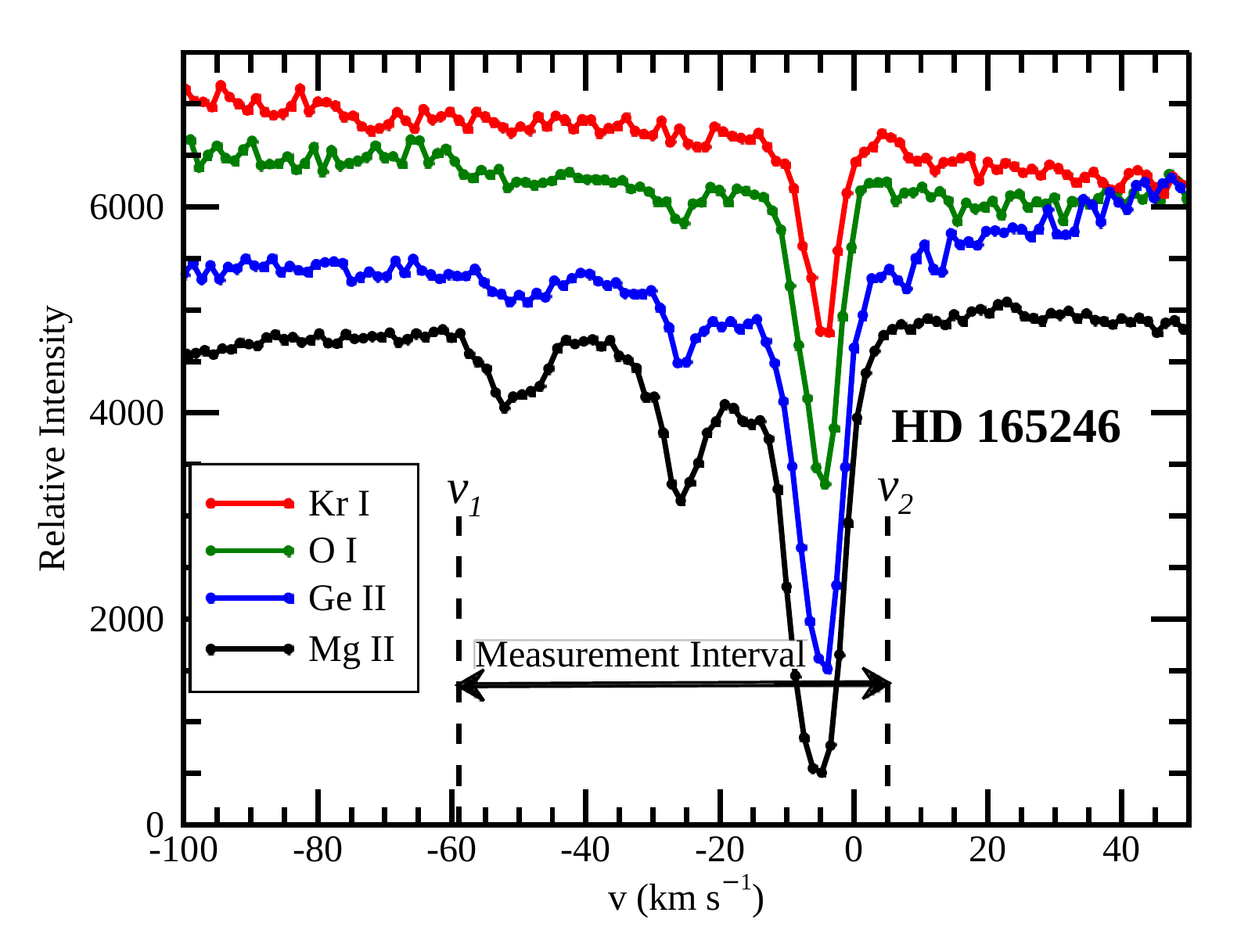}
\caption{Absorption features in the E140H spectrum of HD\,165246 for Kr~I, O~I, Ge~II, 
and Mg~II, which follow a sequence of increasing profile strengths.  The dashed lines 
depict the velocity end points, $v_1=-59\,\kms$ and $v_2=+5\,\kms$ of the Mg~II 
absorption. \label{fig: HD165246}}
\end{figure}

Prior to measuring the absorption features, velocity limits had to be defined.  Unlike the 
clear, sharp absorptions that appear in Figure~\ref{fig: HD170740} for HD\,170740, many 
other stars exhibited additional weak absorptions over large spans in velocity for the 
strongest feature of Mg~II, as illustrated in Figure~\ref{fig: HD165246}.   Without 
reference to the Mg~II profile, one would be inclined to measure the much weaker 
absorptions from other species over narrower velocity intervals.  For instance, evaluations 
of Kr~I and O~I for HD\,165246 might have included only the velocity interval from $-13$ 
to $+2\,\kms$, which would have missed very weak portions of the features that were 
buried in the noise and not apparent to the eye.  For this reason, we adopted a strict policy 
of defining for all species the velocity endpoints $v_1$ and $v_2$ based on the Mg~II 
feature at 1239.925\,\AA, regardless of whether or not the weaker lines showed 
absorptions at the most extreme velocities.   Not only does this tactic prevent us from 
overlooking some absorption, but it also prevents a subtle downward bias in the definition 
of a continuum level (i.e., in some cases a localized depression of intensity in a region 
surrounding an obvious absorption feature could be mistakenly identified as a broad stellar 
absorption feature).   Finally, it is important that we ensure an equitable pairing of the 
heavy elements to H~I and H$_2$, whose measures are independent of velocity.  Overall, 
this rigid definition of the velocity endpoints usually weakens the significance of a 
determination that might otherwise seem by eye to be more confined and easy to measure, 
but we consider it to be a justifiable price to pay for avoiding a bias in the measurements.  
The velocity endpoints $v_1$ and $v_2$ are specified in the header for each target star in 
Table~\ref{tbl: basic_measurements}.

For each absorption feature, a column density was derived by integrating the apparent 
optical depths (AODs) (Savage \& Sembach 1991) over the relevant velocity interval.  An 
apparent column density $N_a$ is then given by the relation
\begin{equation}\label{eqn: N_a}
N_a={m_ec\over \pi e^2f\lambda}\int_{v_1}^{v_2} \tau_a(v)dv~,
\end{equation}
where the AOD $\tau_a(v)= \ln[(I_0(v)/I(v)]$, $I_0(v)$ is the continuum intensity, and 
$I(v)$ is the intensity within the absorption feature.  The expression in front of the integral 
is equal to $3.77\times 10^{14}\,{\rm cm}^{-2}$ if $v$ is expressed in \kms and 
$f\lambda$ is in units of \AA\ (as shown in logarithmic form in Table~\ref{tbl: 
transitions}). 

For each feature, a continuum level $I_0(v)$ was defined using a Legendre polynomial of 
the lowest order that would give a reasonable fit to the spectrum outside of the interval 
spanned by $v_1$ and $v_2$.  Continuum placement errors can have a large influence in the 
uncertainties of weak lines.  We evaluated the expected deviations produced by such errors 
by remeasuring the AODs at the lower and upper bounds for the continua, which were 
derived from the expected formal uncertainties in the polynomial coefficients of the fits as 
described by Sembach \& Savage (1992).  We multiplied these coefficient uncertainties by 
2 in order to make approximate allowances for additional deviations that might arise from 
some freedom in assigning the most appropriate order for the polynomial.  Final 
uncertainty estimates were defined in terms of uncertainties due to noise and those due to 
continuum fitting, both of which were combined in quadrature. Table~\ref{tbl: 
basic_measurements} lists for the different transitions both equivalent width 
measurements and values of $N_a$, along with their respective $1\sigma$ uncertainties.

Table~\ref{tbl: basic_measurements} reveals that there are many duplications in the 
column density derivations.  For some targets, we had access to both high- and
medium-resolution spectra (the E140H and E140M modes, respectively).  For most but not 
all observations we could measure both a strong and a weak line of Mg~II and Mn~II.  In 
making choices on which results to adopt for these two elements, the first level of priority 
was to favor the strong lines over the weaker ones, but with some exceptions described in 
the paragraph that follows.  A secondary consideration (applied to all elements) was that 
measurements taken from the high-resolution spectra were favored over the
medium-resolution ones.

In most instances, the column density outcomes for the two Mg~II and Mn~II lines agreed 
with each other to within their uncertainties.  However, occasionally the weaker line 
yielded a value that was significantly larger than the stronger line.  This condition indicates 
that some saturation of the strong line was occurring over a scale of velocity that was not 
resolved by the instrument, yielding an underestimate of the true opacity averaged over 
velocity.  An example of this condition is shown for the Mg~II doublet shown in 
Fig.~\ref{fig: HD170740} for the star HD\,170740, where both the equivalent widths and 
AODs near the line bottom fall short of the 1.77:1 ratio of transition strengths.  Here the 
AOD outcomes for the strong and weak line are $\log N({\rm Mg~II})=15.75\pm 0.02$ and 
$15.82\pm 0.03$, respectively.  Adopting only the column density of the weaker line is not 
sufficient to overcome the error caused by saturation because even $N_a$ for this line 
could fall short of the true value.  A good approximation for deriving the true column 
density is to correct for the unresolved saturations by subjecting the AOD values at each 
velocity to a standard curve-of-growth analysis, according to the usual convention of 
interpreting equivalent widths for absorption features that are very poorly resolved.  This 
approach is described and justified by Jenkins (1996).  An application of this method to the 
Mg~II lines of HD\,170740 yields $\log N({\rm Mg~II})=15.91\pm 0.04$, which is 
measurably greater than the weak-line result.  For all absorptions considered in this study 
none of the saturations were severe enough to invalidate this approach to correction.  One 
can sense which cases needed the saturation corrections by consulting Table~\ref{tbl: 
basic_measurements} and locating ion entries that have a ``3’’ following them.

We considered a measurement to be marginal if the equivalent width outcome was less 
than the $2\sigma$ level of uncertainty from noise and continuum placement.  For weak 
lines below this uncertainty threshold, we specified an upper limit for the column density 
based on a completely unsaturated line having a strength at the measurement value plus a 
$1\sigma$ positive excursion, but with an allowance for the fact that negative real line 
strengths are not allowed even though we occasionally obtained negative measurement 
outcomes caused by downward noise fluctuations (or a continuum placement that was too 
low).  Details of how we  calculated these $1\sigma$ upper limits are given in Appendix D 
of Bowen et al. (2008).  Such a calculation avoids the absurd conclusion that an upper limit 
for a column density can be nearly zero or negative when the measurement yields an 
outcome that is $\leq-1\sigma$.  It also yields a smooth transition to a conventional 
expression of a measured value plus its $1\sigma$ upper error bar when this value is 
slightly larger than the twice its uncertainty.  A substantial fraction of the Kr 
determinations (49 out of 100 stars) consisted of upper limits.  For O and Ge, there were 
only 11 and 3 such upper limits, respectively.

The adopted column densities for the five elements considered in this study are listed for 
each star in Table~\ref{tbl: element_outcomes}. 

\startlongtable
\begin{deluxetable}{
r	
c	
c	
c	
c	
c	
}
\tabletypesize{\scriptsize}
\tablewidth{0pt}
\tablecaption{Element Measurement Outcomes\tablenotemark{a}\label{tbl: 
element_outcomes}}
\tablehead{
\colhead{Star} & \colhead{$\log N({\rm O~I})$} & \colhead{$\log N({\rm Mg~II})$}
& \colhead{$\log N({\rm Mn~II})$} & \colhead{$\log N({\rm Ge~II})$}
& \colhead{$\log N({\rm Kr~I})$}\\
\colhead{(1)} &
\colhead{(2)} &
\colhead{(3)} &
\colhead{(4)} &
\colhead{(5)} &
\colhead{(6)}\\
}
\startdata
\input{elements_data}
\enddata
\tablenotetext{a}{Values of $N$ are expressed in units of ${\rm cm}^{-2}$.  All 
uncertainties in this table represent possible deviations at a level of $1\sigma$.} 
\end{deluxetable}

\subsection{Atomic and Molecular Hydrogen}\label{sec: H}

All spectra considered here cover the Ly$\alpha$ transition of H~I at 1215.67\AA.
The Ly$\alpha$ absorptions are so broad and well resolved that one may reconstruct how 
a spectrum would appear without this feature by multiplying it by $\exp(+\tau)$ for a line 
with damping wings, as has been done in the past by Jenkins (1971), Bohlin (1975), Savage 
et al. (1977), and Diplas \& Savage (1994), and then inspect which choice of $N({\rm 
H~I})$ yields the most reasonable-looking outcome.  Error limits are defined in terms of 
where the reconstructions appeared by eye to be unreasonable, i.e., showing deviations 
that produced appreciable emission or absorption on either side of the line core.  These 
limits are judged to be at about the $1.5\sigma$ confidence level and will be treated as 
such in later analyses.

For some of the stars with spectral types cooler than B1, one may expect that the stellar
Ly$\alpha$ absorption is strong enough to produce a spurious increase in the inferred 
value for $N({\rm H~I})$ in the foreground.  If this stellar contamination is not too large, a 
corrected value for the hydrogen column density can be derived by subtracting an effective 
$N({\rm H~I})$ based on a prediction for the strength of the stellar contribution.  J09 
described a means for making such corrections, which followed an original prescription of 
Diplas \& Savage (1994).  Cases where such stellar corrections were warranted are listed 
in Table~\ref{tbl: basic_measurements} and are identified with the designation ``H~I~sc’’. 
The result for the original profile fit is shown just above it (without the ``sc’’).

For determinations of the molecular hydrogen abundances, we relied on spectra 
downloaded from the {\it FUSE\/} Virtual Observatory (VO)\footnote{\url{ 
http://archive.stsci.edu/pub/vospectra/fuse2/}} maintained by MAST.  Most of the H$_2$ 
in the ISM is in the two lowest states of rotational excitation $J=0$ and 1 in this molecule’s 
ground vibrational state (Jensen et al. 2010).  Hence, we can ignore the higher $J$ levels 
since their contributions to the total amount of hydrogen $N({\rm H~I})+2N({\rm H_2})$ 
are negligible.  A process similar to the one invoked for interpreting the Ly$\alpha$ 
absorption can be used to derive $N(J=0)$ and $N(J=1)$, but the implementation is more 
complex because one must simultaneously explore values of several free parameters that 
can affect the outcomes.  For instance, most of the intrinsic profile strengths and shapes 
depend not only on the column density but also on the velocity dispersion of the molecular 
gas.  Beyond this, one must account for the influence of such instrumental parameters as 
the line spread function of the {\it FUSE\/} spectrograph and errors in the wavelength 
scale, both of which are not known with much accuracy. The synthesis of intrinsic H$_2$ 
absorptions is greatly facilitated by invoking the optical depth profiles of the Lyman band 
systems provided by McCandliss (2003).  An iterative exploration of the different free 
parameters ultimately revealed the best nulling of the absorptions by the R and P branches 
of the 4-0, 2-0, 1-0 and 0-0 Lyman bands.  The weak
1-0 and 0-0 bands were usually the most useful in constraining the column densities.  As 
with the flux reconstructions for Ly$\alpha$ absorptions, the acceptable ranges for H$_2$ 
column densities were governed by the lack of appreciable emission or absorption after the 
corrections were implemented.

Table~\ref{tbl: H_outcomes} lists for each star the logarithms of the atomic and molecular 
hydrogen column densities (in units of ${\rm cm}^{-2}$).  The table also lists in the last 
three columns the total hydrogen column densities $\log N({\rm H_{tot}})=\log [2N({\rm 
H}_2)+N({\rm H~I})]$, the $J=0$ to 1 rotational temperature $T_{01}=74\,{\rm K}/[\log 
N(J=0)-\log N(J=1)+0.954]$, and $\log f({\rm H_2})=\log [2N({\rm H}_2)/N({\rm 
H_{tot}})]$.  Uncertainty limits in the logarithms of the H$_2$ and H~I column densities are 
usually asymmetrical.  To derive proper offsets and uncertainties for the relevant sums, 
which are needed for calculations of the total $N({\rm H_2})$, $N({\rm H_{tot}})$, 
$T_{01}$, and $f({\rm H_2})$, we implemented the error combination method of Model~2 
of Barlow (2003), which makes use of the fact that the means, variances, and unnormalized 
skewness of distorted Gaussians are additive after convolution.  

The only numbers in Table~\ref{tbl: H_outcomes} that will be incorporated into the study 
of depletions are $\log N({\rm H_{tot}})$ (in Column~6) and $\log f({\rm H_2})$ (Column 
8).  Other quantities are listed in case they are of potential use to other investigations.  
While $T_{01}$ might be a relevant physical quantity to explore for correlations, in fact 
trials to see whether had any influence on depletions failed to reveal any meaningful 
relationships.

\startlongtable
\begin{deluxetable}{
r	
c	
c	
c	
c	
c	
c	
c	
}
\tabletypesize{\footnotesize}
\tablewidth{0pt}
\tablecaption{H$_2$ and H~I Measurement Outcomes\tablenotemark{a}\label{tbl: 
H_outcomes}}
\tablecolumns{8}
\tablehead{
\colhead{Star~~~} & \multicolumn{3}{c}{$\log N({\rm H}_2)$} &
\colhead{$\log N({\rm H~I})$} & \colhead{$\log N({\rm H_{tot}})$} &
\colhead{$T_{01}$} & \colhead{$\log f({\rm H}_2)$}\\
\cline{2-4}
\colhead{} & \colhead{$J=0$} & \colhead{$J=1$} & \colhead{Total} &
\colhead{} & \colhead{(Total H)} & \colhead{} & \colhead{}\\
\colhead{(1)} &
\colhead{(2)} &
\colhead{(3)} &
\colhead{(4)} &
\colhead{(5)} &
\colhead{(6)} &
\colhead{(7)} &
\colhead{(8)}\\
}
\startdata
\input{H_data}
\enddata
\tablenotetext{a}{ Values of $N$ are expressed in units of ${\rm cm}^{-2}$.  All 
uncertainties in this table represent possible deviations at a level estimated to be 
$1.5\sigma$.  The outcomes in columns 4, 6, 7 and 8 have small offsets from the linear 
combinations of quantities expressed in other columns in recognition of shifts that are 
required when quantities with asymmetric errors are utilized, as described in 
Section~\protect\ref{sec: H}.} 
\end{deluxetable}
\section{Derivations of \fstar}\label{sec: derivations_fstar}

As indicated in Section~\ref{sec: design}, comparisons of the column densities of the 
strongly depleted elements Mg and Mn to those of ${\rm H_{tot}}$ indicate the overall 
strengths of depletions \fstar\ for the different sight lines.  For either element, we evaluate 
the equation
\begin{equation}
\fstar={[X_{\rm gas}/{\rm H}]-B_X\over A_X}+z_X~,
\end{equation}
where $X$ denotes either Mg or Mn and the coefficients $A_X$, $B_X$, and $z_X$ are listed 
in Table~4 of J09.\footnote{$B_{\rm Mn}$  was increased by 0.16\,dex to compensate for 
the reduction in the adopted value of $\log f\lambda$ for this element; see Column~5 of 
Table~\protect\ref{tbl: transitions}.}  Column~10 of Table~\ref{tbl: stellar_data} lists 
weighted means of the \fstar\ values derived from these two elements.  Special care must 
be exercised in deriving the values and uncertainties of these means, since both of the 
individual contributions rely in part on a single quantity $N({\rm H_{tot}})$.   Our 
derivations of weighted means for \fstar\ made use of a special treatment that accounted 
for the errors that were common to both measurements (Robinson 2016 pp. 91-95).

\section{Starlight Intensities}\label{sec: starlight_intens}

Ultraviolet and visible photons from stars have a major influence on the physical state and 
chemistry of gas in H~I regions.  They photoionize many of the heavy elements and thus 
produce free electrons, they irradiate dust grains that then eject photoelectrons (which are 
then a source of heating), they may liberate loosely bound atoms from grains via 
photodesorption, and they photodissociate molecular hydrogen and other molecular 
species.  For these reasons, photons may have either direct or indirect influences on the 
atomic species being investigated in this study.  In many instances, the gas can be exposed 
to a strongly elevated radiation field because a large fraction of the interstellar matter can 
reside in a dense gas complex that once formed the target star and its neighbors.  
Conversely, much of the gas near the star may be completely ionized or blown away by 
stellar winds, and thus most of the neutral material along the sight line could be in some 
random location well removed from the UV-bright stars.  For these reasons, we consider 
the intensity of the radiation field a useful parameter to include in our investigation of 
depletions.

About half of the targets in our survey were included in the study of interstellar thermal 
pressures by Jenkins \& Tripp (2011).  They evaluated the characteristic starlight 
intensities for the gas in the sight lines by studying the balance between neutral and
singly ionized carbon atoms and then applying an interpretation based on the competition 
between the photoionizations and the recombinations with free electrons and dust grains.   
They tabulated their calculated values for the intensities of starlight $I$ at energies above 
the ionization potential of neutral carbon (11.26\,eV) in terms of $\log (I/I_0)$, where 
$I_0$ is the average intensity at our location in the Milky Way determined by Mathis et al. 
(1983).  Values of $\log (I/I_0)$ from Jenkins \& Tripp (2011) are listed in Column~10 of 
Table~\ref{tbl: stellar_data}.  All of the values listed here are positive, which reinforces the 
concept that in each case most of the gas is near the target star and is thus subjected to a 
stronger-than-usual radiation field.
\clearpage 
\section{Depletion Trends}\label{sec: depletion_trends}
\subsection{Significance of Kr Depletions}\label{sec: Kr_depl_significance}

\begin{figure}[b]
\plotone{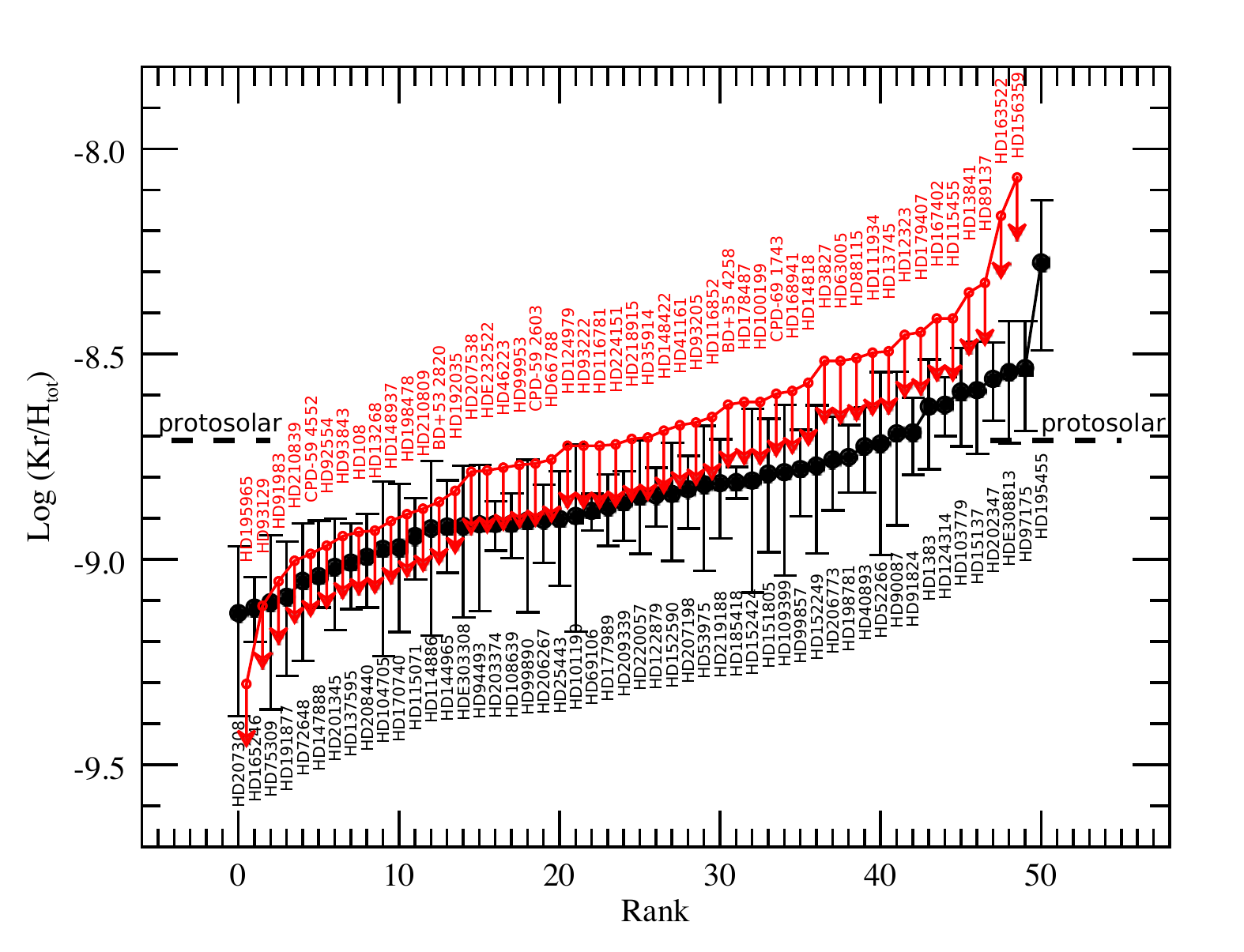}
\caption{Detections (black) and upper limits (red) for values of $\log ({\rm Kr/H_{tot}})$ 
arranged according to their rank.  For comparison, the protosolar level of (Kr/H) is shown 
by the horizontal dashed line.  All error bars represent expected uncertainties at the 
$1\sigma$ level.\label{fig: KrHvsRank}}
\end{figure}

In view of the arguments presented in Section~\ref{sec: krypton_problem} that Kr is an 
element that is least likely to show depletions, it is worthwhile to investigate the reality of 
such depletions --- are they statistically significant?  Figure~\ref{fig: KrHvsRank} presents 
the detections and upper limits for $\log ({\rm Kr/H_{tot}})$, with all cases arranged 
according to their rank in value.   In such a presentation, the error bars are important.  With 
a casual glance they may seem to be symmetrical, but in reality they are not because both 
$N({\rm Kr~I})$ and $N({\rm H_{tot}})$ have asymmetrical error limits.  As we did for 
various quantities related to atomic and molecular hydrogen (Section~\ref{sec: H}), we 
calculated the optimum values of $\log ({\rm Kr/H_{tot}})$ and their uncertainties in each 
direction using the method outlined by Barlow (2003).

The measurable detections of Kr absorption have a median outcome $\log({\rm 
Kr/H_{tot}})=-8.85$.  This abundance level is higher than the general levels of around $-
9.0$ reported by earlier investigators 
(Cardelli \& Meyer 1997 ; Cartledge et al. 2003, 2008 ; Ritchey et al. 2018), 
and this disparity probably arises from a recognition in the present study of additional, but 
not easily apparent, absorptions spanned by the wider velocity intervals defined by the 
limits of the strongest line of Mg~II, as discussed at the beginning of Section~\ref{sec: 
heavy elements} and highlighted in Figure~\ref{fig: HD165246}.  On average, the earlier 
investigations may have missed about 30\% of the total Kr, which may apply especially for 
lines of sight spanning large distances across other spiral arms in the Galaxy.  Ritchey et al. 
(2018) recognized the importance of using a strong absorption feature (in their case, the 
1355\,\AA\ line of O~I) to define the velocity limits and gave specific examples where 
their results gave larger values of $N({\rm Kr~I})$ than previous determinations that were 
based purely on the appearance of the weak line being measured. 

Out of the 51 cases where Kr absorption was detected, there are 28 of them that have 
positive $1\sigma$ error bars for $\log ({\rm Kr/H_{tot}})$ that are below our adopted 
protosolar $\log ({\rm Kr/H})$.  For the upper limits, 24 out of 49 are below the protosolar 
value at the $1\sigma$ level of confidence.  If the interstellar values of (Kr/H) were indeed 
equal to the protosolar value and all of the apparent deficiencies of Kr were due only to 
measurement errors, we would have expected to find a grand total of only 16 such 
examples.   However, it should be noted that the disparity between (28 + 24) and 16 
decreases markedly if one claims that $\log ({\rm Kr/H)_{ref}}$ drops below $-8.71$ 
because many of the upper error bars are just slightly below the dashed line in 
Figure~\ref{fig: KrHvsRank}.  As an added assurance that the lowest outcomes for 
measured detections and upper limits for $\log ({\rm Kr/H_{tot}})$ were real, the spectra 
were reinspected to see whether any anomalies in the observations might have been 
present.  No anomalies were apparent for these cases.

We cannot claim that there is any evidence that (Kr/H) might occasionally be above the 
protosolar value, since there are only four detections that have their lower error bars 
above this level.  These cases could simply be accidental deviations.   However, the 
existence of such outcomes plus others at the high end of the distribution gives us 
confidence that an appropriate reference abundance is probably not much lower than our 
adopted value of $\log ({\rm Kr/H)_{ref}}=-8.71$.

\subsection{Partial Correlations of Depletions with \fstar\ and $\log f(H_2)$}\label{sec: 
correl_fstar_fH2}

We begin our study of depletions by comparing them with two variable parameters, \fstar\ 
and $\log f(H_2)$, that have been determined for all of the sight lines that were available.  
Later (Section~\ref{sec: trends_starlight_intensities} below), we will expand the 
parameters to include $\log (I/I_0)$, but for a smaller number of examples, i.e., only for 
target stars that are in common with those in the investigation by Jenkins \& Tripp (2011).

If the quantities \fstar\ and $\log f(H_2)$ were very strongly correlated with each other for 
the different sight lines, it would be impossible to identify which of the two parameters was 
responsible for any depletion trend.  Fortunately, even though \fstar\ has some 
relationship with molecular fractions (see Figure~16 of J09), the correlation is weak 
enough that we can resolve which parameter is influential on depletions (or if they both 
have an influence).  We do this by determining partial correlations, which address the 
question, if we could hold one of the parameters to a constant value, do variations in the 
other one drive the depletion in one direction or another?

We propose a simple linear relationship
\begin{eqnarray}\label{eqn: coeff2}
[X_{\rm gas}/{\rm H}]&=&B_2\{ X\} + A_2\{ X,\fstar\} \Big(\fstar - z_2\{ X,\fstar\} 
\Big)\nonumber\\
&+&A_2\{ X,\log f({\rm H}_2)\} \Big(\log f({\rm H}_2) - z_2\{ X,\log f({\rm H}_2)\} \Big)
\end{eqnarray}
that describes how, on average, the depletions of element $X$ are influenced by either 
\fstar\ or $\log f({\rm H}_2)$.  (The subscript 2 for the coefficients indicates that there are 
two free environmental parameters.  Later, we will consider three simultaneous 
parameters and attach a subscript 3 to these same coefficients.  Note that the assignments 
$B$, $A$, and $z$ are consistent with the coefficients in Eq.~\ref{eqn: depl_construction} 
for trends with just \fstar.)  As was the case for the $z$ term in Eq.~\ref{eqn: 
depl_construction}, the two $z$ constants in Eq.~\ref{eqn: coeff2} are set to the average 
value of the associated parameters in our sample so that the errors in $B$ and $A$ have a 
near-zero covariance.  This elimination of the covariances makes it simpler to understand 
the uncertainties for other applications that make use of $B$ and $A$. 

Our task of performing a multiparameter regression must accommodate two complicating 
factors.  One is that we must include those determinations that happen to be upper limits 
instead of positive detections.  The proportions of upper limits for measurements of O and 
Ge are low (11\% and 3\%, respectively), but the upper limit fraction for Kr is 49\%.  The 
other complication is that the errors in measuring $N({\rm H_{tot}})$ have a simultaneous 
influence on the two independent variables \fstar\ and $\log f({\rm H_2})$ and also the 
dependent variable $\log (X/{\rm H})$.  Hence, the off-diagonal elements of the covariance 
matrices are nonzero, and this effect can either strengthen or weaken an inferred 
correlation and lead to a bias in the slope coefficients $A_2\{ X,\fstar\}$ and $A_2\{ X,\log 
f({\rm H_2})\}$ (Kelly 2007).  To cope with these two complications, we employ the 
analysis package {\tt mlinmix\_err} developed by Kelly (2007)\footnote{This software is 
available at \url{https://idlastro.gsfc.nasa.gov/ftp/pro/math/}.}, which employs a 
Bayesian approach to linear regression.  One drawback of this routine is that it does not 
accommodate asymmetric errors, so we must convert the errors to a symmetric form equal 
to $(\sigma^++\sigma^-)/2$ for all measured quantities.  The output that shows the joint 
probability distributions for the constants $B_2\{ X\}$, $A_2\{ X,\fstar\}$, and $A_2\{ 
X,\log f({\rm H}_2)\}$ is obtained from a Gibbs sampler that performs random draws from 
the posterior distribution in a Markov chain Monte Carlo (MCMC) method.

\begin{figure}
\epsscale{0.5}
\plotone{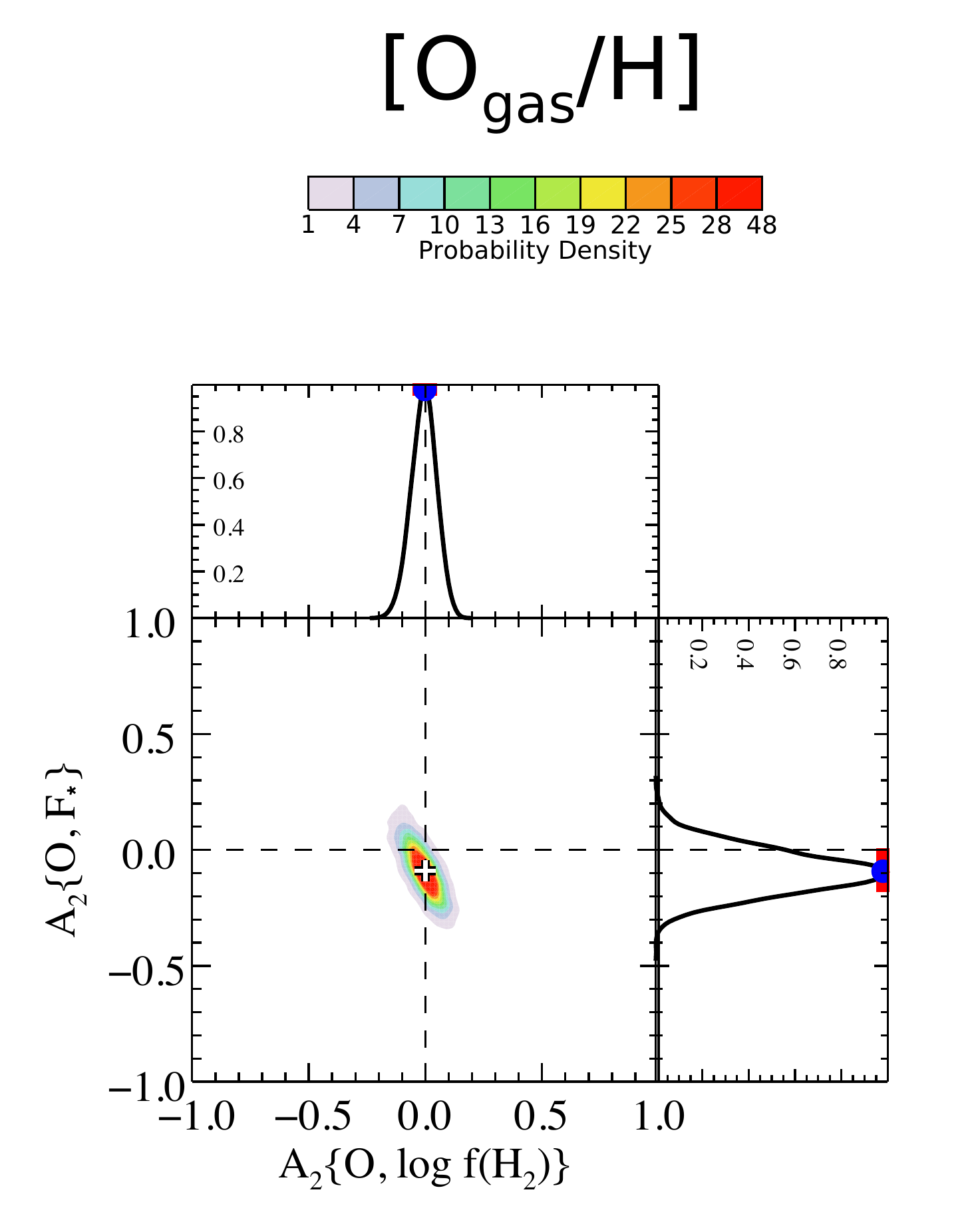}
\caption{The relative joint probabilities of the slope coefficients $A_2\{ {\rm O},\fstar\}$ 
and $A_2\{ {\rm O},\log f({\rm H}_2)\}$ (color display) and their marginalized 
distributions (plots with curves).  At the tops of the distribution plots, the preferred values 
and ranges inside $\pm 1\sigma$ limits are shown with a blue circle and red bar, 
respectively.\label{contour_plot1_o}}
\end{figure}

\begin{figure}
\epsscale{1.}
\plottwo{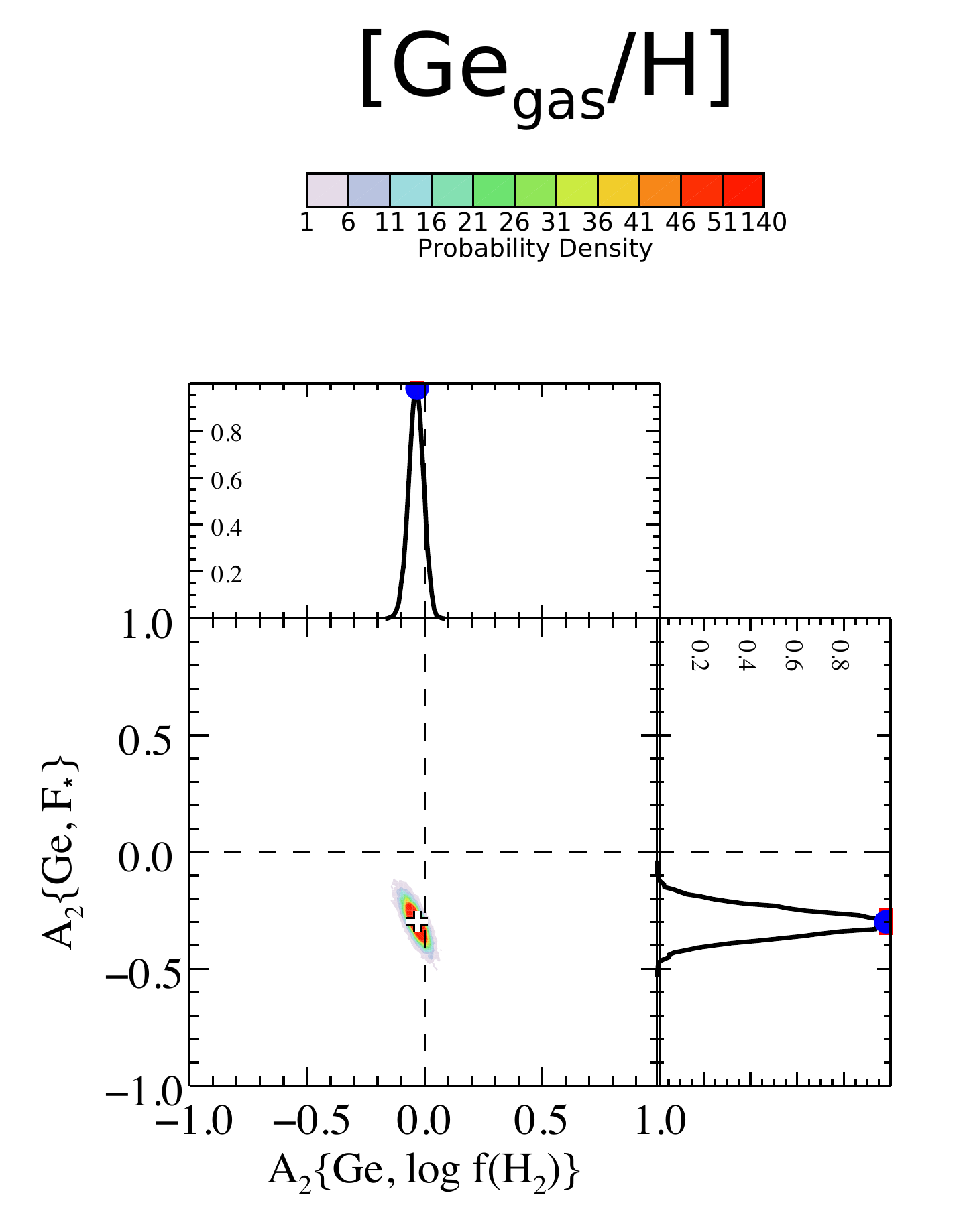}{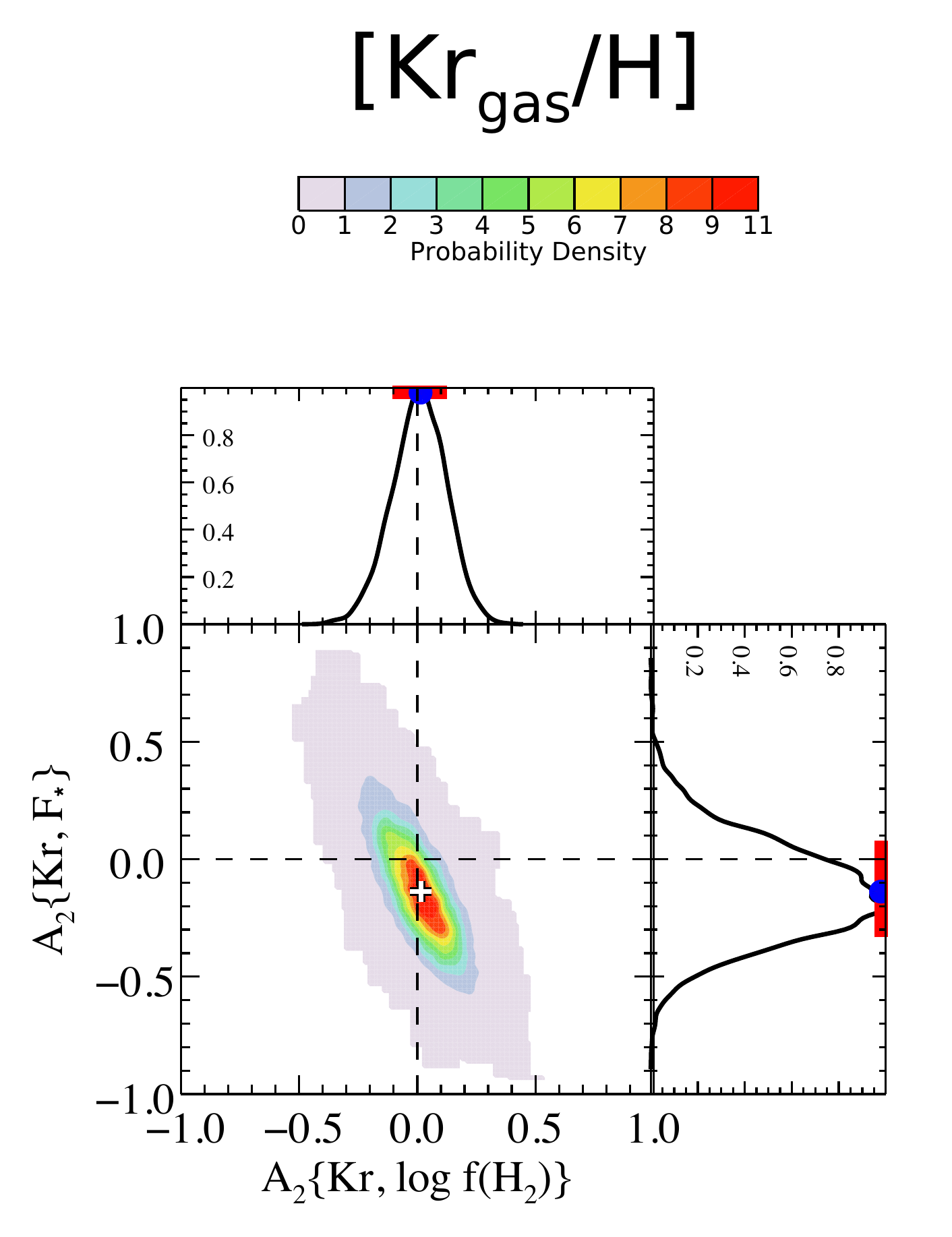}
\caption{Same displays as that shown in Fig.~\protect\ref{contour_plot1_o}, except for the 
elements Ge and Kr.\label{contour_plot1_gekr}}
\end{figure}

The upper portion of Table~\ref{tbl: depletion_coefficients} shows the numerical values 
for $B_2\{ X\}$, $A_2\{ X,\fstar\}$, and $A_2\{ X,\log f({\rm H}_2)\}$ for the three 
elements, along with their uncertainties.  The preferred values were obtained from the 
medians of the marginalized probability functions, and the uncertainties were computed by 
matching the cumulative distributions of these probabilities to that of a Gaussian at the 
$\pm 1\sigma$ points.  The joint probability distributions are shown in 2-dimensional 
form in Figures~\ref{contour_plot1_o} and \ref{contour_plot1_gekr}, along with the 
marginalized outcomes shown as simple graphs.

\begin{deluxetable}{
l	
c	
c	
c	
}[b]
\tablecaption{Depletion Coefficients\label{tbl: depletion_coefficients}}
\tablehead{
\colhead{Coefficient} & \colhead{$X={\rm O}$} & \colhead{$X={\rm Ge}$} & 
\colhead{$X={\rm Kr}$}\\
}
\startdata
\input{Coefficients_data}
\enddata
\tablenotetext{a}{Adopted protosolar abundances on a logarithmic scale where H~=~12.}
\tablenotetext{b}{See the discussion in Section~\protect\ref{sec: oxygen_problem}.}
\tablenotetext{c}{Protosolar abundances generated from the solar photospheric 
abundances compiled by Asplund et al. (2009) plus a 0.04\,dex correction that they 
suggested for gravitational settling.}
\tablenotetext{d}{The uncertainties listed for the $B$ coefficients apply only to those 
associated with the determinations of ISM abundances $(X_{\rm gas}/{\rm H})$ in this 
paper.  They do not include uncertainties in the reference abundances.}
\end{deluxetable}

While the slope coefficients $A_2\{ {\rm O},\fstar\}$ and $A_2\{ {\rm Kr},\fstar\}$ are 
negative, as expected from earlier studies (J09, Ritchey et al. 2018), their significance levels 
are poor.  However, the negative trend for $[{\rm Ge_{gas}/H}]$ against \fstar\ is clearly 
established.  All three elements appear to show no dependence on $\log f({\rm H_2})$, 
which is consistent with the findings of Meyer et al. (1994); (1998) and André et al. (2003) 
for O and Cartledge et al. (2008) for Kr.  The elongations and backward slants of the shapes 
of the joint distributions of \fstar\ and $\log f({\rm H_2})$ for all three elements arise 
from the mild correlation of the two independent variables for the stars that were sampled.  
(If such correlations happened to be stronger, the elongations of the distributions would be 
even more severe.)

The effects of measurement error covariances are important, and they appear to have been 
neglected in previous studies.  As an experiment, if we perform a test run where we ignore 
the covariances by artificially setting them to zero and leave the diagonal error terms as 
they were, we find that that $A_2\{ {\rm O},\fstar\}=-0.20$ (+0.09,$-$0.08), and $A_2\{ 
{\rm Kr},\fstar\} =-0.32$ (+0.19,$-$0.17).  After comparing these outcomes with their 
counterparts in Table~\ref{tbl: depletion_coefficients}, it is clear that they are stronger and 
have a higher level of significance above zero.   When the real depletion trend is 
intrinsically strong, neglecting the covariances is not so important: the value $A_2\{ {\rm 
Ge},\fstar\}=-0.39\pm 0.06$, which is just slightly stronger than the outcome where the 
covariances were included.

\clearpage

\subsection{Trends That Include Starlight Intensities}\label{sec: 
trends_starlight_intensities}

\begin{figure}[b]
\epsscale{0.7}
\plotone{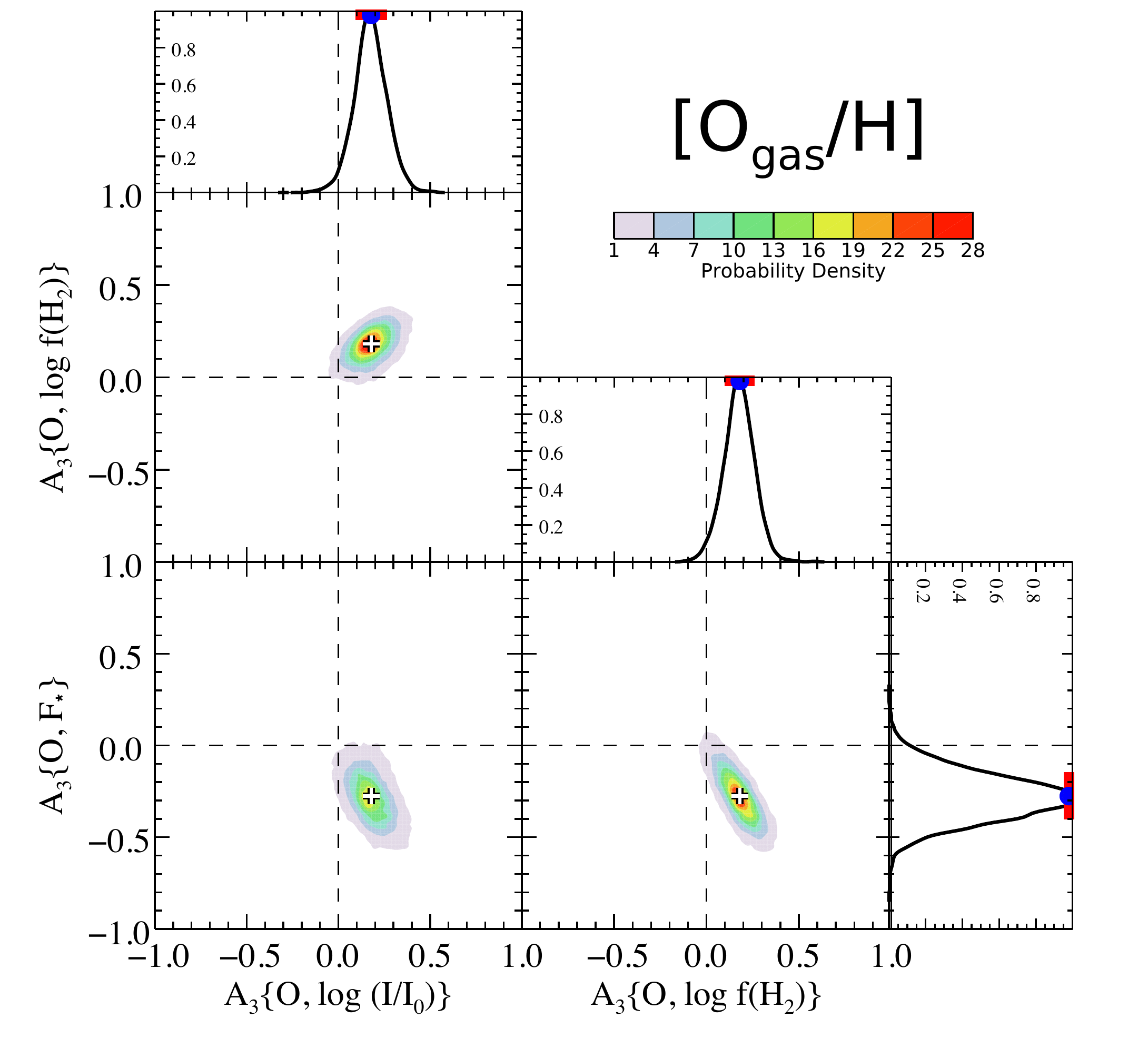}
\caption{Relative joint probabilities of the slope coefficients $A_3\{ {\rm O},\fstar\}$, 
$A_3\{ {\rm O},\log f({\rm H}_2)\}$, and $A_3\{ {\rm O},\log(I/I_0)\}$,  (color displays). 
Each panel depicts the distribution of the probabilities for two coefficients as they would 
appear in projection over the distribution of the third, unspecified coefficient.   The plots on 
the sides showing the curves indicate the probability distributions for a single coefficient 
marginalized over the distributions of the other two quantities.   At the tops of the 
distribution plots, the preferred values and ranges inside $\pm 1\sigma$ limits are shown 
with a blue circle and red bar, respectively.\label{contour_plot3_o}}
\end{figure}

\begin{figure}
\epsscale{0.7}
\plotone{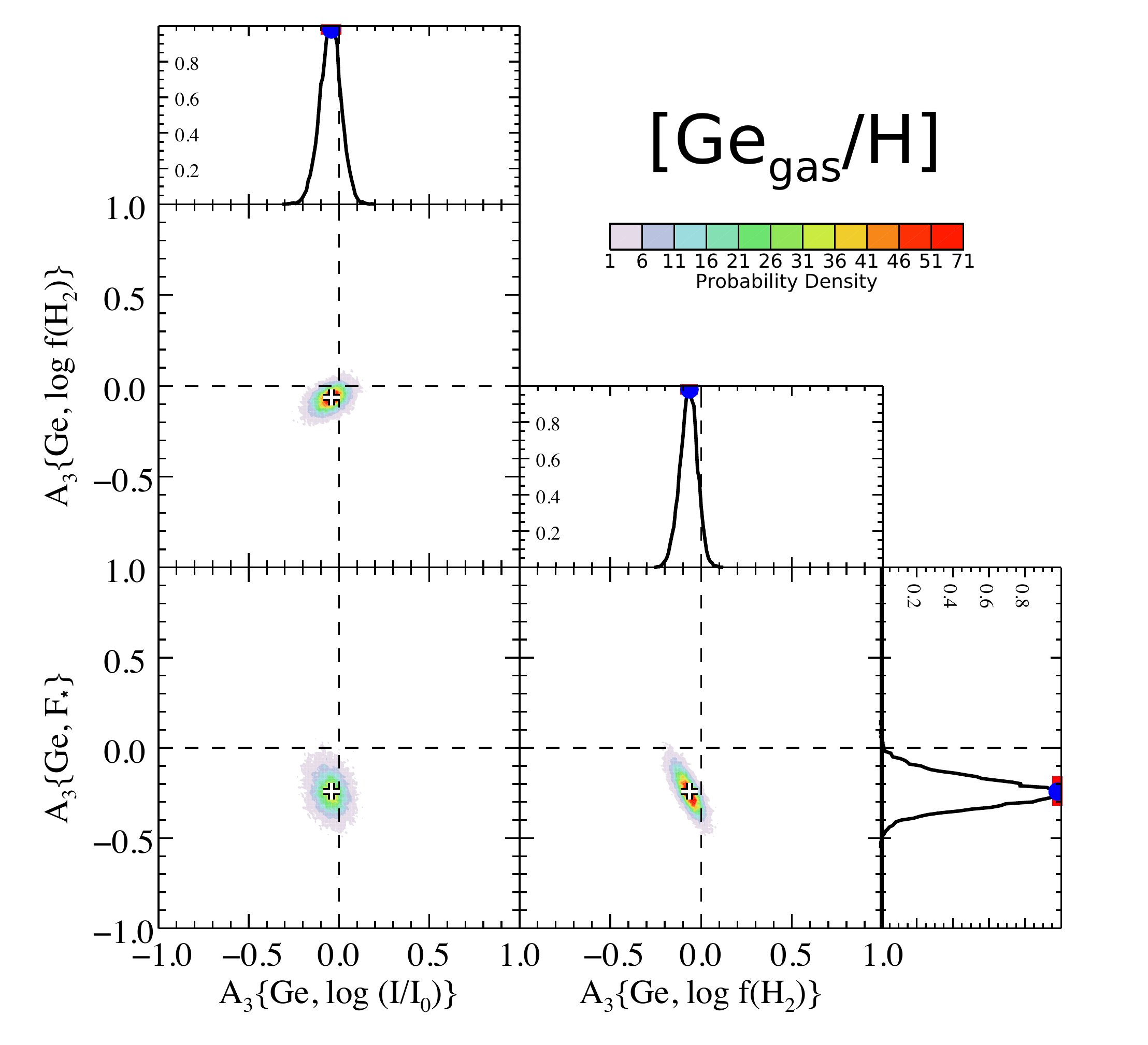}
\caption{Same display as that shown in Fig.~\protect\ref{contour_plot3_o}, except for the 
element Ge.\label{contour_plot3_ge}}
\end{figure}

\begin{figure}
\epsscale{0.7}
\plotone{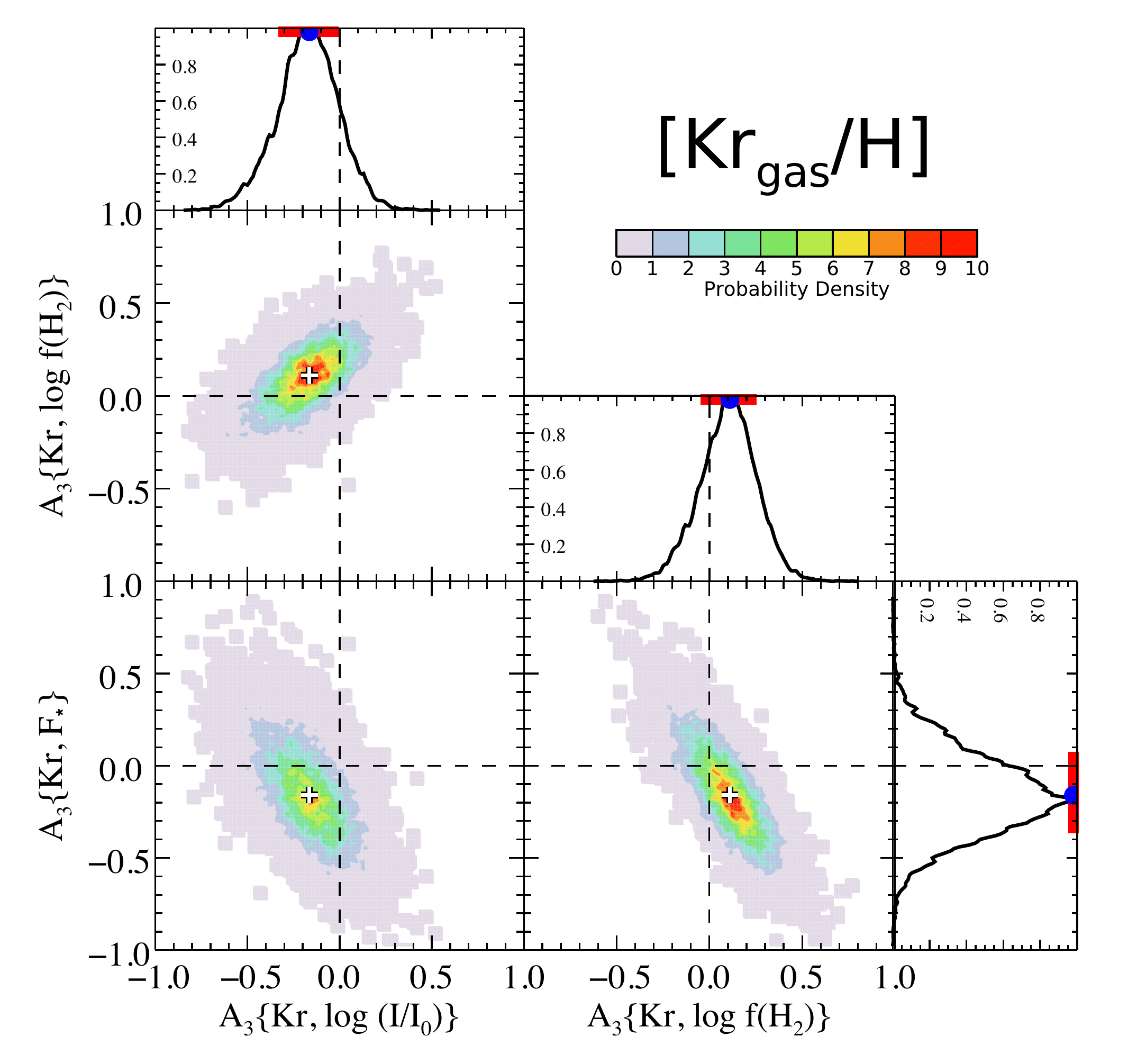}
\caption{Same display as that shown in Fig.~\protect\ref{contour_plot3_o}, except for the 
element Kr.\label{contour_plot3_kr}}
\end{figure}

We now move on to investigate the possible influence of a third physical variable, 
$\log(I/I_0)$ presented in Section~\ref{sec: starlight_intens}, but for a reduced number of 
sight lines.  Using the same construction as that employed for Eq.~\ref{eqn: coeff2}, we 
state that
\begin{eqnarray}\label{eqn: coeff3}
[X_{\rm gas}/{\rm H}]&=&B_3\{ X\} + A_3\{ X,\fstar\} \Big(\fstar - z_3\{ X,\fstar\} \Big) 
\nonumber\\
&+&A_3\{ X,\log f({\rm H}_2)\} \Big(\log f({\rm H}_2) - z_3\{ X,\log f({\rm 
H}_2)\}\Big)\nonumber\\
&+& A_3\{ X,\log (I/I_0)\} \Big(\log(I/I_0) - z_3\{ X,\log(I/I_0)\} \Big)
\end{eqnarray}
The lower portion of Table~\ref{tbl: depletion_coefficients} lists the solutions for the 
coefficients in Eq.~\ref{eqn: coeff3}, and Figs.~\ref{contour_plot3_o}, 
\ref{contour_plot3_ge} and \ref{contour_plot3_kr} show the probability distributions in 
the same style as those presented in the previous section (Figs.~\ref{contour_plot1_o} and 
\ref{contour_plot1_gekr}) .

The recognition that there may be three possible influences on the element depletions has 
resulted in some important changes in the outcomes.  The strength of the trend of the 
depletion of O with \fstar\ now appears to be stronger than before, and its significance has 
improved even though the sample size is smaller and the error in the coefficient is slightly 
larger.  Evidently, our consideration of an extra parameter, the starlight intensity, has 
revealed trends that were otherwise hidden by its influence on the gas-phase abundances.   
It is interesting to note, however, that the value for $ A_3\{ {\rm O},\fstar \}=-0.28\pm 
0.13$ found here is roughly consistent with $A_{\rm O}=-0.225\pm0.053$ derived by J09, 
which was based on a dependence on \fstar\  alone (but for a larger sample size and a 
somewhat different selection of stars).   As we found earlier, the negative slope of the major 
axis of the probability cloud projected on the $A_3\{ {\rm O},\log f({\rm H_2})\}$ ---$ 
A_3\{ {\rm O},\fstar \}$ plane arises from the mild correlation between $\log f({\rm 
H_2})$ and \fstar.  A positive slope seen for the cloud on the $A_3\{ {\rm O},\log(I/I_0)\}$ 
--- $A_3\{ {\rm O},\log f({\rm H_2})\}$ plane arises from a weak anticorrelation between 
the molecular abundances and starlight intensity.  These two effects are simply a 
consequence of the conditions that exist in our sample. 

An important outcome is the fact that the oxygen $A_3$ slope coefficients for both the 
molecular fractions and starlight intensities exhibit significant positive values.
If we multiply the best values for the $A_3$ coefficients for O by the spans over which their 
accompanying parameters vary in our sample, we find that both \fstar\ and $\log ({\rm 
H_2})$ each account for depletion changes of about $\pm 0.15$\,dex.  For $\log (I/I_0)$ 
the changes amount to about $\pm 0.07\,$dex.  Typical uncertainties in $\log ({\rm 
O_{gas}/H})$ range from 0.05 to 0.20\,dex, but 70\% of the cases are below 0.10\,dex.  One 
can appreciate that the effects of any of the three environmental conditions on typical 
single cases are usually difficult to identify with much confidence.

The distribution for $A_3\{ {\rm Kr},\fstar \}$ seems to be almost the same as that for 
$A_2\{ {\rm Kr},\fstar \}$.  All three of the $A_3$ coefficients for Kr differ from zero, but 
only at about or slightly less than the $1\sigma$ level of significance.  The result for 
$A_3\{\rm Kr,\fstar \}$ is not as definitive as the determination of $A_{\rm Kr}$ in the 
study by Ritchey et al. (2018), who found that this coefficient differed from zero at the 
$2.8\sigma$ significance level.  However,  the preferred value of $A_3\{ {\rm Kr},\fstar \}$ 
is identical to the value of $A_{\rm Kr}$ listed in J09 and not far from that of Ritchey et al. 
(2018).  Our general finding for the strength of the depletions, as indicated by $B_3\{ {\rm 
Kr} \}$, is less than all previous determinations.  We attribute the apparent weaker 
depletions of Kr to two important differences in analysis methodologies.  First, as discussed 
in Section~\ref{sec: heavy elements}, our adoption of generously wide velocity limits from 
the appearance of strong lines often increased the outcomes for the values of the column 
densities (and also their uncertainties because more noise and stretches of continuum had 
to be incorporated into the measurements).  Second, as pointed out in Section~\ref{sec: 
correl_fstar_fH2}, any neglect of error covariances can produce misleading results.  This 
brought about a need to construct a probability distribution for the depletion trends with 
\fstar\ that acknowledged and corrected for the effect that errors in $N({\rm H_{tot}})$ 
induce a spurious strengthening of the trend (i.e., any erroneous increase in $N({\rm 
H_{tot}})$ creates an increase in \fstar\ at the same time that it lowers the inferred $[{\rm 
Kr_{gas}/H}]$, which then strengthens a negative correlation).  Larger uncertainties may 
also arise from the fact that we are examining the dependence of depletions on more than 
one parameter.

The negative trend for [Ge$_{\rm gas}$/H] against \fstar\ is still significant, although its 
magnitude has decreased slightly from that expressed by the $A_2$ coefficient, and it 
seems much less steep than the relationships $A_{\rm Ge}=-0.615\pm 0.083$ and $-
0.371\pm 0.081$ found by J09 and Ritchey et al. (2018), respectively.  It is probable that 
the diminished response to \fstar\ compared to the findings of the previous studies arises 
from the effects discussed above for Kr.  The magnitudes of the behaviors of $[{\rm 
Ge_{gas}/H}]$ against $\log f({\rm H_2})$ and $\log(I/I_0)$ are similar to those of $[{\rm 
O_{gas}/H}]$, but with a reversal of sign.
\clearpage
\section{Interpretation}\label{sec: interpretation}

The consideration of partial correlations of the abundances of free O, Ge, and Kr atoms with 
two environmental conditions other than \fstar\ should provide clues on secondary 
processes that may either enhance or reduce the atomic depletions.  At this stage, we can 
offer speculations about the meanings of these other trends, which might encourage future 
investigators to follow up on specific approaches, either theoretical or observational.   In 
the following subsections, we explore factors beyond \fstar\ derived from comparing Mg 
and Mn to H, which we acknowledge to be an important factor for driving the depletions of 
the elements O, Ge, and Kr.

\subsection{Influences on [O$_{\rm gas}$/H]}\label{sec: influences on [O/H]}

One might expect that as $f({\rm H}_2)$ increases we would see less O in atomic form 
because higher molecular hydrogen fractions could be more favorable for creating a chain 
of reactions that could involve other molecules, some of which could contain oxygen.  The 
near-zero outcome for $A_2\{ {\rm O},\log f({\rm H}_2)\}$ indicates that this concept is 
overly simplistic and probably not important.   Instead, after we test for the independent 
actions of $f({\rm H}_2)$ and starlight intensity $I$, a subtler picture emerges, where 
[O$_{\rm gas}$/H] shows positive correlations with both of these environmental 
conditions.

We first address a possible reason why $A_3\{ {\rm O},\log f({\rm H}_2)\}$ is significantly 
above zero.  We know that a concentration of molecular hydrogen is governed mostly by a 
balance between the local rate of formation on dust grains and the photodissociation of the 
free molecules by starlight.  Moreover, the formation rate in any given volume should be 
proportional to the product of the local densities of atomic hydrogen and dust grains.  Since 
\fstar\ is well correlated with local hydrogen densities, as well as the proportion of 
elements in the form of dust, we can assert that to first order it can be used as a measure of 
the H$_2$ formation rate.  This concept, along with the principle that partial correlations 
are intended to indicate what should happen if we vary one parameter while the others are 
held constant, can lead to an intriguing conjecture based on the following thought 
experiment.   If we had the ability to peg the destruction rate (proportional to $I$) to a 
constant value, and we could do likewise for \fstar, from the arguments just presented we 
would simplistically expect to find $f({\rm H}_2)$ to remain constant.   By virtue of our 
detecting changes in [O$_{\rm gas}$/H] with residual deviations in $f({\rm H}_2)$,  we 
discover that there must be variations in the H$_2$ concentrations caused by some factors 
(or a single factor) other than \fstar\ or $I$, and the gas-phase O-to-H ratio seems to 
respond to such factors.  Possible secondary effects could include variations in grain 
composition or size at fixed values of \fstar, which might influence the production rate of 
H$_2$ (Snow 1983).  In turn, changes in these grain properties could partly influence the 
concentration of free O atoms for any fixed value of dissociating radiation intensity and 
dust-to-gas mass ratio.  For instance, we could speculate that a dust population that is more 
strongly dominated by relatively small grains may produce more H$_2$, given the 
increased average surface-to-volume ratio, but these same grains may more effectively lose 
any embedded H$_2$O ice via photodesorption (Cruz-Diaz et al. 2018), and the free water 
molecules would subsequently be broken up by photodissociation.  One might check this 
proposal by learning more about the extinction properties of dust on the sight lines to see if 
grain sizes really matter, but that examination is a complex undertaking that is beyond the 
scope of this study.

Two possibilities may explain the positive value for $A_3\{ {\rm O},\log (I/I_0)\}$.  One is 
simply a different approach to the argument presented earlier about the positive outcome 
for $A_3\{ {\rm O},\log f({\rm H}_2)\}$.  If \fstar\ and $f({\rm H}_2)$ could be held 
constant, one would expect that $I$ would need to increase to compensate for any greater 
efficiency of H$_2$ production by some differing grain population, and, as proposed before, 
those same grains could have secondary properties that would make them less able to 
sequester O atoms.  An alternative (or additional) means for explaining the increase in the 
free oxygen atoms for an increase in $I$ may simply be an enhanced liberation of O arising 
from the more rapid photodesorption and subsequent photodissociation of H$_2$O ice that 
may have resided on the surfaces of dust grains.  However, this proposal is disfavored by 
the observed lack of infrared ice-band features in sight lines with extinctions as low as 
those in our sample, unless, of course, many of the grains are so large that they are 
completely opaque in the infrared (Poteet et al. 2015).

\subsection{Influences on [Kr$_{\rm gas}$/H]}\label{sec: influences on [Kr/H]}

We return to the topics discussed in Section~\ref{sec: krypton_problem} that addressed 
possible causes for the depletions of Kr in the ISM.   The level of significance of the three 
$A_3$ coefficients determined in this study for Kr is not high.  Nevertheless, it is useful to 
explore the possible interpretations of these coefficients if the inferred values are indeed 
real.

We had dismissed the possibility that positively charged radicals in the gas phase could 
play an important role in depleting Kr, but from the meteoritic studies and laboratory 
experiments there were indications that heavier noble gases could bind to solid materials 
in space.  Of special relevance to Kr depletions studied here are the experiments conducted 
by Nichols et al. (1992) that indicated that enhanced adsorption with stronger binding than 
a Van der Walls attraction could occur if a surface was irradiated by energetic protons, 
which create free radicals and broken bonds capable of maintaining strong chemical 
attractions for the noble gas atoms (Hohenberg et al. 2002).\footnote{These authors and 
others often refer to this type of binding as ``anomalous adsorption.’’}  Ionizing radiation 
may also be able to create such enhanced binding sites on the surfaces of materials.  These 
findings may account for our observation that Kr depletions not only depend on the 
presence of solids, as indicated by the negative value of $A_3\{{\rm Kr},\fstar \}$, but they 
also respond to increased levels of UV irradiation that can create more chemically active 
binding sites, as suggested by the negative value of $A_3\{{\rm Kr},\log (I/I_0)\}$.  An 
alternative explanation for a negative outcome for $A_3\{{\rm Kr},\log (I/I_0)\}$ is that 
there is a stronger tendency for Kr to be photoionized relative to H in low-density regions 
where some infiltration of Kr-ionizing radiation may occur.

We still face the challenge of interpreting the positive outcome for $A_3\{ {\rm Kr},\log 
f({\rm H}_2)\}$.  We speculate that this dependence, if real, may arise from the more 
effective neutralization of the active sites by gases that have richer concentrations of 
molecules.

\section{Regional Variations }\label{sec: regional_variations}

Now that we have characterized the responses of [O$_{\rm gas}$/H] and [Ge$_{\rm 
gas}$/H] with respect to certain environmental parameters, we can touch on the topic of 
possible changes in the abundances of these two elements with location.  In doing so, it can 
be instructive to discount the effects of \fstar\ and $f({\rm H}_2)$, so that the shifts 
attributable only to location are more apparent.   In principle, we could also compensate for 
the influence of $I/I_0$, but doing so would significantly limit the number of available 
samples.  We do not address changes in [Kr$_{\rm gas}$/H] because a high proportion of 
the measurements are upper limits.

It is important to acknowledge that two considerations might disfavor our being able to 
detect meaningful changes in abundances with location.  First, Roy \& Kunth (1995) 
pointed out that there are a number of hydrodynamical processes that should efficiently 
mix the interstellar gases at many different distance scales, although hydrodynamical 
simulations by de Avillez \& Mac Low (2002) indicate that the characteristic decay rate for 
localized chemical inhomogeneities may be as long as tens of megayears.  From an 
observational standpoint, we note that Esteban \& García-Rojas (2018) found that the 
abundances of O relative to H in H~II regions in our Galaxy deviate by less than $\pm 
0.05$\,dex (which is of order their measurement uncertainties) from a best fit to an overall 
abundance gradient with galactocentric distance.   Nevertheless, while mixing effects may 
be important, we must acknowledge the possibility that infalling, low-metallicity gas 
impacting the Galactic plane could create temporary dilutions of the existing material in 
specific regions.  The second consideration that may reduce abundance contrasts is that 
most of our measurements sample gas that is likely to be distributed over many 
kiloparsecs, which would tend to wash out any fluctuations that we may be looking for.  
While this could apply in most cases, there may still be instances where much of the 
material that is sensed happens to be concentrated in the progenitorial star-forming clouds 
in the immediate vicinity of the target stars. 

\subsection{Changes in Corrected [O$_{\rm gas}$/H] with Location}\label{sec: location 
[O/H]}

From the investigation presented in Section~\ref{sec: correl_fstar_fH2}, we found that 
$A_2\{{\rm O},\log f({\rm H}_2)\}=0.00\pm 0.05$, which indicates that no correction for 
$\log f({\rm H}_2)$ is needed (unless we were to include $\log (I/I_0)$ as a relevant 
parameter).  However, we do need to compensate for changes in \fstar.  Thus, we can 
define a quantity
\begin{eqnarray}\label{eqn: DeltaO/H}
\Delta ({\rm O/H})&=&\log N({\rm O~I})-\log N({\rm H_{tot}})-({\rm O/H})_{\rm 
ref}+12-B_2\{ {\rm O}\}\nonumber\\
&-&A_2\{ {\rm O},\fstar\}\Big(\fstar-z_2\{ {\rm O},\fstar\} \Big),
\end{eqnarray}
 which indicates residual deviations from the average behavior of oxygen gas-phase 
abundances.

\begin{figure}
\plotone{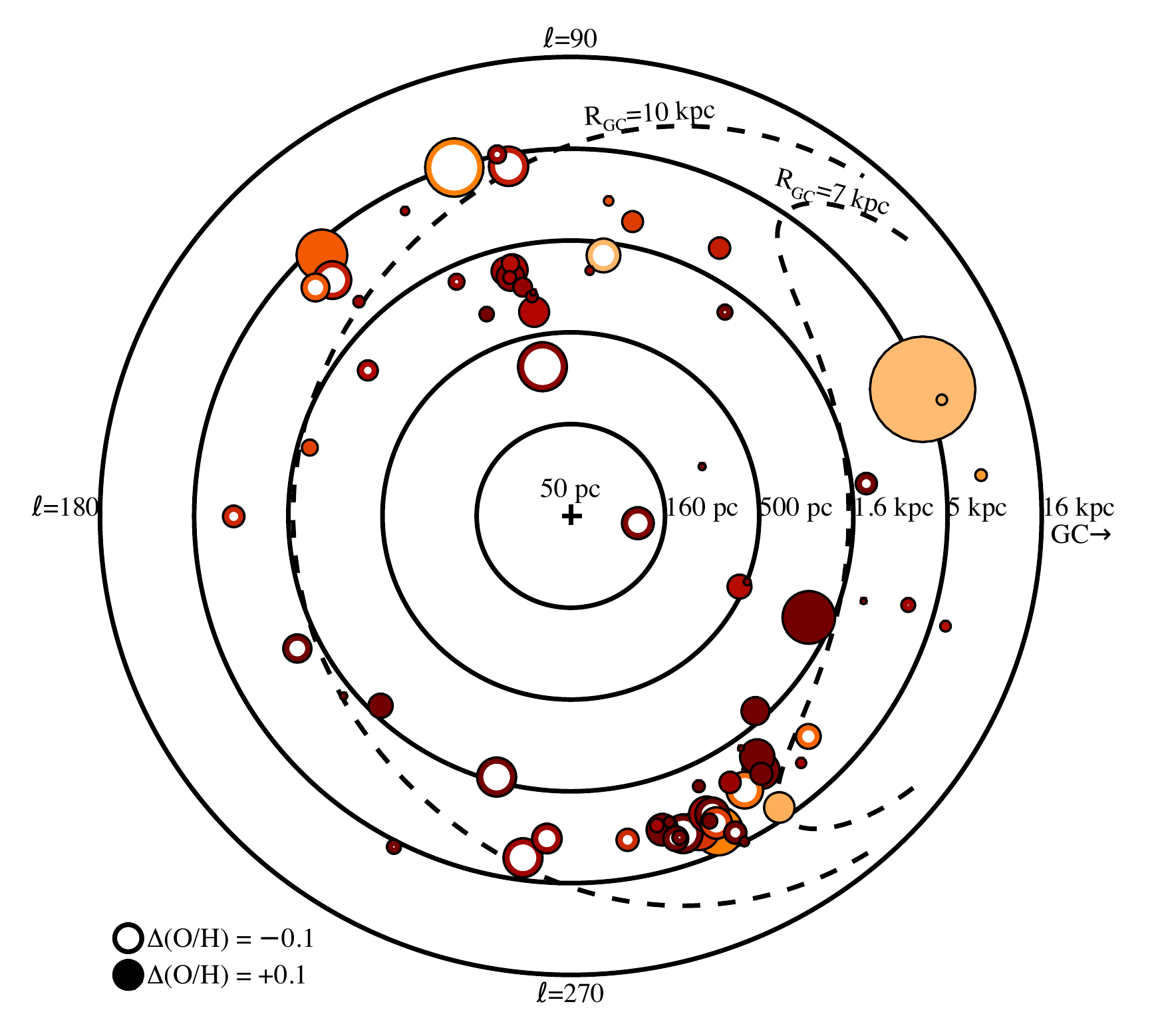}
\caption{Depiction of the results for $\Delta ({\rm O/H})$ at the locations of 75 targets 
projected on the plane of the Galaxy, but with distances from the Sun laid out in concentric 
rings having equal logarithmic intervals.  The diameter of each circle that represents a 
target is proportional to the absolute value of $\Delta ({\rm O/H})$, with solid interiors 
depicting positive values and open ones depicting negative values.  The colors of the 
symbols (or their borders) represent distances from the Galactic plane, ranging from dark 
brown for $z\sim 0$ to orange for targets at large distances either above or below the 
plane.  Distances from the Galactic center $R_{\rm GC}$ are indicated by the dashed 
lines.\label{fig: planview_o}}
\end{figure}

Figure~\ref{fig: planview_o} depicts the results for $\Delta ({\rm O/H})$ for 75 sight lines 
that have uncertainties in this quantity that are less than 0.15\,dex.  This choice for the 
cutoff in uncertainties was governed by a desire to have a balance between a respectably 
large number of samples and yet show results that were not too unreliable.  Unfortunately, 
there are too few targets at small distances to test the assertions by
Meyer et al. (1994, 1998), André et al. (2003) and Cartledge et al. (2004) that the 
interstellar gases within about 800\,pc of the Sun may have been diluted by the infall of 
low-metallicity, extragalactic gas near our location .\footnote{The other studies made use 
of observations taken with the GHRS on HST, which could tolerate the high count rates for 
the spectra of nearby, bright stars.  The permissible upper limits on count rate for the 
detectors on STIS are low enough to force the selection of rather distant O- and B-type 
stars, except for a few observations that used neutral density filters.}  Another prospect for 
detecting dilution by the infall of high-velocity, metal-poor gas might have been Complex~C 
(Richter et al. 2001 ; Tripp et al. 2003), which extends down to the Galactic plane at 
longitudes $40\arcdeg < \ell < 80\arcdeg$.  Unfortunately, the distance to Complex~C is 
about $10\pm 2.5\,{\rm kpc}$ (Thom et al. 2007 ; Wakker et al. 2007), which is not 
covered by our sample. 

With one exception (HD\,12323 located at $r,\ell=[4.37\,{\rm kpc},133\arcdeg]$ with 
$\Delta ({\rm O/H})=+0.17\pm 0.07\,{\rm dex}$), the measurements for stars with 
galactocentric distances $R_{\rm GC}>10\,{\rm kpc}$ seem to be deficient in oxygen, with 
$\Delta ({\rm O/H})\sim -0.05$ to $-$0.2\, dex.  A majority of the stars near or inside the 
$R_{\rm GC}=7\,{\rm kpc}$ boundary show an excess of oxygen.  This behavior is 
consistent with the abundance gradients $-0.07\,{\rm dex~kpc}^{-1}$ for light elements in 
early B-type main-sequence stars found by Rolleston et al. (2000) and $-0.051\,{\rm 
dex~kpc}^{-1}$ for $R_{\rm GC}>7\,{\rm kpc}$ determined for H~II regions by Esteban 
\& García-Rojas (2018).  $\Delta ({\rm O/H})$ could also be influenced in part by possible 
systematic differences in starlight intensity and molecular fractions with galactocentric 
distance. The outstanding case in the diagram for an overabundance of O, ($\Delta ({\rm 
O/H})=0.37\pm 0.13\,{\rm dex}$) located at $r,\ell=[4.9\,{\rm kpc},20.3\arcdeg]$, 
corresponds to the sight line toward the star HD\,195455, which is at $z=-3.0\,{\rm kpc}$ 
below the Galactic plane.

\subsection{Changes in Corrected [Ge$_{\rm gas}$/O$_{\rm gas}$] with 
Location}\label{sec: location [Ge/O]}

\begin{figure}
\plotone{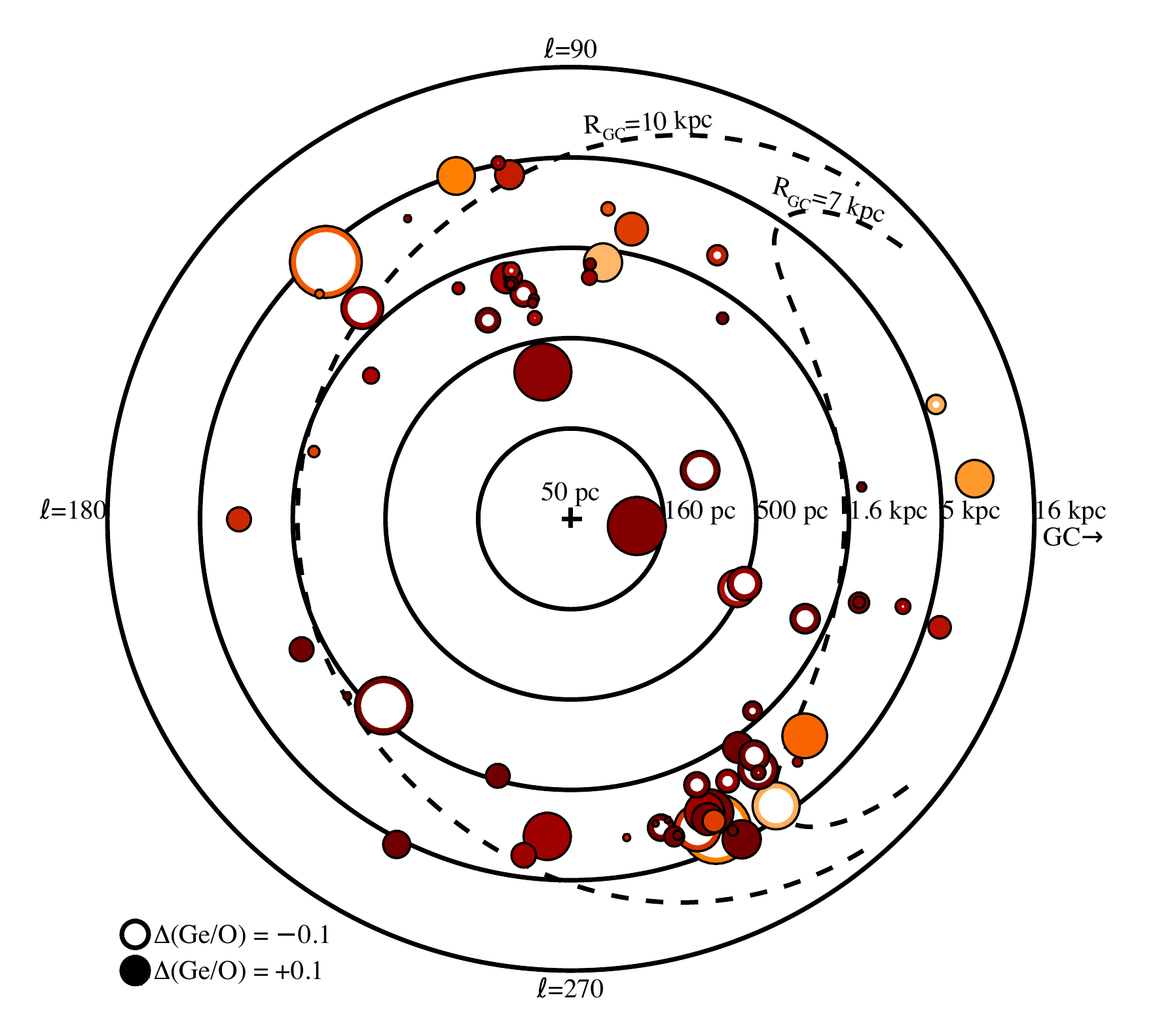}
\caption{Diagram indicating 72 values of $\Delta ({\rm Ge/O})$, using the same 
representation as in Figure~\protect\ref{fig: planview_o}.  As with the choices for 
examples in Figure~\protect\ref{fig: planview_o}, determinations with uncertainties 
greater than 0.15\,dex were rejected. \label{fig: planview_ge}}
\end{figure}

To check on possible local enhancements of neutron-capture elements relative to
$\alpha$-process ones, we can use Ge and O as proxies and examine the distribution with 
location of the quantity
\begin{eqnarray}\label{eqn: DeltaGe/O}
\Delta ({\rm Ge/O})&=&\log N({\rm Ge~II})-\log N({\rm O~I})-({\rm Ge/H})_{\rm 
ref}+({\rm O/H})_{\rm ref} - B_2\{ {\rm Ge}\}+B_2\{ {\rm O}\}\nonumber\\
&-& A_2\{ {\rm Ge},\fstar\}\Big(\fstar-z_2\{ {\rm Ge},\fstar\} \Big)
+A_2\{ {\rm O},\fstar\}\Big(\fstar-z_2\{ {\rm O},\fstar\} \Big)\nonumber\\
&–&A_2\{{\rm Ge},\log f({\rm H}_2)\}\Big(\log f({\rm H}_2)-z_2\{{\rm Ge},\log f({\rm 
H}_2)\}\Big)~,
\end{eqnarray}
which again represents abundances that have been corrected for the general trends in 
\fstar\ and $\log f({\rm H}_2)$.  One advantage of this measure is that the outcomes are 
not sensitive to errors in $N({\rm H_{tot}})$, as is the case for $\Delta({\rm O/H})$.  
Moreover, the influences of measurement errors in either \fstar\ or $\log f({\rm H}_2)$ 
are reduced in proportion to the respective differences in the $A_2$ coefficients for the two 
elements.  A depiction of the values of $\Delta ({\rm Ge/O})$ vs. location is given in 
Figure~\ref{fig: planview_ge}.  Changes in $\Delta ({\rm Ge/O})$ seem to show no 
coherent patterns beyond what one would expect to find from chance, and in particular 
there does not seem to be a duplication of the pattern for the Kr/H vs. distance trend 
shown by Cartledge et al. (2008).  There appears to be an approximate anticorrelation in 
the outcomes shown in Figures~\ref{fig: planview_o} and \ref{fig: planview_ge}; this is 
simply a consequence of the fact that $\log N({\rm O~I})$ appears in both expressions 
shown in Eqs.~\ref{eqn: DeltaO/H} and \ref{eqn: DeltaGe/O}, but with opposite sign.  
Random changes in $\log N({\rm O~I})$ due to either measurement errors or real 
variations will contribute to this anticorrelation.
\vspace{1cm}
\section{Summary}\label{sec: summary}

The aim of his investigation was to probe to greater depth the various factors that can 
influence the interstellar gas-phase abundances of the elements O, Ge, and Kr relative to 
hydrogen in both atomic and molecular forms.  These three elements were selected 
because they have weak (and thus unsaturated) absorption features in a wavelength region 
that is covered by many medium- and high-resolution echelle STIS spectra in the public 
archive.  In addition, the depletions of O and Kr below their stellar reference abundance 
levels are of special interest.  For instance, O seems more depleted than one would expect 
from the formation of silicates and oxides (as outlined by J09), and we would expect Kr to 
show virtually no depletion because it is chemically inert.  A more definitive investigation 
that addressed different responses of such depletions was anticipated to yield more 
insights on these puzzling cases.

One hundred stars were found to satisfy the criteria that the interstellar features of these 
elements could be measured with reasonable accuracy, although many cases revealed only 
upper limits, but ones that were still meaningful.  The profiles of Ly$\alpha$ absorption 
reveal the column densities of H~I, and Lyman band absorptions of H$_2$ recorded by {\it 
FUSE\/} indicate the column densities of this molecule in its two lowest rotational states.  
Two different transitions of both Mn~II and Mg~II were used to derive column densities of 
these two elements, which, after comparisons against total hydrogen densities (both 
atomic and molecular), were employed to derive a measure of the generalized strength of 
depletions of atoms onto dust grains, as characterized by the parameter \fstar\ defined by 
J09.  In addition to \fstar, two other environmental parameters of interest in this study are 
the fraction of H atoms in molecular form $f({\rm H}_2)$, and the local intensity of UV 
starlight $I$ determined for about half of our stars by Jenkins \& Tripp (2011).  Initially, 
the rotation temperatures of the $J=0$ and 1 levels of H$_2$ were thought to be of interest, 
but attempts to discover meaningful relationships with this parameter failed to show any 
significant effects.

All line measurements were performed using a uniform protocol for determining the 
outcomes and their uncertainties.  For instance, the velocity intervals over which all of the 
spectral intensities were sampled were strictly defined by the limits of absorption for the 
strongest and most easily measurable features, those of the Mg~II doublet at 1240\,\AA.  
In most cases this conservative definition of velocity endpoints seemed wider than the 
apparent span of a weak line, such as that for O~I or Kr~I, with the consequence that both 
the measurement uncertainties and the column density outcomes were increased.  
Nevertheless, these increased velocity spans guaranteed a more complete pairing with all 
of the hydrogen gas (for which velocity limits were unavailable) and may explain why the 
overall krypton-to-hydrogen and germanium-to-hydrogen ratios generally came out to be 
slightly larger in the present work than in previous surveys.

The study of partial correlations of atomic abundances relative to hydrogen (atomic plus 
molecular) was divided into two parts.  One part included all of the stars but was limited to 
exploring relationships with only \fstar\ and $\log f({\rm H}_2)$.  The other part 
expanded the parameter list to include the starlight intensity $\log( I/I_0)$ (where $I_0$ is 
an average value for our region of the Galaxy), but the sample size was about half as large.  
One important feature in our derivation of the dependencies of the abundances against the 
parameters was a recognition of error covariances, which tended to change the coefficient 
values and their estimated uncertainties.  For O and Kr, including the covariances, as 
opposed to ignoring them, resulted in a reduction of the magnitudes of the coefficients and 
an increase in their expected uncertainties.  For Ge, recognizing covariances had little 
effect.  Another feature in the analysis was that we incorporated upper limits for the 
column densities into the coefficient derivations instead of simply ignoring them.

All three elements show atomic abundances that decrease with increasing \fstar, which is 
in accord with previous studies.  In the investigation where $\log(I/I_0)$ was not 
considered, there seemed to be no changes with $\log f({\rm H}_2)$ for any of the 
elements.  However, when this extra parameter was included, the negative trend of O with 
\fstar\ seemed to strengthen, and significant positive correlations with $\log f({\rm H}_2)$ 
and $\log (I/I_0)$ emerged.  Speculations on the possible causes of these relationships are 
discussed in Section~\ref{sec: influences on [O/H]}.  Trends of Kr depletions onto solids 
might be explained by variations in the ability of grain surfaces to provide 
enhanced absorption sites, as suggested in Section~\ref{sec: influences on [Kr/H]}.

Finally, attempts were made to look for regional variations in abundance ratios in two 
forms: O/H and Ge/O.  A special feature of this investigation was that these two ratios were 
expressed in the forms of residual deviations away from the general trends that were found 
for \fstar\ and $\log f({\rm H}_2)$.  The only variation that seemed to go beyond just 
chance fluctuations was a tendency for the corrected O/H to decrease with galactocentric 
distance, which is in accord with many previous studies of elemental abundances in our 
Galaxy.

\acknowledgements

This research was supported by an archival research grant nr. HST-AR-14570.001-A provided by NASA through the 
Space Telescope Science Institute (STScI), which is operated by the Associations of Universities for Research in 
Astronomy, Incorporated, under NASA contract NAS5-26555.  All of the spectroscopic data analyzed in this paper 
were obtained from the {\it Mikulski Archive for Space Telescopes\/} (MAST) maintained by the STScI.  Specific 
observations can be accessed via the following collections on MAST: STIS E140H data: 
\dataset[https://doi.org/10.17909/t9-dre3-6g69]{https://doi.org/10.17909/t9-dre3-6g69}  ; 
STIS E140M data: 
\dataset[https://doi.org/10.17909/t9-0mtk-2a73]{https://doi.org/10.17909/t9-0mtk-2a73} ; FUSE 
data:  \dataset[https://doi.org/10.17909/t9-537q-4w41]{https://doi.org/10.17909/t9-537q-4w41}.  
The author is grateful to Sarah Weissman at the STScI for her assistance in creating the DOIs.  This research has 
made use of the SIMBAD database, operated at CDS, Strasbourg, France. For some theoretical insights, the author 
benefited from useful discussions with Bruce Draine.

\facilities{HST (STIS), FUSE}
\software{mlinmix\_err (Kelly 2007)}

\clearpage
\appendix
\section{Basic Measurement Outcomes}\label{sec: basic_measurements}

Table~\ref{tbl: basic_measurements} lists for each star the results of analyzing the 
interstellar absorption features of H~I, O~I, Mg~II, Mn~II, Ge~II, and Kr~I, based on the 
analysis methods discussed in Section~\ref{sec: column densities}.   Measurements of 
H$_2$ are not included in this table but do appear in Table~\ref{tbl: H_outcomes}.

\startlongtable
\begin{deluxetable}{
l	
c	
c	
}
\tablecaption{STIS Measurement Outcomes\tablenotemark{a}
\label{tbl: basic_measurements}}
\tablehead{
\colhead{Line\tablenotemark{b}} & \colhead{$W_\lambda$ (m\AA)} & \colhead{$\log 
N_a({\rm cm}^{-2})$\tablenotemark{c}}
}
\tablecolumns{3}
\startdata
\input{A1_data}
\enddata
\tablenotetext{a}{The subheadings list the name of the star, the observation mode [H = 
high-resolution mode (E140H), M = medium-resolution mode (E140M)], and the endpoints 
$v_1$ and $v_2$ for the heliocentric velocities that span the integration intervals for 
equivalent widths $W_\lambda$ and apparent column densities $N_a$.}
\tablenotetext{b}{For the heavy elements, the transitions that match the entries here are 
listed in Table~\protect\ref{tbl: transitions}.  If the numeral 3 follows an ion, it means that 
a correction for line saturation needed to be applied.  See Section~\protect\ref{sec: heavy 
elements} for details.  The entry ``H~I’’ refers to a direct measurement of the Ly$\alpha$ 
absorption, whereas for the few cases that needed a correction for stellar Ly$\alpha$ 
absorption ``H~I sc’’ shows the modified value for $\log N({\rm H~I})$.}
\tablenotetext{c}{See Eq.~\protect\ref{eqn: N_a}.}
\tablenotetext{d}{Not included in the study of element abundance behaviors because no 
{\it FUSE\/} observations that revealed H$_2$ column densities are available for this star.  
The outcomes for H~I and heavy elements are shown for the benefit of other investigations 
that could make use of these results.}
\end{deluxetable}

\end{document}

%% file: stellar_data.tex
  \object{BD+35 4258} 	 &  77.19&$-4.74$& 2.9& 2&$-0.2$&9.42 &9.46&  0.22&$   0.38\pm   0.06$&\nodata&	B0.5 Vn          	 \\
  \object{BD+53 2820} 	 & 101.24&$-1.69$& 5.1& 2&$-0.1$&10.02&9.96&  0.28&$   0.44\pm   0.07$&\nodata&	B0 IV:n           	\\
  \object{CPD-59 2603}	 & 287.59&$-0.69$& 3.5& 1&$-0.0$&8.93 &8.75&  0.43&$   0.51\pm   0.05$&0.31&	O7 Vf	             \\
  \object{CPD-59 4552}	 & 303.22&$ 2.47$& 2.0& 4&$0.08$&8.337&8.24&  0.31&$   0.40\pm   0.05$&\nodata&	B1 III           	 \\
  \object{CPD-69 1743}	 & 303.71&$-7.35$& 5.5& 2&$-0.7$&9.43 &9.46&  0.19&$   0.40\pm   0.07$&\nodata&	B0.5 IIIn        	 \\
  \object{HD108}      	 & 117.93&$ 1.25$& 3.8& 1&$0.08$&7.58 &7.40&  0.42&$   0.53\pm   0.04$&0.30&	O6 pe         	    \\
  \object{HD1383}     	 & 119.02&$-0.89$& 2.9& 2&$-0.0$&7.89 &7.63&  0.37&$   0.48\pm   0.05$&0.27&	B1 II            	 \\
  \object{HD3827}     	 & 120.79&$-23.2$& 1.8& 4&$-0.7$&7.76 &7.95&  0.05&$   0.41\pm   0.07$&0.16&	B0.7 Vn      	     \\
  \object{HD12323}    	 & 132.91&$-5.87$& 4.4& 1&$-0.4$&8.87 &8.92&  0.23&$   0.51\pm   0.05$&\nodata&	O9 V          \\
  \object{HD13268}   	 & 133.96&$-4.99$& 2.1& 1&$-0.1$&8.24 &8.18&  0.35&$   0.45\pm   0.07$&\nodata&	O8 Vnn       	  \\
  \object{HD13745}    	 & 134.58&$-4.96$& 3.2& 1&$-0.2$&7.99 &7.90&  0.34&$   0.55\pm   0.05$&\nodata&	O9.7 IIn       \\
  \object{HD13841}    	 & 134.38&$-3.93$& 2.9& 4&$-0.2$&7.60 &7.37&  0.33&$   0.58\pm   0.07$&\nodata&	B1.5Ib        \\
  \object{HD14818}    	 & 135.62&$-3.93$& 2.8& 4&$-0.1$&6.56 &6.26&  0.37&$   0.39\pm   0.07$&\nodata&	B2 Ia            	 \\
  \object{HD15137}    	 & 137.46&$-7.58$& 3.5& 1&$-0.4$&7.92 &7.86&  0.24&$   0.36\pm   0.06$&0.32&	O9.5 II-IIIn     	 \\
  \object{HD25443}    	 & 143.68&$ 7.35$& 1.1& 4&$0.14$&6.998&6.76&  0.44&$  0.72\pm   0.05$&\nodata&	B0.5 III         	 \\
  \object{HD35914}    	 & 215.21&$-24.2$& 5.4& 4&$-2.2$&9.78 &13.0&  0.37&$   0.46\pm   0.06$&\nodata&	O7fp            	  \\
  \object{HD40893}    	 & 180.09&$ 4.34$& 3.1& 2&$0.23$&9.05 &8.99&  0.31&$   0.58\pm   0.04$&0.37&	B0 IV           	  \\
  \object{HD41161}    	 & 164.97&$12.89$& 1.4& 1&$0.31$&6.658&6.76&  0.19&$ 0.35\pm   0.04$&\nodata& O8 Vn            	  \\
  \object{HD46223}    	 & 206.44&$-2.07$& 2.1& 4&$-0.0$&7.50 &7.28&  0.44&$   0.86\pm   0.03$&\nodata&	O4 Vf        	     \\
  \object{HD52266}    	 & 219.13&$-0.68$& 1.8& 2&$-0.0$&7.22 &7.23&  0.22&$   0.56\pm   0.05$&0.48&	O9 IVn        	    \\
  \object{HD53975}    	 & 225.68&$-2.32$& 1.4& 2&$-0.0$&6.40 &6.50&  0.16&$   0.38\pm   0.03$&\nodata&	O7.5 V          	  \\
  \object{HD63005}    	 & 242.47&$-0.93$& 5.4& 1&$-0.0$&9.12 &9.13&  0.22&$   0.68\pm   0.04$&\nodata&	O6 Vf            	 \\
  \object{HD66788}    	 & 245.43&$ 2.05$& 4.3& 1&$0.15$&9.35 &9.43&  0.20&$   0.55\pm   0.04$&\nodata&	O9 V        	      \\
  \object{HD69106}    	 & 254.52&$-1.33$& 1.5& 1&$-0.0$&7.020&7.13&  0.14&$   0.56\pm   0.04$&0.45& B0.5 IVnn        	  \\
  \object{HD72648}    	 & 262.23&$-2.48$& 3.8& 4&$-0.1$&7.75 &7.62&  0.21&$   0.83\pm   0.08$&\nodata&	B1 Ib          	   \\
  \object{HD75309}    	 & 265.86&$-1.90$& 2.9& 1&$-0.1$&7.82 &7.84&  0.18&$   0.58\pm   0.04$&0.46&	B1 IIp         	   \\
  \object{HD88115}    	 & 285.32&$-5.53$& 3.7& 1&$-0.3$&8.26 &8.31&  0.12&$   0.39\pm   0.08$&0.51& B1.5 Iin            \\
  \object{HD89137}    	 & 279.69&$ 4.45$& 3.1& 1&$0.24$&7.93 &7.97&  0.17&$   0.45\pm   0.04$&\nodata&	O9.7 IIInp      	  \\
  \object{HD90087}    	 & 285.16&$-2.13$& 2.8& 1&$-0.1$&7.80 &7.80&  0.22&$   0.36\pm   0.07$&\nodata&	O9 IIIn     	      \\
  \object{HD91824}    	 & 285.70&$ 0.07$& 3.0& 2&$0.00$&8.09 &8.15&  0.22&$   0.33\pm   0.04$&0.78&	O7 V      	        \\
  \object{HD91983}    	 & 285.88&$ 0.05$& 3.0& 2&$0.00$&8.62 &8.58&  0.14&$   0.31\pm   0.05$&0.51&	B1 III           	 \\
  \object{HD92554}    	 & 287.60&$-2.02$& 6.9& 1&$-0.2$&9.60 &9.50&  0.34&$   0.15\pm   0.10$&\nodata&	O9.5 IIn           \\
  \object{HD93129}    	 & 287.41&$-0.57$& 2.8& 1&$-0.0$&7.06 &6.90&  0.46&$   0.49\pm   0.05$&\nodata&	O2 If	             \\
  \object{HD93205}    	 & 287.57&$-0.71$& 3.3& 1&$-0.0$&7.80 &7.75&  0.44&$   0.33\pm   0.05$&0.58&	O3 Vf  	           \\
  \object{HD93222}    	 & 287.74&$-1.02$& 3.6& 1&$-0.0$&8.15 &8.10&  0.32&$   0.33\pm   0.04$&0.82&	O7 IIIf       	    \\
  \object{HD93843}    	 & 288.24&$-0.90$& 3.5& 1&$-0.0$&7.29 &7.33&  0.24&$   0.50\pm   0.04$&0.66&	O5 IIIf       	    \\
  \object{HD94493}    	 & 289.01&$-1.18$& 3.4& 1&$-0.0$&7.27 &7.27&  0.15 &$   0.32\pm   0.05$&0.49&	B1 Ib       	      \\
  \object{HD97175}    	 & 294.53&$-9.17$& 3.9& 4&$-0.6$&8.79 &8.87&  0.16&$   0.49\pm   0.05$&\nodata&	B0.5 III       	   \\
  \object{HD99857}    	 & 294.78&$-4.94$& 3.5& 1&$-0.3$&7.68 &7.47&  0.27&$   0.45\pm   0.05$&0.55&	B0.5 Ib            \\
  \object{HD99890}    	 & 291.75&$ 4.43$& 3.5& 1&$0.27$&8.23 &8.31&  0.15&$   0.16\pm   0.05$&\nodata&	B0 IIIn          	 \\
  \object{HD99953}    	 & 293.93&$-2.13$& 3.0& 4&$-0.1$&6.88 &6.57&  0.39&$   0.67\pm   0.06$&\nodata&	B2 Ia            	 \\
  \object{HD100199}   	 & 293.94&$-1.49$& 3.3& 1&$-0.0$&8.147&8.17&  0.19&$   0.41\pm   0.09$&\nodata&	B0 IIIne   	       \\
  \object{HD101190}   	 & 294.78&$-1.49$& 2.1& 1&$-0.0$&7.37 &7.33&  0.30&$   0.51\pm   0.05$&\nodata&	O6 Vf       	      \\
  \object{HD103779}   	 & 296.85&$-1.02$& 4.3& 2&$-0.0$&7.185&7.22&  0.17&$   0.29\pm   0.05$&0.33& B0.5 Iab       	    \\
  \object{HD104705}   	 & 297.45&$-0.34$& 5.0& 1&$-0.0$&7.78 &7.83&  0.17&$   0.40\pm   0.05$&0.53&	B0 Ib        	     \\
  \object{HD108639}   	 & 300.22&$ 1.95$& 2.4& 2&$0.08$&7.89 &7.81&  0.26&$   0.34\pm   0.04$&0.55& B0.2 III           \\
  \object{HD109399}   	 & 301.71&$-9.88$& 2.9& 2&$-0.5$&7.67 &7.67&  0.19&$   0.44\pm   0.05$&0.59&	B0.7 II         	  \\
  \object{HD111934}   	 & 303.20&$ 2.51$& 2.3& 2&$0.10$&7.12 &6.92&  0.32&$   0.33\pm   0.10$&0.63&	B1.5 Ib            \\
  \object{HD114886}   	 & 305.52&$-0.83$& 1.8& 2&$-0.0$&6.98 &6.89&  0.32&$   0.47\pm   0.05$&0.37&	O9 IIIn       	    \\
  \object{HD115071}   	 & 305.76&$ 0.15$& 2.7& 2&$0.00$&8.13 &7.97&  0.40&$   0.50\pm   0.06$&0.74&	B0.5 Vn    	       \\
  \object{HD115455}   	 & 306.06&$ 0.22$& 2.6& 1&$0.01$&8.17 &7.97&  0.40&$   0.41\pm   0.06$&0.58&	O7.5 III     	     \\
  \object{HD116781}   	 & 307.05&$-0.07$& 2.2& 1&$-0.0$&7.74 &7.62&  0.31&$   0.51\pm   0.04$&0.47&	B0 IIIne       	   \\
  \object{HD116852}   	 & 304.88&$-16.1$& 4.5& 2&$-1.2$&8.38 &8.47&  0.14&$   0.48\pm   0.04$&0.80&	O9 III 	           \\
  \object{HD122879}   	 & 312.26&$ 1.79$& 3.3& 1&$0.10$&6.64 &6.50&  0.29&$   0.51\pm   0.04$&0.49&	B0 Ia            	 \\
  \object{HD124314}   	 & 312.67&$-0.42$& 1.4& 1&$-0.0$&6.85 &6.64&  0.43&$   0.47\pm   0.05$&0.58&	O6 Vnf     	       \\
  \object{HD124979}   	 & 316.40&$ 9.08$& 2.8& 1&$0.44$&8.61 &8.51&  0.30&$   0.46\pm   0.07$&\nodata& O8 Vf               \\
  \object{HD137595}   	 & 336.72&$18.86$&0.50& 2&$0.16$&7.52 &7.49&  0.18&$   0.85\pm   0.05$&\nodata&	B3 Vn        	     \\
  \object{HD144965}   	 & 339.04&$ 8.42$&0.51& 2&$0.07$&7.206&7.11&  0.27&$   1.07\pm   0.09$&0.76& B2 Vne           	  \\
  \object{HD147888}   	 & 353.65&$17.71$&0.12& 2&$0.03$&7.05 &6.74&  0.42&$   1.15\pm   0.12$&1.20&	B3 V           	   \\
  \object{HD148422}   	 & 329.92&$-5.60$&  10& 1&$-0.9$&8.69 &8.64&  0.23&$   0.21\pm   0.09$&\nodata&	B1 Ia            	 \\
  \object{HD148937}   	 & 336.37&$-0.22$& 1.2& 4&$-0.0$&7.12 &6.71&  0.52&$   0.56\pm   0.07$&0.67&	O6 fp           	  \\
  \object{HD151805}   	 & 343.20&$ 1.59$& 6.0& 1&$0.17$&9.01 &9.01&  0.19&$   0.54\pm   0.04$&\nodata&	B1 Ib            	 \\
  \object{HD152249}   	 & 343.45&$ 1.16$& 2.1& 4&$0.04$&6.65 &6.45&  0.42&$   0.45\pm   0.06$&\nodata&	O9 Iab          	  \\
  \object{HD152424}   	 & 343.36&$ 0.89$& 2.1& 4&$0.03$&6.69 &6.27&  0.57&$   0.66\pm   0.06$&\nodata&	O9.2 Ia         	  \\
  \object{HD152590}   	 & 344.84&$ 1.83$& 3.6& 2&$0.11$&8.56 &8.48&  0.37&$   0.52\pm   0.05$&0.77&	O7 V          	    \\
  \object{HD156359}   	 & 328.68&$-14.5$&  17& 4&$-4.2$&9.52 &9.72&  0.06&$   0.24\pm   0.09$&\nodata&	B0 Ia         	    \\
  \object{HD163522}   	 & 349.57&$-9.09$& 9.9& 2&$-1.5$&8.43 &8.43&  0.16&$   0.34\pm   0.08$&\nodata&	B1 Ia            	 \\
  \object{HD165246}   	 &   6.40&$-1.56$& 1.9& 2&$-0.0$&7.70 &7.60&  0.33&$   0.78\pm   0.04$&0.79&	O8 Vn          	   \\
  \object{HD167402}   	 &   2.26&$-6.39$& 7.0& 3&$-0.7$&8.94 &8.95&  0.21&$   0.25\pm   0.07$&\nodata&	B0 II/B0.5 Ib    	 \\
  \object{HD168941}   	 &   5.82&$-6.31$& 7.8& 2&$-0.8$&9.36 &9.37&  0.24&$   0.68\pm   0.06$&\nodata&	O9.5 IIp         	 \\
  \object{HD170740}   	 &  21.06&$-0.53$&0.28& 2&$-0.0$&5.96 &5.72&  0.38&$   0.87\pm   0.09$&\nodata&	B2 IV-V          	 \\
  \object{HD177989}   	 &  17.81&$-11.8$& 6.0& 1&$-1.2$&9.22 &9.34&  0.11&$   0.66\pm   0.05$&0.35&	B0 III            	\\
  \object{HD178487}   	 &  25.78&$-8.56$& 5.7& 1&$-0.8$&8.78 &8.69&  0.29&$   0.74\pm   0.07$&\nodata& B0.5 Ib    	        \\
  \object{HD179407}   	 &  24.02&$-10.4$& 9.2& 2&$-1.6$&9.44 &9.44&  0.23&$   0.63\pm   0.09$&\nodata& B0.5 Ib          	  \\
  \object{HD185418}   	 &  53.60&$-2.17$& 1.2& 1&$-0.0$&7.639&7.49&  0.38&$   0.74\pm   0.04$&0.23& B0.5 V           	  \\
  \object{HD191877}   	 &  61.57&$-6.45$& 2.3& 1&$-0.2$&6.217&6.27&  0.14&$   0.53\pm   0.05$&\nodata& B1 Ib            	  \\
  \object{HD192035}   	 &  83.33&$ 7.76$& 2.7& 1&$0.36$&8.26 &8.22&  0.28&$   0.74\pm   0.07$&\nodata&	B0 III-IVn     	   \\
  \object{HD195455}   	 &  20.27&$-32.1$& 5.8& 2&$-3.0$&9.02 &9.20&  0.07&$   0.39\pm   0.07$&\nodata&	B0.5 III        	  \\
  \object{HD195965}   	 &  85.71&$ 5.00$& 1.1& 1&$0.10$&6.899&6.97&  0.19&$   0.45\pm   0.05$&0.32& B0 V             	  \\
  \object{HD198478}   	 &  85.75&$ 1.49$& 1.3& 2&$0.03$&5.28 &4.86&  0.43&$   0.75\pm   0.14$&0.43&	B3 Ia            	 \\
  \object{HD198781}   	 &  99.94&$12.61$&0.69& 2&$0.15$&6.472&6.45&  0.26&$   0.79\pm   0.05$&0.37& B0.5 V           	  \\
  \object{HD201345}   	 &  78.44&$-9.54$& 2.2& 1&$-0.3$&7.611&7.76&  0.14&$   0.34\pm   0.06$&0.32& O9 V       	        \\
  \object{HD202347}   	 &  88.22&$-2.08$& 1.0& 1&$-0.0$&7.41 &7.50&  0.11&$   0.66\pm   0.08$&0.20& B1.5 V             	\\
  \object{HD203374}   	 & 100.51&$ 8.62$&0.34& 2&$0.05$&6.908&6.67&  0.43&$   0.66\pm   0.04$&0.31& B2 Vne          	   \\
  \object{HD206267}   	 &  99.29&$ 3.74$&0.86& 2&$0.05$&5.83 &5.62&  0.45&$   0.80\pm   0.05$&0.30&	O6.5 V 	           \\
  \object{HD206773}   	 &  99.80&$ 3.62$&0.82& 2&$0.05$&7.10 &6.87&  0.39&$   0.60\pm   0.04$&0.19&	B0 V:nnep          \\
  \object{HD207198}   	 & 103.14&$ 6.99$& 1.3& 1&$0.16$&6.25 &5.94&  0.47&$   0.80\pm   0.05$&0.20&	O9.5 Ib-II         \\
  \object{HD207308}   	 & 103.11&$ 6.82$& 1.2& 2&$0.14$&7.74 &7.49&  0.44&$   0.82\pm   0.05$&\nodata&	B0.7 III-IV(n)     \\
  \object{HD207538}   	 & 101.60&$ 4.67$&0.94& 2&$0.07$&7.55 &7.30&  0.51&$   0.85\pm   0.05$&\nodata&	O9.5 V          	  \\
  \object{HD208440}   	 & 104.03&$ 6.44$& 1.1& 2&$0.12$&7.93 &7.91&  0.27&$   0.65\pm   0.05$&0.35&	B1 V             	 \\
  \object{HD209339}   	 & 104.58&$ 5.87$& 1.2& 2&$0.12$&6.733&6.73&  0.24&$   0.50\pm   0.05$&0.34& B0 IV          	    \\
  \object{HD210809}   	 &  99.85&$-3.13$& 4.3& 1&$-0.2$&7.61 &7.56&  0.28&$   0.31\pm   0.06$&0.29&	O9 Iab           	 \\
  \object{HD210839}   	 & 103.83&$ 2.61$& 1.1& 1&$0.05$&5.29 &5.05&  0.49&$   0.78\pm   0.04$&0.47&	O6 Infp    	       \\
  \object{HD218915}   	 & 108.06&$-6.89$& 5.0& 1&$-0.6$&7.22 &7.20&  0.21&$   0.39\pm   0.06$&0.33&	O9.5 Iabe         	\\
  \object{HD219188}   	 &  83.03&$-50.1$& 2.1& 2&$-1.6$&6.90 &7.06&  0.09 &$   0.37\pm   0.06$&0.01&	B0.5 IIIn        	 \\
  \object{HD220057}   	 & 112.13&$ 0.21$&0.77& 2&$0.00$&6.948&6.94&  0.17&$   0.69\pm   0.13$&0.35& B3 IV               \\
  \object{HD224151}   	 & 115.44&$-4.64$& 1.3& 2&$-0.1$&6.21 &6.00&  0.34&$   0.49\pm   0.05$&0.16&	B0.5 II-III      	 \\
  \object{HDE232522}  	 & 130.70&$-6.71$& 6.1& 1&$-0.7$&8.65 &8.70&  0.14&$   0.41\pm   0.04$&0.35&	B1 II            	 \\
  \object{HDE303308}  	 & 287.59&$-0.61$& 3.8& 1&$-0.0$&8.30 &8.17&  0.33&$   0.42\pm   0.05$&0.59&	O3 Vf     	        \\
  \object{HDE308813}  	 & 294.79&$-1.61$& 3.1& 1&$-0.0$&9.30 &9.32&  0.26&$   0.52\pm   0.06$&\nodata&	O9.5 V       	     \\

%% file: elements_data.tex
 BD+35 4258&17.55 (+0.14,$-0.21$)&16.26 (+0.04,$-0.03$)&13.68$\pm 0.04$&12.49$\pm 0.05$&$<12.59$\\
 BD+53 2820&17.99 (+0.09,$-0.11$)&16.37 (+0.06,$-0.04$)&13.78$\pm 0.04$&12.55 (+0.08,$-0.09$)&$<12.50$\\
CPD-59 2603&18.02 (+0.05,$-0.06$)&16.31$\pm 0.01$&13.91$\pm 0.04$&12.66 (+0.04,$-0.05$)&$<12.66$\\
CPD-59 4552&17.90 (+0.13,$-0.19$)&16.36$\pm 0.03$&13.70$\pm 0.06$&12.57 (+0.02,$-0.03$)&$<12.37$\\
CPD-69 1743&$<17.59$&16.20$\pm 0.04$&13.45 (+0.06,$-0.07$)&12.26 (+0.09,$-0.12$)&$<12.55$\\
      HD108&18.09 (+0.06,$-0.07$)&16.30$\pm 0.02$&13.69$\pm 0.05$&12.66$\pm 0.03$&$<12.51$\\
     HD1383&$<18.30$&16.42$\pm 0.02$&13.75$\pm 0.04$&12.66$\pm 0.04$&12.91 (+0.11,$-0.15$)\\
     HD3827&$<17.25$&15.48$\pm 0.03$&13.23$\pm 0.05$&$<11.95$&$<12.01$\\
    HD12323&18.08 (+0.06,$-0.07$)&16.14$\pm 0.03$&13.57 (+0.05,$-0.06$)&12.40 (+0.06,$-0.07$)&$<12.80$\\
    HD13268&18.10 (+0.08,$-0.11$)&16.44 (+0.06,$-0.04$)&13.80$\pm 0.03$&12.54 (+0.09,$-0.11$)&$<12.47$\\
    HD13745&17.97 (+0.10,$-0.14$)&16.28$\pm 0.03$&13.76 (+0.03,$-0.04$)&12.62 (+0.08,$-0.10$)&$<12.94$\\
    HD13841&$<18.29$&16.31$\pm 0.02$&13.76 (+0.10,$-0.13$)&12.75 (+0.04,$-0.05$)&$<13.13$\\
    HD14818&18.16 (+0.14,$-0.21$)&16.49$\pm 0.04$&13.78 (+0.08,$-0.10$)&12.62 (+0.06,$-0.08$)&$<12.90$\\
    HD15137&17.88$\pm 0.05$&16.22 (+0.02,$-0.01$)&13.59$\pm 0.02$&12.48$\pm 0.06$&12.74 (+0.11,$-0.15$)\\
    HD25443&18.12 (+0.05,$-0.06$)&16.17$\pm 0.01$&13.63 (+0.03,$-0.04$)&12.67$\pm 0.02$&12.66 (+0.11,$-0.16$)\\
    HD35914&\nodata&16.12$\pm 0.03$&13.31 (+0.08,$-0.09$)&12.52 (+0.06,$-0.08$)&$<12.48$\\
    HD40893&18.12$\pm 0.04$&16.30$\pm 0.01$&13.74$\pm 0.02$&12.74$\pm 0.02$&12.83 (+0.09,$-0.11$)\\
    HD41161&17.84 (+0.05,$-0.06$)&16.08$\pm 0.01$&13.44 (+0.01,$-0.02$)&12.41$\pm 0.03$&$<12.46$\\
    HD46223&18.11 (+0.04,$-0.05$)&16.44$\pm 0.08$&13.66 (+0.01,$-0.02$)&12.67$\pm 0.02$&$<12.80$\\
    HD52266&17.89 (+0.08,$-0.10$)&16.07$\pm 0.01$&13.70$\pm 0.03$&12.45$\pm 0.02$&12.55 (+0.17,$-0.27$)\\
    HD53975&17.81 (+0.06,$-0.07$)&16.06$\pm 0.01$&13.59$\pm 0.05$&12.24 (+0.03,$-0.04$)&12.28 (+0.14,$-0.21$)\\
    HD63005&17.89 (+0.07,$-0.08$)&15.98$\pm 0.02$&13.56 (+0.06,$-0.07$)&12.51 (+0.05,$-0.06$)&$<12.76$\\
    HD66788&$<17.98$&16.07$\pm 0.02$&13.30 (+0.10,$-0.12$)&$<12.27$&$<12.49$\\
    HD69106&17.61$\pm 0.05$&15.89 (+0.02,$-0.01$)&13.36$\pm 0.01$&12.24$\pm 0.02$&12.23 (+0.03,$-0.04$)\\
    HD72648&17.89 (+0.12,$-0.16$)&15.94$\pm 0.03$&13.43 (+0.03,$-0.04$)&12.45$\pm 0.03$&12.36 (+0.13,$-0.19$)\\
    HD75309&17.72 (+0.09,$-0.12$)&15.94$\pm 0.01$&13.43$\pm 0.02$&12.43$\pm 0.03$&12.08 (+0.16,$-0.26$)\\
    HD88115&17.69 (+0.14,$-0.21$)&16.04$\pm 0.01$&13.76 (+0.07,$-0.05$)&12.42$\pm 0.03$&$<12.47$\\
    HD89137&17.70 (+0.08,$-0.10$)&16.02$\pm 0.01$&13.71$\pm 0.04$&12.29 (+0.03,$-0.04$)&$<12.77$\\
    HD90087&17.90 (+0.04,$-0.05$)&16.19$\pm 0.02$&13.67$\pm 0.01$&12.50$\pm 0.02$&12.53 (+0.14,$-0.22$)\\
    HD91824&$<17.70$&16.20$\pm 0.02$&13.61$\pm 0.02$&12.48$\pm 0.02$&12.47 (+0.08,$-0.10$)\\
    HD91983&17.97 (+0.06,$-0.08$)&16.28 (+0.03,$-0.02$)&13.64 (+0.03,$-0.04$)&12.50$\pm 0.03$&$<12.14$\\
    HD92554&17.89 (+0.15,$-0.24$)&16.48$\pm 0.03$&13.80$\pm 0.03$&12.67$\pm 0.05$&$<12.31$\\
    HD93129&18.18 (+0.04,$-0.05$)&16.37$\pm 0.01$&13.85$\pm 0.05$&12.76$\pm 0.02$&$<12.38$\\
    HD93205&\nodata&16.40$\pm 0.02$&14.07$\pm 0.02$&12.65 (+0.03,$-0.04$)&$<12.68$\\
    HD93222&\nodata&16.50$\pm 0.02$&14.22$\pm 0.07$&12.71$\pm 0.02$&$<12.74$\\
    HD93843&17.93 (+0.09,$-0.12$)&16.25$\pm 0.01$&13.88$\pm 0.01$&12.55 (+0.02,$-0.03$)&$<12.35$\\
    HD94493&17.71 (+0.12,$-0.16$)&16.22$\pm 0.02$&13.68$\pm 0.02$&12.45 (+0.08,$-0.09$)&12.27 (+0.14,$-0.21$)\\
    HD97175&17.87 (+0.09,$-0.12$)&15.93$\pm 0.02$&13.62$\pm 0.09$&12.20 (+0.06,$-0.07$)&12.54 (+0.11,$-0.15$)\\
    HD99857&17.90 (+0.09,$-0.11$)&16.26$\pm 0.02$&13.70$\pm 0.02$&12.54$\pm 0.03$&12.58 (+0.09,$-0.11$)\\
    HD99890&17.92 (+0.09,$-0.12$)&16.19$\pm 0.01$&13.63$\pm 0.01$&12.41$\pm 0.03$&12.23 (+0.14,$-0.22$)\\
    HD99953&18.07 (+0.11,$-0.14$)&16.24$\pm 0.02$&13.78 (+0.04,$-0.05$)&12.73$\pm 0.02$&$<12.75$\\
   HD100199&17.92 (+0.10,$-0.13$)&16.23$\pm 0.04$&13.61$\pm 0.03$&12.61 (+0.05,$-0.06$)&$<12.57$\\
   HD101190&18.00 (+0.06,$-0.07$)&16.29$\pm 0.04$&13.64$\pm 0.03$&12.49$\pm 0.03$&12.45 (+0.17,$-0.28$)\\
   HD103779&17.80 (+0.11,$-0.16$)&16.27$\pm 0.02$&13.62 (+0.03,$-0.04$)&12.44 (+0.05,$-0.06$)&12.62 (+0.10,$-0.13$)\\
   HD104705&17.84 (+0.07,$-0.08$)&16.15$\pm 0.01$&13.67$\pm 0.02$&12.56 (+0.07,$-0.08$)&12.24 (+0.16,$-0.26$)\\
   HD108639&18.11 (+0.04,$-0.05$)&16.41$\pm 0.02$&13.97$\pm 0.05$&12.66$\pm 0.02$&12.49 (+0.07,$-0.08$)\\
   HD109399&17.69 (+0.12,$-0.17$)&16.09$\pm 0.02$&13.54 (+0.02,$-0.03$)&12.34 (+0.05,$-0.06$)&12.38 (+0.16,$-0.25$)\\
   HD111934&17.96 (+0.14,$-0.22$)&16.45 (+0.05,$-0.04$)&13.82 (+0.06,$-0.07$)&12.67$\pm 0.02$&$<12.84$\\
   HD114886&18.05 (+0.11,$-0.14$)&16.28 (+0.01,$-0.02$)&13.76$\pm 0.02$&12.72$\pm 0.02$&12.49 (+0.16,$-0.26$)\\
   HD115071&18.20 (+0.05,$-0.06$)&16.36$\pm 0.02$&13.96$\pm 0.04$&12.74$\pm 0.01$&12.56 (+0.08,$-0.10$)\\
   HD115455&18.24 (+0.09,$-0.12$)&16.44 (+0.03,$-0.02$)&13.77$\pm 0.04$&12.70$\pm 0.02$&$<13.03$\\
   HD116781&18.01 (+0.07,$-0.09$)&16.11 (+0.01,$-0.02$)&13.65$\pm 0.02$&12.49$\pm 0.04$&$<12.52$\\
   HD116852&17.74 (+0.10,$-0.13$)&15.90$\pm 0.01$&13.59$\pm 0.03$&12.17$\pm 0.04$&$<12.33$\\
   HD122879&18.04$\pm 0.05$&16.20$\pm 0.01$&13.73$\pm 0.01$&12.62$\pm 0.02$&12.55 (+0.06,$-0.07$)\\
   HD124314&18.21$\pm 0.02$&16.37 (+0.02,$-0.01$)&13.93$\pm 0.03$&12.74$\pm 0.02$&12.87 (+0.06,$-0.07$)\\
   HD124979&17.94 (+0.08,$-0.10$)&16.30$\pm 0.01$&13.88$\pm 0.02$&12.67$\pm 0.04$&$<12.60$\\
   HD137595&17.91 (+0.05,$-0.06$)&15.74$\pm 0.02$&13.56$\pm 0.05$&12.28$\pm 0.04$&12.23 (+0.09,$-0.11$)\\
   HD144965&17.89 (+0.06,$-0.08$)&15.57$\pm 0.02$&13.10 (+0.05,$-0.06$)&12.22$\pm 0.04$&12.38 (+0.08,$-0.10$)\\
   HD147888&18.21$\pm 0.02$&16.00 (+0.03,$-0.02$)&13.77 (+0.06,$-0.04$)&12.85$\pm 0.07$&12.66 (+0.08,$-0.03$)\\
   HD148422&$<18.23$&16.47$\pm 0.06$&13.88 (+0.10,$-0.12$)&$<12.54$&$<12.59$\\
   HD148937&18.40$\pm 0.03$&16.40$\pm 0.02$&14.12$\pm 0.07$&12.87$\pm 0.02$&$<12.64$\\
   HD151805&18.06 (+0.08,$-0.09$)&16.22$\pm 0.02$&13.85$\pm 0.04$&12.69$\pm 0.03$&12.62 (+0.13,$-0.19$)\\
   HD152249&18.09 (+0.08,$-0.09$)&16.35$\pm 0.02$&13.75 (+0.20,$-0.05$)&12.70$\pm 0.02$&12.68 (+0.14,$-0.21$)\\
   HD152424&18.21 (+0.08,$-0.09$)&16.31$\pm 0.01$&13.96$\pm 0.03$&12.81 (+0.03,$-0.04$)&12.78 (+0.17,$-0.27$)\\
   HD152590&18.08 (+0.12,$-0.16$)&16.30$\pm 0.02$&13.72 (+0.04,$-0.05$)&12.60$\pm 0.01$&12.64 (+0.12,$-0.16$)\\
   HD156359&$<17.79$&15.92$\pm 0.05$&13.17 (+0.14,$-0.21$)&12.02 (+0.15,$-0.24$)&$<12.69$\\
   HD163522&$<18.00$&16.18$\pm 0.03$&13.50 (+0.14,$-0.21$)&12.50 (+0.06,$-0.07$)&$<12.96$\\
   HD165246&18.01 (+0.04,$-0.05$)&16.05$\pm 0.01$&13.80$\pm 0.03$&12.55 (+0.05,$-0.06$)&12.34 (+0.07,$-0.08$)\\
   HD167402&$<18.04$&16.34$\pm 0.06$&13.72 (+0.08,$-0.10$)&12.51 (+0.10,$-0.12$)&$<12.76$\\
   HD168941&17.84 (+0.08,$-0.10$)&15.92$\pm 0.04$&13.51 (+0.06,$-0.08$)&12.50 (+0.06,$-0.08$)&$<12.62$\\
   HD170740&18.01 (+0.05,$-0.06$)&15.91$\pm 0.04$&13.29 (+0.06,$-0.07$)&12.36$\pm 0.04$&12.43 (+0.14,$-0.20$)\\
   HD177989&17.75$\pm 0.05$&15.84$\pm 0.01$&13.44$\pm 0.01$&12.24$\pm 0.03$&12.24 (+0.07,$-0.09$)\\
   HD178487&$<17.83$&15.97$\pm 0.04$&13.59 (+0.10,$-0.13$)&12.38 (+0.11,$-0.15$)&$<12.69$\\
   HD179407&17.82 (+0.15,$-0.24$)&15.98 (+0.07,$-0.09$)&13.60 (+0.06,$-0.08$)&12.46 (+0.13,$-0.19$)&$<12.78$\\
   HD185418&18.00$\pm 0.03$&16.03 (+0.02,$-0.01$)&13.72$\pm 0.03$&12.49$\pm 0.01$&12.61$\pm 0.03$\\
   HD191877&17.80 (+0.07,$-0.09$)&15.92$\pm 0.01$&13.46$\pm 0.02$&12.31$\pm 0.04$&12.02 (+0.13,$-0.19$)\\
   HD192035&18.01 (+0.05,$-0.06$)&16.07 (+0.05,$-0.04$)&13.40 (+0.05,$-0.06$)&12.49$\pm 0.04$&$<12.51$\\
   HD195455&17.62 (+0.10,$-0.14$)&15.58$\pm 0.02$&13.10 (+0.05,$-0.06$)&11.70 (+0.11,$-0.14$)&12.33 (+0.14,$-0.21$)\\
   HD195965&17.74 (+0.05,$-0.06$)&16.00 (+0.03,$-0.02$)&13.43 (+0.05,$-0.06$)&12.35$\pm 0.03$&$<11.76$\\
   HD198478&18.19 (+0.11,$-0.15$)&16.13$\pm 0.05$&13.83 (+0.09,$-0.08$)&12.68 (+0.03,$-0.04$)&$<12.56$\\
   HD198781&17.87 (+0.04,$-0.05$)&15.74$\pm 0.01$&13.35$\pm 0.02$&12.34$\pm 0.02$&12.42 (+0.07,$-0.08$)\\
   HD201345&17.72 (+0.06,$-0.07$)&16.02$\pm 0.02$&13.47$\pm 0.01$&12.44 (+0.04,$-0.05$)&12.00 (+0.11,$-0.15$)\\
   HD202347&17.43 (+0.14,$-0.22$)&15.63$\pm 0.02$&13.18 (+0.06,$-0.07$)&12.16$\pm 0.04$&12.38 (+0.07,$-0.09$)\\
   HD203374&17.87$\pm 0.05$&16.09$\pm 0.01$&13.62$\pm 0.02$&12.59$\pm 0.02$&12.50 (+0.05,$-0.06$)\\
   HD206267&18.10$\pm 0.03$&16.04$\pm 0.02$&13.62$\pm 0.02$&12.58$\pm 0.03$&12.59 (+0.08,$-0.10$)\\
   HD206773&17.89$\pm 0.03$&15.98$\pm 0.01$&13.53 (+0.01,$-0.02$)&12.45$\pm 0.03$&12.49 (+0.10,$-0.12$)\\
   HD207198&18.15$\pm 0.03$&16.06$\pm 0.01$&13.66$\pm 0.02$&12.61$\pm 0.02$&12.67 (+0.07,$-0.09$)\\
   HD207308&17.97 (+0.08,$-0.10$)&16.01$\pm 0.04$&13.55$\pm 0.04$&12.51$\pm 0.03$&12.33 (+0.16,$-0.25$)\\
   HD207538&18.17 (+0.04,$-0.05$)&16.02$\pm 0.02$&13.56 (+0.09,$-0.12$)&12.58 (+0.05,$-0.06$)&$<12.70$\\
   HD208440&17.98$\pm 0.04$&16.04$\pm 0.01$&13.69$\pm 0.04$&12.51$\pm 0.02$&12.34 (+0.10,$-0.12$)\\
   HD209339&17.82$\pm 0.05$&16.11 (+0.03,$-0.02$)&13.73$\pm 0.05$&12.49$\pm 0.02$&12.41 (+0.07,$-0.09$)\\
   HD210809&17.88 (+0.08,$-0.10$)&16.39$\pm 0.03$&13.96$\pm 0.06$&12.59$\pm 0.04$&$<12.44$\\
   HD210839&18.17$\pm 0.03$&16.04$\pm 0.01$&13.59$\pm 0.02$&12.64$\pm 0.03$&$<12.45$\\
   HD218915&17.72 (+0.11,$-0.15$)&16.22$\pm 0.02$&13.60$\pm 0.03$&12.43$\pm 0.05$&$<12.53$\\
   HD219188&17.30 (+0.08,$-0.10$)&15.72$\pm 0.02$&13.10$\pm 0.02$&12.02$\pm 0.04$&11.95 (+0.10,$-0.13$)\\
   HD220057&17.76 (+0.05,$-0.06$)&15.80 (+0.04,$-0.02$)&13.44$\pm 0.05$&12.22$\pm 0.03$&12.26 (+0.09,$-0.12$)\\
   HD224151&18.06 (+0.07,$-0.09$)&16.30$\pm 0.01$&13.75$\pm 0.03$&12.65$\pm 0.04$&$<12.71$\\
  HDE232522&18.02 (+0.13,$-0.19$)&16.15 (+0.01,$-0.02$)&13.61$\pm 0.04$&12.45$\pm 0.03$&$<12.40$\\
  HDE303308&\nodata&16.39$\pm 0.02$&14.13$\pm 0.01$&12.70$\pm 0.03$&12.53 (+0.14,$-0.22$)\\
  HDE308813&17.81 (+0.12,$-0.16$)&16.12 (+0.03,$-0.04$)&13.62 (+0.04,$-0.05$)&12.53$\pm 0.04$&12.74 (+0.12,$-0.17$)\\

%% file: H_data.tex
&&\\
BD+35 4258&$ 19.09^{+0.14}_{-0.19}$&$ 19.45^{+0.13}_{-0.18}$
&$ 19.61^{+0.11}_{-0.13}$&$ 21.24^{+0.03}_{-0.07}$&$ 21.26^{+0.03}_{-0.07}$
&$ 124^{+  83}_{-  35}$&$ -1.34^{+0.12}_{-0.14}$\\
BD+53 2820&$ 19.70^{+0.11}_{-0.26}$&$ 19.89\pm 0.15$&$ 20.10^{+0.11}_{-0.13}$
&$ 21.35^{+0.05}_{-0.07}$&$ 21.40^{+0.05}_{-0.06}$&$ 100^{+  58}_{-  24}$
&$ -0.99^{+0.12}_{-0.14}$\\
CPD-59 2603&$ 19.51^{+0.25}_{-0.08}$&$ 19.82\pm 0.11$&$ 20.02^{+0.12}_{-0.10}$
&$ 21.43^{+0.04}_{-0.05}$&$ 21.46^{+0.04}_{-0.05}$&$ 112^{+  41}_{-  31}$
&$ -1.14^{+0.13}_{-0.10}$\\
CPD-59 4552&$ 20.13^{+0.11}_{-0.09}$&$ 20.15^{+0.10}_{-0.14}$&$ 20.45\pm 0.08$
&$ 21.28^{+0.05}_{-0.06}$&$ 21.39^{+0.04}_{-0.05}$&$  78^{+  16}_{-  11}$
&$ -0.64\pm 0.09$\\
CPD-69 1743&$ 19.58^{+0.23}_{-0.16}$&$ 19.49^{+0.13}_{-0.21}$
&$ 19.85^{+0.15}_{-0.13}$&$ 21.16^{+0.04}_{-0.09}$&$ 21.20^{+0.04}_{-0.08}$
&$  68^{+  22}_{-  13}$&$ -1.04^{+0.17}_{-0.14}$\\
HD108&$ 20.17^{+0.12}_{-0.07}$&$ 20.10\pm 0.09$&$ 20.45^{+0.08}_{-0.06}$
&$ 21.38\pm 0.05$&$ 21.47\pm 0.04$&$  72^{+  11}_{-   8}$
&$ -0.72^{+0.09}_{-0.08}$\\
HD1383&$ 20.05\pm 0.11$&$ 20.28^{+0.09}_{-0.10}$&$ 20.49\pm 0.07$
&$ 21.46^{+0.05}_{-0.06}$&$ 21.54^{+0.04}_{-0.05}$&$ 103^{+  26}_{-  17}$
&$ -0.75^{+0.08}_{-0.09}$\\
HD3827&$ 17.71^{+0.18}_{-0.50}$&$ 18.24\pm 0.30$&$ 18.35^{+0.25}_{-0.23}$
&$ 20.55^{+0.07}_{-0.05}$&$ 20.56^{+0.07}_{-0.05}$&$ 203^{+\infty}_{- 113}$
&$ -1.92^{+0.25}_{-0.24}$\\
HD12323&$ 19.87^{+0.12}_{-0.15}$&$ 20.04^{+0.09}_{-0.17}$
&$ 20.26^{+0.08}_{-0.11}$&$ 21.19^{+0.04}_{-0.05}$&$ 21.28\pm 0.04$
&$  93^{+  30}_{-  18}$&$ -0.72^{+0.09}_{-0.12}$\\
HD13268&$ 20.03\pm 0.13$&$ 20.24\pm 0.10$&$ 20.46\pm 0.08$&$ 21.34\pm 0.07$
&$ 21.44\pm 0.06$&$  99^{+  29}_{-  18}$&$ -0.68\pm 0.10$\\
HD13745&$ 20.25\pm 0.07$&$ 20.20^{+0.12}_{-0.14}$&$ 20.53\pm 0.07$
&$ 21.34\pm 0.05$&$ 21.46\pm 0.04$&$  74^{+  12}_{-  10}$&$ -0.63\pm 0.08$\\
HD13841&$ 20.28^{+0.08}_{-0.10}$&$ 20.23\pm 0.18$&$ 20.56\pm 0.10$
&$ 21.44^{+0.06}_{-0.12}$&$ 21.54^{+0.05}_{-0.09}$&$  74^{+  19}_{-  13}$
&$ -0.66^{+0.14}_{-0.12}$\\
HD14818&$ 20.22\pm 0.18$&$ 20.19^{+0.11}_{-0.08}$&$ 20.52^{+0.11}_{-0.10}$
&$ 21.40\pm 0.07$&$ 21.51\pm 0.06$&$  77^{+  19}_{-  14}$
&$ -0.68^{+0.13}_{-0.12}$\\
HD15137&$ 19.81^{+0.08}_{-0.11}$&$ 20.03^{+0.11}_{-0.21}$
&$ 20.23^{+0.08}_{-0.12}$&$ 21.24^{+0.08}_{-0.06}$&$ 21.32^{+0.07}_{-0.05}$
&$ 101^{+  29}_{-  24}$&$ -0.80^{+0.10}_{-0.14}$\\
HD25443&$ 20.62^{+0.07}_{-0.10}$&$ 20.61\pm 0.13$&$ 20.92\pm 0.08$
&$ 21.29^{+0.06}_{-0.05}$&$ 21.56\pm 0.05$&$  78^{+  15}_{-  11}$
&$ -0.34\pm 0.09$\\
HD35914&$ 19.54\pm 0.07$&$ 19.06\pm 0.14$&$ 19.67\pm 0.06$
&$ 21.18^{+0.07}_{-0.04}$&$ 21.21^{+0.07}_{-0.04}$&$  52^{+   6}_{-   5}$
&$ -1.24^{+0.08}_{-0.09}$\\
HD40893&$ 20.28^{+0.02}_{-0.04}$&$ 20.30^{+0.05}_{-0.06}$&$ 20.59\pm 0.03$
&$ 21.46^{+0.04}_{-0.06}$&$ 21.56^{+0.03}_{-0.05}$&$  79\pm   5$
&$ -0.67^{+0.06}_{-0.05}$\\
HD41161&$ 19.60\pm 0.12$&$ 19.76^{+0.12}_{-0.11}$&$ 19.99^{+0.09}_{-0.08}$
&$ 21.09\pm 0.05$&$ 21.16\pm 0.04$&$  94^{+  25}_{-  16}$&$ -0.86\pm 0.10$\\
HD46223&$ 20.37^{+0.04}_{-0.09}$&$ 20.37^{+0.10}_{-0.06}$
&$ 20.67^{+0.06}_{-0.05}$&$ 21.46^{+0.04}_{-0.02}$&$ 21.59^{+0.03}_{-0.02}$
&$  80^{+  11}_{-   8}$&$ -0.62^{+0.07}_{-0.06}$\\
HD52266&$ 19.68\pm 0.09$&$ 19.70^{+0.13}_{-0.12}$&$ 20.00\pm 0.08$
&$ 21.22^{+0.04}_{-0.05}$&$ 21.27^{+0.04}_{-0.05}$&$  79^{+  16}_{-  11}$
&$ -0.97\pm 0.09$\\
HD53975&$ 18.75^{+0.16}_{-0.19}$&$ 18.90^{+0.11}_{-0.07}$&$ 19.15\pm 0.09$
&$ 21.08^{+0.04}_{-0.02}$&$ 21.09^{+0.04}_{-0.02}$&$  94^{+  31}_{-  19}$
&$ -1.65\pm 0.10$\\
HD63005&$ 19.78^{+0.14}_{-0.21}$&$ 19.92^{+0.10}_{-0.04}$&$ 20.17\pm 0.09$
&$ 21.24^{+0.03}_{-0.06}$&$ 21.31^{+0.03}_{-0.05}$&$  94^{+  31}_{-  18}$
&$ -0.83\pm 0.10$\\
HD66788&$ 19.31^{+0.21}_{-0.03}$&$ 19.43^{+0.17}_{-0.12}$
&$ 19.72^{+0.14}_{-0.10}$&$ 21.23^{+0.04}_{-0.02}$&$ 21.26^{+0.04}_{-0.02}$
&$  85^{+  30}_{-  17}$&$ -1.24^{+0.14}_{-0.11}$\\
HD69106&$ 19.40\pm 0.10$&$ 19.49\pm 0.10$&$ 19.75\pm 0.07$&$ 21.07\pm 0.04$
&$ 21.11\pm 0.04$&$  86^{+  17}_{-  12}$&$ -1.06\pm 0.08$\\
HD72648&$ 20.30\pm 0.19$&$ 20.45\pm 0.20$&$ 20.70\pm 0.15$
&$ 21.19^{+0.07}_{-0.06}$&$ 21.41\pm 0.07$&$  92^{+  48}_{-  23}$
&$ -0.42\pm 0.16$\\
HD75309&$ 19.77^{+0.10}_{-0.09}$&$ 19.93^{+0.08}_{-0.15}$
&$ 20.16^{+0.06}_{-0.09}$&$ 21.10^{+0.03}_{-0.06}$&$ 21.19^{+0.03}_{-0.05}$
&$  91^{+  20}_{-  15}$&$ -0.72^{+0.08}_{-0.10}$\\
HD88115&$ 18.81\pm 0.18$&$ 19.02\pm 0.18$&$ 19.25\pm 0.14$
&$ 21.03^{+0.06}_{-0.09}$&$ 21.04^{+0.06}_{-0.09}$&$  99^{+  53}_{-  26}$
&$ -1.49^{+0.16}_{-0.15}$\\
HD89137&$ 19.62^{+0.14}_{-0.08}$&$ 19.75^{+0.12}_{-0.07}$
&$ 20.02^{+0.09}_{-0.07}$&$ 21.03^{+0.07}_{-0.02}$&$ 21.11^{+0.06}_{-0.02}$
&$  91^{+  20}_{-  15}$&$ -0.81^{+0.10}_{-0.09}$\\
HD90087&$ 19.53^{+0.11}_{-0.06}$&$ 19.61\pm 0.10$&$ 19.88\pm 0.07$
&$ 21.19^{+0.05}_{-0.09}$&$ 21.23^{+0.05}_{-0.08}$&$  83^{+  15}_{-  11}$
&$ -1.04^{+0.11}_{-0.09}$\\
HD91824&$ 19.59\pm 0.13$&$ 19.43^{+0.07}_{-0.21}$&$ 19.81^{+0.09}_{-0.11}$
&$ 21.12\pm 0.04$&$ 21.16\pm 0.04$&$  65^{+  13}_{-  10}$
&$ -1.05^{+0.10}_{-0.11}$\\
HD91983&$ 19.86^{+0.11}_{-0.06}$&$ 19.74^{+0.03}_{-0.13}$&$ 20.10\pm 0.07$
&$ 21.15^{+0.06}_{-0.05}$&$ 21.22^{+0.05}_{-0.04}$&$  66^{+   9}_{-   7}$
&$ -0.82\pm 0.08$\\
HD92554&$ 18.97^{+0.11}_{-0.07}$&$ 18.90^{+0.08}_{-0.21}$
&$ 19.24^{+0.07}_{-0.10}$&$ 21.34^{+0.09}_{-0.11}$&$ 21.35^{+0.09}_{-0.11}$
&$  70^{+  12}_{-  11}$&$ -1.81\pm 0.13$\\
HD93129&$ 19.85^{+0.09}_{-0.05}$&$ 19.94^{+0.10}_{-0.07}$
&$ 20.21^{+0.07}_{-0.05}$&$ 21.47^{+0.07}_{-0.04}$&$ 21.52^{+0.06}_{-0.04}$
&$  86^{+  14}_{-  10}$&$ -1.01\pm 0.08$\\
HD93205&$ 19.34\pm 0.17$&$ 19.47^{+0.15}_{-0.10}$&$ 19.73^{+0.12}_{-0.10}$
&$ 21.36\pm 0.05$&$ 21.38\pm 0.05$&$  92^{+  32}_{-  20}$
&$ -1.35^{+0.13}_{-0.12}$\\
HD93222&$ 19.49^{+0.13}_{-0.10}$&$ 19.43^{+0.13}_{-0.18}$
&$ 19.77^{+0.09}_{-0.10}$&$ 21.47^{+0.03}_{-0.04}$&$ 21.49^{+0.03}_{-0.04}$
&$  72^{+  16}_{-  11}$&$ -1.41\pm 0.10$\\
HD93843&$ 19.14^{+0.17}_{-0.12}$&$ 19.42\pm 0.13$&$ 19.62\pm 0.10$
&$ 21.30^{+0.05}_{-0.04}$&$ 21.32^{+0.05}_{-0.04}$&$ 108^{+  42}_{-  24}$
&$ -1.40\pm 0.11$\\
HD94493&$ 19.78\pm 0.07$&$ 19.80^{+0.09}_{-0.17}$&$ 20.09^{+0.06}_{-0.09}$
&$ 21.10^{+0.06}_{-0.05}$&$ 21.18^{+0.05}_{-0.04}$&$  79^{+  13}_{-  12}$
&$ -0.79^{+0.08}_{-0.10}$\\
HD97175&$ 19.80^{+0.12}_{-0.09}$&$ 19.75^{+0.14}_{-0.07}$
&$ 20.10^{+0.09}_{-0.07}$&$ 20.96^{+0.06}_{-0.05}$&$ 21.07^{+0.05}_{-0.04}$
&$  74^{+  14}_{-  10}$&$ -0.67^{+0.10}_{-0.09}$\\
HD99857&$ 19.97\pm 0.07$&$ 20.02^{+0.06}_{-0.18}$&$ 20.30^{+0.05}_{-0.09}$
&$ 21.27^{+0.07}_{-0.03}$&$ 21.35^{+0.06}_{-0.03}$&$  80\pm  12$
&$ -0.77^{+0.07}_{-0.11}$\\
HD99890&$ 19.18^{+0.14}_{-0.09}$&$ 19.30\pm 0.12$&$ 19.56\pm 0.09$
&$ 21.12^{+0.05}_{-0.07}$&$ 21.14^{+0.05}_{-0.07}$&$  88^{+  23}_{-  15}$
&$ -1.27^{+0.11}_{-0.10}$\\
HD99953&$ 20.39\pm 0.17$&$ 20.12^{+0.22}_{-0.20}$&$ 20.60\pm 0.14$
&$ 21.45\pm 0.07$&$ 21.56\pm 0.06$&$  61^{+  17}_{-  11}$&$ -0.66\pm 0.15$\\
HD100199&$ 19.81\pm 0.09$&$ 19.87^{+0.13}_{-0.08}$&$ 20.15^{+0.08}_{-0.07}$
&$ 21.18^{+0.06}_{-0.13}$&$ 21.26^{+0.05}_{-0.11}$&$  84^{+  16}_{-  11}$
&$ -0.79^{+0.13}_{-0.09}$\\
HD101190&$ 20.22^{+0.05}_{-0.13}$&$ 20.03^{+0.04}_{-0.10}$
&$ 20.43^{+0.04}_{-0.08}$&$ 21.24^{+0.03}_{-0.06}$&$ 21.35^{+0.03}_{-0.05}$
&$  64^{+  10}_{-   6}$&$ -0.62^{+0.07}_{-0.09}$\\
HD103779&$ 19.46\pm 0.15$&$ 19.68\pm 0.08$&$ 19.90\pm 0.08$&$ 21.17\pm 0.05$
&$ 21.21\pm 0.05$&$ 101^{+  32}_{-  19}$&$ -1.02\pm 0.09$\\
HD104705&$ 19.67^{+0.09}_{-0.10}$&$ 19.76^{+0.09}_{-0.17}$
&$ 20.02^{+0.07}_{-0.10}$&$ 21.15^{+0.06}_{-0.05}$&$ 21.21^{+0.05}_{-0.04}$
&$  85^{+  18}_{-  14}$&$ -0.90^{+0.08}_{-0.11}$\\
HD108639&$ 19.68\pm 0.14$&$ 19.72^{+0.13}_{-0.19}$&$ 20.01^{+0.10}_{-0.11}$
&$ 21.36^{+0.04}_{-0.03}$&$ 21.40^{+0.04}_{-0.03}$&$  80^{+  24}_{-  15}$
&$ -1.09^{+0.10}_{-0.12}$\\
HD109399&$ 19.76\pm 0.12$&$ 19.62\pm 0.12$&$ 20.01\pm 0.09$
&$ 21.11^{+0.05}_{-0.06}$&$ 21.17^{+0.04}_{-0.05}$&$  67^{+  12}_{-   9}$
&$ -0.86\pm 0.10$\\
HD111934&$ 20.11\pm 0.21$&$ 19.98^{+0.17}_{-0.13}$&$ 20.37^{+0.15}_{-0.14}$
&$ 21.32^{+0.06}_{-0.14}$&$ 21.41^{+0.06}_{-0.11}$&$  69^{+  21}_{-  14}$
&$ -0.72^{+0.18}_{-0.16}$\\
HD114886&$ 19.90\pm 0.11$&$ 20.07^{+0.08}_{-0.12}$&$ 20.30^{+0.07}_{-0.08}$
&$ 21.34^{+0.06}_{-0.05}$&$ 21.41^{+0.05}_{-0.04}$&$  92^{+  21}_{-  14}$
&$ -0.82^{+0.08}_{-0.09}$\\
HD115071&$ 20.37^{+0.04}_{-0.11}$&$ 20.29^{+0.10}_{-0.14}$
&$ 20.63^{+0.06}_{-0.08}$&$ 21.39^{+0.04}_{-0.09}$&$ 21.51^{+0.04}_{-0.07}$
&$  73^{+  11}_{-  10}$&$ -0.58\pm 0.09$\\
HD115455&$ 20.24\pm 0.06$&$ 20.24^{+0.12}_{-0.07}$&$ 20.55^{+0.07}_{-0.05}$
&$ 21.38^{+0.05}_{-0.09}$&$ 21.49^{+0.04}_{-0.07}$&$  78^{+  12}_{-   8}$
&$ -0.63^{+0.10}_{-0.07}$\\
HD116781&$ 19.66^{+0.09}_{-0.11}$&$ 19.86\pm 0.12$&$ 20.08^{+0.09}_{-0.08}$
&$ 21.21\pm 0.05$&$ 21.27\pm 0.04$&$  99^{+  27}_{-  18}$&$ -0.89\pm 0.10$\\
HD116852&$ 19.48^{+0.15}_{-0.11}$&$ 19.40\pm 0.09$&$ 19.75^{+0.09}_{-0.07}$
&$ 20.96\pm 0.04$&$ 21.01\pm 0.04$&$  71^{+  12}_{-  10}$
&$ -0.96^{+0.10}_{-0.08}$\\
HD122879&$ 19.88^{+0.11}_{-0.08}$&$ 20.12\pm 0.09$&$ 20.33^{+0.07}_{-0.06}$
&$ 21.31^{+0.06}_{-0.05}$&$ 21.39^{+0.05}_{-0.04}$&$ 103^{+  22}_{-  16}$
&$ -0.77\pm 0.08$\\
HD124314&$ 20.13\pm 0.09$&$ 20.06^{+0.16}_{-0.14}$&$ 20.41^{+0.09}_{-0.08}$
&$ 21.41^{+0.06}_{-0.04}$&$ 21.49^{+0.05}_{-0.04}$&$  72^{+  15}_{-  10}$
&$ -0.79\pm 0.10$\\
HD124979&$ 20.08\pm 0.07$&$ 20.12^{+0.05}_{-0.15}$&$ 20.39^{+0.05}_{-0.08}$
&$ 21.27\pm 0.09$&$ 21.37\pm 0.07$&$  79\pm  10$&$ -0.69^{+0.09}_{-0.11}$\\
HD137595&$ 20.26^{+0.09}_{-0.07}$&$ 20.31\pm 0.08$&$ 20.59\pm 0.06$
&$ 20.97^{+0.06}_{-0.08}$&$ 21.24^{+0.04}_{-0.05}$&$  81^{+  12}_{-   9}$
&$ -0.34^{+0.08}_{-0.07}$\\
HD144965&$ 20.47^{+0.04}_{-0.05}$&$ 20.39^{+0.13}_{-0.06}$
&$ 20.74^{+0.07}_{-0.04}$&$ 20.97^{+0.09}_{-0.30}$&$ 21.31^{+0.06}_{-0.12}$
&$  72^{+  10}_{-   6}$&$ -0.25^{+0.13}_{-0.08}$\\
HD147888&$ 20.36^{+0.05}_{-0.08}$&$ 19.73\pm 0.13$&$ 20.45^{+0.05}_{-0.07}$
&$ 21.68^{+0.08}_{-0.19}$&$ 21.73^{+0.07}_{-0.17}$&$  47^{+   5}_{-   4}$
&$ -0.97^{+0.18}_{-0.11}$\\
HD148422&$ 19.92\pm 0.13$&$ 19.71^{+0.19}_{-0.10}$&$ 20.15^{+0.11}_{-0.10}$
&$ 21.24^{+0.09}_{-0.06}$&$ 21.31^{+0.08}_{-0.05}$&$  65^{+  14}_{-   9}$
&$ -0.87^{+0.12}_{-0.13}$\\
HD148937&$ 20.40^{+0.08}_{-0.04}$&$ 20.35\pm 0.09$&$ 20.68^{+0.06}_{-0.05}$
&$ 21.48^{+0.06}_{-0.11}$&$ 21.60^{+0.05}_{-0.08}$&$  73^{+   9}_{-   7}$
&$ -0.61^{+0.10}_{-0.08}$\\
HD151805&$ 20.15^{+0.08}_{-0.11}$&$ 19.91^{+0.14}_{-0.19}$
&$ 20.35^{+0.07}_{-0.09}$&$ 21.33\pm 0.05$&$ 21.41\pm 0.04$
&$  62^{+  11}_{-   9}$&$ -0.76^{+0.09}_{-0.10}$\\
HD152249&$ 20.02\pm 0.11$&$ 19.90\pm 0.14$&$ 20.28\pm 0.09$&$ 21.38\pm 0.07$
&$ 21.45\pm 0.06$&$  69^{+  14}_{-  10}$&$ -0.87\pm 0.11$\\
HD152424&$ 20.27\pm 0.06$&$ 20.41^{+0.06}_{-0.19}$&$ 20.65^{+0.05}_{-0.11}$
&$ 21.48^{+0.06}_{-0.07}$&$ 21.59^{+0.05}_{-0.06}$&$  90^{+  14}_{-  17}$
&$ -0.65^{+0.08}_{-0.12}$\\
HD152590&$ 20.21\pm 0.12$&$ 20.08^{+0.17}_{-0.15}$&$ 20.47\pm 0.10$
&$ 21.37^{+0.07}_{-0.03}$&$ 21.47^{+0.06}_{-0.03}$&$  69^{+  16}_{-  11}$
&$ -0.72\pm 0.11$\\
HD156359&$ 17.74\pm 0.27$&$ 17.70\pm 0.30$&$ 18.07^{+0.21}_{-0.23}$
&$ 20.80^{+0.10}_{-0.06}$&$ 20.80^{+0.10}_{-0.06}$&$  75^{+  51}_{-  22}$
&$ -2.45^{+0.22}_{-0.25}$\\
HD163522&$ 19.26^{+0.26}_{-0.11}$&$ 19.36\pm 0.25$&$ 19.66^{+0.19}_{-0.18}$
&$ 21.14^{+0.08}_{-0.07}$&$ 21.17^{+0.08}_{-0.07}$&$  82^{+  52}_{-  21}$
&$ -1.21\pm 0.20$\\
HD165246&$ 19.78\pm 0.12$&$ 19.87^{+0.09}_{-0.06}$&$ 20.14^{+0.07}_{-0.06}$
&$ 21.41^{+0.03}_{-0.04}$&$ 21.46^{+0.03}_{-0.04}$&$  87^{+  16}_{-  13}$
&$ -1.01^{+0.08}_{-0.07}$\\
HD167402&$ 19.80^{+0.14}_{-0.07}$&$ 19.69\pm 0.11$&$ 20.06^{+0.10}_{-0.07}$
&$ 21.13^{+0.05}_{-0.04}$&$ 21.20\pm 0.04$&$  69^{+  12}_{-   9}$
&$ -0.84^{+0.10}_{-0.08}$\\
HD168941&$ 19.84\pm 0.07$&$ 19.76^{+0.14}_{-0.11}$&$ 20.11^{+0.08}_{-0.07}$
&$ 21.18^{+0.05}_{-0.07}$&$ 21.25^{+0.04}_{-0.06}$&$  72^{+  13}_{-   8}$
&$ -0.84^{+0.10}_{-0.08}$\\
HD170740&$ 20.67^{+0.08}_{-0.17}$&$ 20.40\pm 0.15$&$ 20.86^{+0.08}_{-0.12}$
&$ 21.09^{+0.06}_{-0.17}$&$ 21.41^{+0.06}_{-0.10}$&$  62^{+  12}_{-   9}$
&$ -0.25^{+0.13}_{-0.14}$\\
HD177989&$ 19.99\pm 0.12$&$ 19.61^{+0.24}_{-0.22}$&$ 20.16\pm 0.12$
&$ 20.99^{+0.05}_{-0.06}$&$ 21.11\pm 0.05$&$  56^{+  14}_{-   9}$
&$ -0.64\pm 0.13$\\
HD178487&$ 20.25\pm 0.11$&$ 20.14\pm 0.15$&$ 20.51\pm 0.09$
&$ 21.22^{+0.04}_{-0.10}$&$ 21.36^{+0.04}_{-0.08}$&$  70^{+  15}_{-  10}$
&$ -0.54^{+0.12}_{-0.11}$\\
HD179407&$ 19.97\pm 0.14$&$ 19.90^{+0.08}_{-0.19}$&$ 20.23^{+0.09}_{-0.11}$
&$ 21.20^{+0.06}_{-0.10}$&$ 21.28^{+0.05}_{-0.08}$&$  70^{+  17}_{-  12}$
&$ -0.75\pm 0.12$\\
HD185418&$ 20.34^{+0.11}_{-0.07}$&$ 20.47\pm 0.08$&$ 20.72^{+0.07}_{-0.06}$
&$ 21.19^{+0.05}_{-0.04}$&$ 21.42^{+0.04}_{-0.03}$&$  89^{+  15}_{-  12}$
&$ -0.40\pm 0.07$\\
HD191877&$ 19.73^{+0.27}_{-0.10}$&$ 19.68^{+0.03}_{-0.17}$
&$ 20.00^{+0.17}_{-0.10}$&$ 21.03\pm 0.05$&$ 21.11\pm 0.05$
&$  68^{+  19}_{-  12}$&$ -0.81^{+0.17}_{-0.11}$\\
HD192035&$ 20.40^{+0.07}_{-0.03}$&$ 20.18^{+0.17}_{-0.03}$
&$ 20.63^{+0.09}_{-0.05}$&$ 21.20^{+0.04}_{-0.10}$&$ 21.39^{+0.04}_{-0.07}$
&$  63^{+  12}_{-   4}$&$ -0.44^{+0.11}_{-0.07}$\\
HD195455&$ 17.96^{+0.12}_{-0.31}$&$ 18.24^{+0.26}_{-0.21}$
&$ 18.42^{+0.19}_{-0.16}$&$ 20.61^{+0.04}_{-0.09}$&$ 20.62^{+0.04}_{-0.09}$
&$ 122^{+ 146}_{-  44}$&$ -1.87^{+0.21}_{-0.17}$\\
HD195965&$ 19.99^{+0.11}_{-0.12}$&$ 19.95^{+0.10}_{-0.11}$&$ 20.28\pm 0.08$
&$ 20.92\pm 0.05$&$ 21.09\pm 0.04$&$  74^{+  14}_{-  10}$&$ -0.51\pm 0.09$\\
HD198478&$ 20.43^{+0.33}_{-0.19}$&$ 20.21^{+0.53}_{-0.23}$
&$ 20.76^{+0.33}_{-0.35}$&$ 21.32^{+0.12}_{-0.13}$&$ 21.53^{+0.13}_{-0.12}$
&$  63^{+  63}_{-  17}$&$ -0.48^{+0.35}_{-0.37}$\\
HD198781&$ 20.27^{+0.08}_{-0.07}$&$ 20.04^{+0.16}_{-0.28}$
&$ 20.48^{+0.08}_{-0.10}$&$ 20.93^{+0.07}_{-0.03}$&$ 21.17\pm 0.05$
&$  61^{+  12}_{-  11}$&$ -0.40^{+0.10}_{-0.11}$\\
HD201345&$ 18.83^{+0.06}_{-0.20}$&$ 19.22\pm 0.16$&$ 19.36\pm 0.12$
&$ 21.00^{+0.05}_{-0.06}$&$ 21.02^{+0.05}_{-0.06}$&$ 141^{+ 100}_{-  43}$
&$ -1.35^{+0.14}_{-0.13}$\\
HD202347&$ 19.62^{+0.17}_{-0.15}$&$ 19.71^{+0.09}_{-0.11}$
&$ 19.98^{+0.10}_{-0.09}$&$ 20.83^{+0.08}_{-0.10}$&$ 20.94^{+0.07}_{-0.08}$
&$  85^{+  24}_{-  15}$&$ -0.66^{+0.12}_{-0.11}$\\
HD203374&$ 20.36^{+0.09}_{-0.05}$&$ 20.35^{+0.10}_{-0.06}$
&$ 20.67^{+0.07}_{-0.05}$&$ 21.20^{+0.05}_{-0.04}$&$ 21.41^{+0.04}_{-0.03}$
&$  76^{+  10}_{-   7}$&$ -0.44^{+0.07}_{-0.06}$\\
HD206267&$ 20.59^{+0.10}_{-0.06}$&$ 20.47^{+0.14}_{-0.12}$
&$ 20.85^{+0.08}_{-0.07}$&$ 21.22^{+0.06}_{-0.04}$&$ 21.49^{+0.05}_{-0.04}$
&$  69^{+  13}_{-   8}$&$ -0.35^{+0.09}_{-0.08}$\\
HD206773&$ 20.01^{+0.09}_{-0.03}$&$ 20.18^{+0.07}_{-0.15}$
&$ 20.41^{+0.06}_{-0.09}$&$ 21.09^{+0.07}_{-0.03}$&$ 21.24^{+0.05}_{-0.03}$
&$  91^{+  18}_{-  13}$&$ -0.54^{+0.07}_{-0.10}$\\
HD207198&$ 20.59\pm 0.05$&$ 20.34^{+0.10}_{-0.08}$&$ 20.79\pm 0.05$
&$ 21.28^{+0.07}_{-0.09}$&$ 21.50^{+0.05}_{-0.06}$&$  61^{+   6}_{-   5}$
&$ -0.41\pm 0.07$\\
HD207308&$ 20.66^{+0.07}_{-0.09}$&$ 20.23\pm 0.12$&$ 20.80^{+0.06}_{-0.07}$
&$ 21.20^{+0.06}_{-0.05}$&$ 21.46\pm 0.04$&$  54^{+   6}_{-   5}$
&$ -0.36^{+0.07}_{-0.08}$\\
HD207538&$ 20.60^{+0.07}_{-0.08}$&$ 20.49\pm 0.11$&$ 20.85\pm 0.06$
&$ 21.27^{+0.06}_{-0.07}$&$ 21.52^{+0.04}_{-0.05}$&$  70^{+  10}_{-   8}$
&$ -0.36\pm 0.08$\\
HD208440&$ 20.04^{+0.15}_{-0.10}$&$ 19.87^{+0.16}_{-0.12}$
&$ 20.28^{+0.11}_{-0.09}$&$ 21.24^{+0.06}_{-0.04}$&$ 21.33^{+0.05}_{-0.04}$
&$  66^{+  14}_{-  10}$&$ -0.75^{+0.12}_{-0.11}$\\
HD209339&$ 19.78^{+0.07}_{-0.17}$&$ 19.89^{+0.08}_{-0.18}$
&$ 20.13^{+0.06}_{-0.11}$&$ 21.20\pm 0.04$&$ 21.27\pm 0.04$
&$  87^{+  26}_{-  17}$&$ -0.84^{+0.08}_{-0.12}$\\
HD210809&$ 19.54\pm 0.09$&$ 19.75^{+0.17}_{-0.14}$&$ 19.97^{+0.12}_{-0.09}$
&$ 21.31^{+0.06}_{-0.05}$&$ 21.35^{+0.06}_{-0.05}$&$ 100^{+  34}_{-  19}$
&$ -1.08^{+0.13}_{-0.11}$\\
HD210839&$ 20.54\pm 0.06$&$ 20.46\pm 0.09$&$ 20.80\pm 0.05$&$ 21.24\pm 0.05$
&$ 21.48\pm 0.04$&$  72^{+   9}_{-   7}$&$ -0.38^{+0.06}_{-0.07}$\\
HD218915&$ 19.79^{+0.12}_{-0.10}$&$ 19.91^{+0.06}_{-0.07}$&$ 20.16\pm 0.06$
&$ 21.20^{+0.07}_{-0.06}$&$ 21.27^{+0.06}_{-0.05}$&$  88^{+  15}_{-  12}$
&$ -0.81\pm 0.08$\\
HD219188&$ 18.82^{+0.21}_{-0.07}$&$ 19.20^{+0.12}_{-0.09}$
&$ 19.38^{+0.11}_{-0.09}$&$ 20.72^{+0.07}_{-0.05}$&$ 20.76^{+0.06}_{-0.05}$
&$ 126^{+  49}_{-  34}$&$ -1.09^{+0.12}_{-0.11}$\\
HD220057&$ 20.06^{+0.12}_{-0.06}$&$ 19.84^{+0.11}_{-0.14}$
&$ 20.27^{+0.09}_{-0.07}$&$ 20.95^{+0.14}_{-0.23}$&$ 21.11^{+0.10}_{-0.15}$
&$  62^{+  10}_{-   7}$&$ -0.52^{+0.17}_{-0.13}$\\
HD224151&$ 20.21^{+0.03}_{-0.08}$&$ 20.31^{+0.09}_{-0.19}$
&$ 20.55^{+0.06}_{-0.10}$&$ 21.35^{+0.05}_{-0.08}$&$ 21.47^{+0.04}_{-0.06}$
&$  88^{+  13}_{-  18}$&$ -0.61^{+0.09}_{-0.11}$\\
HDE232522&$ 19.88^{+0.13}_{-0.11}$&$ 19.90^{+0.03}_{-0.17}$
&$ 20.18^{+0.08}_{-0.10}$&$ 21.12^{+0.04}_{-0.05}$&$ 21.21\pm 0.04$
&$  76^{+  18}_{-  12}$&$ -0.73^{+0.09}_{-0.11}$\\
HDE303308&$ 19.81\pm 0.08$&$ 19.97^{+0.07}_{-0.11}$&$ 20.20^{+0.05}_{-0.07}$
&$ 21.41^{+0.03}_{-0.08}$&$ 21.46^{+0.03}_{-0.07}$&$  92^{+  16}_{-  12}$
&$ -0.95^{+0.09}_{-0.08}$\\
HDE308813&$ 19.95^{+0.08}_{-0.12}$&$ 19.88^{+0.10}_{-0.05}$&$ 20.23\pm 0.07$
&$ 21.20^{+0.06}_{-0.04}$&$ 21.28^{+0.05}_{-0.04}$&$  74^{+  11}_{-   9}$
&$ -0.76^{+0.08}_{-0.09}$\\

%% file: Coefficients_data.tex
$(X/{\rm H})_{\rm ref}$\tablenotemark{a}&8.76\tablenotemark{b,c}&3.69\tablenotemark{c}&3.29\tablenotemark{c}\\[5pt]
$B_2\{ X\} $\tablenotemark{d}&$-0.13\pm 0.11$&$-0.49\pm0.006$&$-0.16$ (+0.023,$-$0.027)\\
$A_2\{ X,F_*\} $&$-0.09$ (+0.10,$-$0.09)&$-0.30\pm 0.06$&$-0.14$ (+0.22,$-$0.19)\\
$z_2\{ X,F_*\} $&0.538&0.531&0.531\\
$A_2\{ X,\log f({\rm H}_2)\} $&$0.00\pm 0.05$&$-0.03\pm 0.03$&$ 0.01$ (+0.11,$-$0.12)\\
$z_2\{ X,\log f({\rm H}_2)\} $&$-0.826$&$-0.843$&$-0.843$\\[5pt]
$B_3\{ X\} $\tablenotemark{d}&$-0.14\pm 0.14$&$-0.49\pm 0.008$&$-0.13\pm 0.034$\\
$A_3\{ X,F_*\} $&$-0.28\pm 0.13$&$-0.24$ (+0.09,$-$0.08)&$-0.16$ (+0.23,$-$0.21)\\
$z_3\{ X,F_*\} $&0.543&0.531&0.531\\
$A_3\{ X,\log f({\rm H}_2)\} $&$ 0.18\pm 0.08$&$-0.06\pm 0.05$&$ 0.11$ (+0.14,$-$0.16)\\
$z_3\{ X,\log f({\rm H}_2)\} $&$-0.801$&$-0.825$&$-0.825$\\
$A_3\{ X,\log (I/I_0)\} $&$ 0.18\pm 0.08$&$-0.04\pm 0.06$&$-0.16\pm 0.16$\\
$z_3\{ X,\log (I/I_0)\} $&0.449&0.460&0.460\\

%% file: A1_data.tex
&&\\
\cutinhead{BD+35 4258 M  $-42$ to  13 \kms}
H~I~~~~&\nodata&  21.24 ($+0.03,-0.07$)\\                      
O~I~~~~&$   6.3\pm  2.6 $&  17.55 ($+0.14,-0.21$)\\            
Mg~II~1&$  71.3\pm  3.1 $&$  16.07\pm 0.02$\\                  
Mg~II~2&$  58.9\pm  3.8 $&$  16.18\pm 0.03$\\                  
Mg~II~3&\nodata&  16.26 ($+0.04,-0.03$)\\                      
Mn~II~1&$  66.9\pm  6.2 $&$  13.68\pm 0.04$\\                  
Mn~II~2&$  45.9\pm 11.2 $&  13.71 ($+0.10,-0.13$)\\

%% file: Kr.bbl
\begin{references}

\reference{9373} Amari, S., Matsuda, J.-i., Stroud, R. M., \& Chisholm, M. F. 2013, ApJ, 778, 37
\reference{9366} Amberg, C. H., Spencer, W. B., \& Beebe, R. A. 1955, CaJCh, 33, 305
\reference{5144} André, M., Oliveira, C., Howk, J. C., et al. 2003, ApJ, 591, 1000
\reference{7052} Asplund, M., Grevesse, N., Sauval, A. J., \& Scott, P. 2009, ARA\&A, 47, 481
\reference{9154} Ayres, T. R., Lyons, J. R., Ludwig, H. G., Caffau, E., \& Wedemeyer-Böhm, S. 2013, ApJ, 765, 46
\reference{8092} Barlow, M. J., Swinyard, B. M., Owen, P. J., et al. 2013, Sci, 342, 1343
\reference{8998} Barlow, R. 2003,  ArXiv Physics e-prints.0306138
\reference{1771} Bohlin, R. C. 1975, ApJ, 200, 402
\reference{6557} Bowen, D. V., Jenkins, E. B., Tripp, T. M., et al. 2008, ApJS, 176, 59
\reference{3346} Cardelli, J. A., Meyer, D. M., Jura, M., \& Savage, B. D. 1996, ApJ, 467, 334
\reference{4254} Cardelli, J. A., \& Meyer, D. M. 1997, ApJ, 477, L57
\reference{5288} Cartledge, S. I. B., Meyer, D. M., \& Lauroesch, J. T. 2003, ApJ, 597, 408
\reference{5599} Cartledge, S. I. B., Lauroesch, J. T., Meyer, D. M., \& Sofia, U. J. 2004, ApJ, 613, 1037
\reference{6140} ---. 2006, ApJ, 641, 327
\reference{6759} Cartledge, S. I. B., Lauroesch, J. T., Meyer, D. M., Sofia, U. J., \& Clayton, G. C. 2008, ApJ, 687, 
1043
\reference{8953} Cashman, F. H., Kulkarni, V. P., Kisielius, R., Ferland, G. J., \& Bogdanovich, P. 2017, ApJS, 230, 8
\reference{1354} Chan, W. F., Cooper, G., Guo, X., Burton, G. R., \& Brion, C. E. 1992, PhRvA, 46, 149
\reference{9220} Cruz-Diaz, G. A., Martín-Doménech, R., Moreno, E., Muñoz Caro, G. M., \& Chen, Y.-J. 2018, 
MNRAS, 474, 3080
\reference{5036} de Avillez, M. A., \& Mac Low, M. M. 2002, ApJ, 581, 1047
\reference{2712} Diplas, A., \& Savage, B. D. 1994, ApJS, 93, 211
\reference{9195} Esteban, C., \& García-Rojas, J. 2018, MNRAS, 478, 2315
\reference{5138} Gondhalekar, P. M. 1985, ApJ, 293, 230
\reference{1117} Harris, A. W., Gry, C., \& Bromage, G. E. 1984, ApJ, 284, 157
\reference{9038} Heidarian, N., Irving, R. E., Federman, S. R., et al. 2017, JPhB, 50, 155007
\reference{9368} Hohenberg, C. M., Thonnard, N., \& Meshik, A. 2002, Meteoritics and Planetary Science, 37, 257
\reference{600} Holloway, J. 1968, Noble-Gas Chemistry (London: Methuen Publishing)
\reference{1312} Jenkins, E. B. 1971, ApJ, 169, 25
\reference{1063} Jenkins, E. B., Savage, B. D., \& Spitzer, L. 1986, ApJ, 301, 355
\reference{3184} Jenkins, E. B. 1996, ApJ, 471, 292
\reference{6999} ---. 2009, ApJ, 700, 1299
\reference{7294} Jenkins, E. B., \& Tripp, T. M. 2011, ApJ, 734, 65
\reference{7760} Jenkins, E. B. 2013, ApJ, 764, 25
\reference{1388} ---. 1987, in Interstellar Processes, eds. D. J. Hollenbach, \& H. A. Thronson, Jr. (Dordrecht: 
Reidel), 533-559
\reference{6559} Jensen, A. G., \& Snow, T. P. 2007a, ApJ, 669, 401
\reference{6558} ---. 2007b, ApJ, 669, 378
\reference{7108} Jensen, A. G., Snow, T. P., Sonneborn, G., \& Rachford, B. L. 2010, ApJ, 711, 1236
\reference{6496} Kelly, B. C. 2007, ApJ, 665, 1489
\reference{604} Lewis, R. S., Srinivasan, B., \& Anders, E. 1975, Sci, 190, 1251
\reference{5604} Lodders, K. 2003, ApJ, 591, 1220
\reference{9350} Marrocchi, Y., Razafitianamaharavo, A., Michot, L. J., \& Marty, B. 2005, GeCoA, 69, 2419
\reference{3271} Mathis, J. S., Mezger, P. G., \& Panagia, N. 1983, A\&A, 128, 212
\reference{5172} McCandliss, S. R. 2003, PASP, 115, 651
\reference{2858} Meyer, D. M., Jura, M., Hawkins, I., \& Cardelli, J. A. 1994, ApJ, 437, L59
\reference{4196} Meyer, D. M., Jura, M., \& Cardelli, J. A. 1998, ApJ, 493, 222
\reference{9372} Mitchell, J. B. A. 1990, PhR, 186, 215
\reference{9371} Mitchell, J. B. A., Novotny, O., LeGarrec, J. L., et al. 2005, JPhB, 38, L175
\reference{3719} Morton, D. C. 2000, ApJS, 130, 403
\reference{5404} Morton, D. C. 2003, ApJS, 149, 205
\reference{1053} Murray, M. J., Dufton, P. L., Hibbert, A., \& York, D. G. 1984, ApJ, 282, 481
\reference{9367} Nichols, R. H., Jr., Nuth, J. A., III, Hohenberg, C. M., Olinger, C. T., \& Moore, M. H. 1992, Metic, 
27, 555
\reference{7593} Nieva, M. F., \& Przybilla, N. 2012, A\&A, 539, 143N
\reference{6880} Pauzat, F., \& Ellinger, Y. 2007, JChPh, 127, 014308
\reference{8714} Pauzat, F., Ellinger, Y., Pilmé, J., \& Mousis, O. 2009, JChPh, 130, 174313
\reference{8390} Poteet, C. A., Whittet, D. C. B., \& Draine, B. T. 2015, ApJ, 801, 110
\reference{3980} Richter, P., Sembach, K. R., Wakker, B. P., et al. 2001, ApJ, 559, 318
\reference{9153} Ritchey, A. M., Federman, S. R., \& Lambert, D. L. 2018, ApJS, 236, 36
\reference{8841} Robinson, E. L. 2016, Data Analysis for Scientists and Engineers (Princeton: Princeton U. Press)
\reference{6831} Rolleston, W. R. J., Smartt, S. J., Dufton, P. L., \& Ryans, R. S. I. 2000, A\&A, 363, 537
\reference{2896} Roy, J. R., \& Kunth, D. 1995, A\&A, 294, 432
\reference{1141} Savage, B. D., Bohlin, R. C., Drake, J. F., \& Budich, W. 1977, ApJ, 216, 291
\reference{1142} Savage, B. D., \& Bohlin, R. C. 1979, ApJ, 229, 136
\reference{110} Savage, B. D., \& Sembach, K. R. 1991, ApJ, 379, 245
\reference{328} ---. 1996, ARA\&A, 34, 279
\reference{9025} Savage, B. D., Kim, T.-S., Fox, A. J., et al. 2017, ApJS, 232, 25
\reference{9352} Schelhaas, N., Ott, U., \& Begemann, F. 1990, GeCoA, 54, 2869
\reference{8689} Schilke, P., Neufeld, D. A., Müller, H. S. P., et al. 2014, A\&A, 566, A29
\reference{181} Sembach, K. R., \& Savage, B. D. 1992, ApJS, 83, 147
\reference{9365} Snow, T. P. 1983, ApJ, 269, L57
\reference{4971} Snow, T. P., Rachford, B. L., \& Figoski, L. 2002, ApJ, 573, 662
\reference{8560} Steffen, M., Prakapavičius, D., Caffau, E., et al. 2015, A\&A, 583, A57
\reference{9152} Sterling, N. C. 2011, A\&A, 533, A62
\reference{8267} Theis, R. A., Morgan, W. J., \& Fortenberry, R. C. 2015, MNRAS, 446, 195
\reference{6599} Thom, C., Peek, J. E. G., Putman, M. E., et al. 2007, ApJ, 684, 364
\reference{9155} Toner, A., \& Hibbert, A. 2005, MNRAS, 361, 673
\reference{5160} Tripp, T. M., Wakker, B. P., Jenkins, E. B., et al. 2003, AJ, 125, 3122
\reference{7130} Voshchinnikov, N. V., \& Henning, T. 2010, A\&A, 517, A45
\reference{9351} Wacker, J. F. 1989, GeCoA, 53, 1421
\reference{6552} Wakker, B. P., York, D. G., Howk, J. C., et al. 2007, ApJ, 670, L113
\reference{8566} Wang, S., Li, A., \& Jiang, B. W. 2015, MNRAS, 454, 569
\reference{3471} Welsh, B. Y., Sasseen, T., Craig, N., Jelinsky, S., \& Albert, C. E. 1997, ApJS, 112, 507
\reference{7092} Whittet, D. C. B. 2010, ApJ, 710, 1009
\reference{601} Wyatt, J. R., Strattan, L. W., Snyder, S. C., \& Hierl, P. M. 1975, JChPh, 62, 2555
\reference{3352} Yang, J., \& Anders, E. 1982a, GeCoA, 46, 861
\reference{3351} Yang, J., Lewis, R. S., \& Anders, E. 1982, GeCoA, 46, 841
\reference{3353} Yang, J., \& Anders, E. 1982b, GeCoA, 46, 877
\end{references}
